\documentclass[referee]{raa}            

\usepackage{graphicx,times}             
\usepackage{natbib}
\usepackage{tabularx}
\usepackage{siunitx}
\usepackage{amsmath,amssymb,amsfonts}
\usepackage{fontspec}
\bibpunct{(}{)}{;}{a}{}{,}

\usepackage[pagebackref=true]{hyperref}

\begin{document}
  \title{CPI-C: Cool Planet Imaging Coronagraph on Chinese Space Station Survey Telescope
}

   \volnopage{Vol.0 (20xx) No.0, 000--000}      
   \setcounter{page}{1}          

   \author{
   Jiangpei Dou  \inst{1, 2} \and
   Xi Zhang \inst{1, 2} \and
   Gang Zhao \inst{1, 2} \and
   Mingming Xu \inst{1, 2} \and
   Zhen Wu \inst{1, 2} \and
   Gang Wang \inst{1, 2} \and
   Baoning Yuan \inst{1, 2} \and
   Lingyi Kong \inst{1, 2} \and
   Yiming Zhu \inst{1, 2} \and
   Bingli Niu \inst{1, 2} \and  
   Zhonghua Lv \inst{1, 2} \and   
   Yongjun Qi \inst{1, 2} \and
   Shu Jiang \inst{1, 2} \and
   Bo Chen \inst{1, 2} \and
   Wei Guo \inst{1, 2} \and
   Di Wang \inst{1, 2} \and
   Yinglu Lin \inst{1, 2} \and
   Liping Zheng \inst{1, 2} \and
   Jing Guo \inst{1, 2} \and
   Ruokun Li \inst{1, 2} \and
   Liyan Xu \inst{1, 2} \and
   Huihai Wu \inst{1, 2} \and
   Cheng Wen \inst{1, 2} \and
   Shuwei Miao \inst{1, 2} \and
   Boyang Lv \inst{1, 2} \and
   Weimiao Li \inst{1, 2}
   }

   \institute{Nanjing Institute of Astronomical Optics $\&$ Technology, Chinese Academy of Sciences, Nanjing 210042, China; 
 {\it jpdou@niaot.ac.cn; xzhang@niaot.ac.cn; gzhao@niaot.ac.cn; mingxu@niaot.ac.cn}\\
        \and
             CAS Key Laboratory of Astronomical Optics $\&$ Technology, Nanjing Institute of Astronomical Optics $\&$ Technology, Nanjing 210042,China\\
\vs\no
   {\small Received 20xx month day; accepted 20xx month day}}

\abstract{Cool Planet Imaging Coronagraph (CPI-C) on Chinese Space Station Survey Telescope (CSST) is proposed to direct image the cool planets around nearby solar-type stars (within 40 pc). The core scientific objective of CPI-C is to conduct high-contrast directly imaging surveys of exoplanets ranging in size from Neptune-like to Jupiter-like, located at separations of 0.5 to 5 AU from their host stars, and to perform systematic spectroscopic analysis of the detected planets through high-precision multi-band photometry. CPI-C employs a step-transmission apodization technique to suppress the diffraction noises from the telescope pupil and a precise phase correction technique to eliminate the speckle noises due to imperfections of the optical surfaces. The contrast requirement is better than $10^{-8}$ at an inner working angle (IWA) of $3-4\lambda/D$, in the visible wavelength from 600 nm to 900 nm. CPI-C will be the first space-based instrument capable of directly imaging the reflection light from the cool exoplanets in the visible wavelength enabling the measurement of key physical parameters such as the effective temperature, surface gravity, radius, mass, and other key parameters. The potential observation results will significantly contribute to further understand the formation and evolution mechanisms of planets, which will also lay a solid foundation for future confirmation of the Earth-twins in the next generation space flagship missions.
}

   \authorrunning{J.-P. Dou \it{et al.} }            
   \titlerunning{CPI-C on CSST}  

   \maketitle
%
%
\section{Executive Summary}
\subsection{CPI-C’s Core Scientific Goals}
Since the 1990s, more than six thousand exoplanets have been discovered, and related research has become a major focus in contemporary astronomy \citep{1995Natur.378..355M,2000ApJ...529L..45C,2004ApJ...606L.155B}. The primary methods for exoplanet detection include indirect techniques such as radial velocity (RV) measurements, transit photometry, microlensing, timing, and astrometry, as well as the direct imaging (DI) method~\citep{liu20, 2021ARA&A..59..291Z,2022RAA....22g2003J,2024ChJSS..44..193J,cjss2024012024yg01}. Direct imaging, which can truly ``see'' the planets, makes it possible to study the atmospheric signatures of planets and thus becomes a crucial technology to confirm the extraterrestrial life signals, such as the  $\mathrm{O_2}$ and the reduced gases like $\mathrm{CH_4}$ or $\mathrm{N_2O}$  ~\citep{1975RSPSB.189..167L,1965Natur.207..568L,2007ApJ...658..598K}. This work ultimately aims to address one of the most fundamental scientific questions: ``Are we alone in the universe?''. 

To date, ground-based exoplanet imagers have discovered approximately forty exoplanets, mostly with a mass above 3 $M_\mathrm{J}$ and a semi-major axis over 10 AU. The vast majority of these planetary systems are very young (<200 Myr), with effective temperatures approaching or exceeding 1000 K~\citep{2008Sci...322.1348M,2010Natur.468.1080M,2015Sci...350...64M,2015ApJ...802...12D,2021RAA....21...82Z}. These young planets still emit strong infrared radiation, enabling detection under moderate contrast conditions, typically ranging from $10^{−4}$ to $10^{−6}$ . However, current ground-based facilities remain unable to detect cooler or more mature planets due to limitations in achievable imaging contrast. This contrast limitation is primarily caused by residual wavefront errors from adaptive optics (AO) systems induced by atmospheric turbulence and the constraints of available observational wavelength windows, thereby restricting both the number and diversity of exoplanets detectable via direct imaging~\citep{2010RAA....10..189D}. 
CPI-C on CSST will fully take the advantage of the space environment, which is free from atmospheric turbulence and offers broader observational wavelength windows. By integrating extreme high-contrast imaging techniques\citep{2007Natur.446..771T,2016ApJ...832...84D}, CPI-C is expected to overcome current limitations in observational contrast. This enhanced contrast capability will enable, for the first time, the detection and characterization of "cool" exoplanets orbiting solar-type stars (F, G, K). 

To achieve the contrast on the order of $10^{−8}$ and beyond, it is essential to mitigate diffraction-induced photon noise originating from the telescope pupil, as well as speckle noise caused by imperfections in both the telescope optics and the coronagraph~\citep{2011RAA....11..198D}. To this end, CPI-C employs a step-transmission apodized pupil design to suppress diffraction~\citep{2006ApJS..167...81G,ren07,ren10,2010RAA....10..189D,2011PASP..123..341R,2012RAA....12..591Z,2016ApJ...832...84D,Shen20} and a ``one-step''  precise phase aberration control technique to eliminate speckles in the focal plane point spread function (PSF)~\citep{2011RAA....11..198D,2012PASP..124..247R,dou14,2016ApJ...832...84D,2019OptEn..58a4102R,2020PASJ...72...30R,2021RAA....21...82Z,zhang22}. 

CPI-C will improve the imaging contrast by two orders of magnitude compared to current ground-based performance, enabling the first direct imaging of the reflection light from ``cool'' exoplanets in the visible wavelength orbiting nearby solar-type stars. With total integrated exposure times ranging from 30 seconds to one hour or longer, a sufficient signal-to-noise ratio (S/N$\gtrsim5\sigma$) can be achieved for planets with magnitudes between 20th and 25th orbiting stars with magnitudes between 0th and 5th in the V-band. Specifically, a 20th magnitude planet requires approximately 30 seconds of exposure, while fainter planets approaching 25th magnitude require exposures of one hour or longer. Once planet candidates reach adequate S/N, multi-band photometry from the visible to the near-infrared will be conducted, followed by spectroscopic analysis to derive their physical properties. Multi-band photometric data will allow reconstruction of the reflected light energy distribution from the target planet. By fitting these data with reflection spectra,  key physical parameters such as the cloud properties, atmospheric abundances, radius, effective temperature, surface gravity, and mass can be retrieved \citep{lacy19,2016ApJ...832...84D}. Additionally, through repeated observations, the planet's orbital parameters can be determined, allowing further constraints on its dynamical mass\citep{bowler18}. Based on CPI-C’s detection capabilities, the mission is expected to detect planetary systems ranging from Jupiter-sized to Neptune-sized planets at separations of 0.5 to 5 AU from their host stars. In addition to detection sensitivity, approximately 700 nearby stars have been selected as observational targets for CPI-C, taking into account satellite Control Moment Gyroscope (CMG) operational constraints and potential scientific return.
  
The synergistic use of multiple detection methods is essential for confirming the exoplanets and achieving a comprehensive understanding of their physical properties. RV and transit methods are particularly effective for detecting planets with short orbits (on the order of one hundredth of an AU). CPI-C will for the first time offer a high contrast on the order of $10^{−8}$ , thus enabling the follow-up observations of several RV detected exoplanets, while it is impossible for current ground observations due to the contrast limit. With more astrometric information obtained through direct imaging, we can obtain the three-dimensional orbital information, which will further constrain the mass of these planetary systems, when combined with the RV data. We have selected approximately twenty exoplanets within the detectable range of CPI-C. Upon confirmation of these candidates, additional astrometric observations will be conducted to derive three-dimensional orbital parameters, thereby tightly constraining the masses of these planetary systems. Notably, combining RV and DI measurements resolves the $\sin i$ degeneracy, yielding true dynamical masses and more precise orbital constraints: RV data anchor long-term accelerations, while DI provides inclination and sky-plane geometry ~\citep{2019AJ....158..140B}. 

CPI-C will also conduct direct imaging observations of circumstellar disks, including protoplanetary disk and debris disks. Using CPI-C, direct imaging observations of circumstellar disks can be carried out to obtain high-precision images of scattered light from the disks in the optical and near-infrared bands. Structures created by the planetary perturbations on the disk can be identified in the scattered light images. By studying these structures, we can potentially infer and directly image the presence of planets in the system. By obtaining information on the composition of the disk, such as dust and planetesimals, we can systematically understand the main components of planets, the structural characteristics of exoplanetary systems, and information about the mass of potential planets within the disk.

The transit timing variation (TTV) has proven to be a powerful technique to discover exoplanets massively. In recent research, Kepler-725 c, a super-earth exoplanet within the HZ of the late G-type dwarf Star is firstly discovered by using the TTV~\citep{2025NatAs...9.1184S}. The short-period exoplanets that are easily to be detected by the transit method are likely to co-exist with the cold giant planets in orbits of several AU~\citep{2021ARA&A..59..291Z}. When these outer cold giants lie in or near high-order mean-motion resonances (MMRs) with the inner transiting planets, they can induce significant TTV signals~\citep{2022MNRAS.512.3113B,2025AJ....169..342W}. This makes TTV-identified systems containing both inner and outer planets highly suitable candidates for observation with CPI-C.A key scientific goal is to perform the first simultaneous atmospheric characterization of such systems: obtaining reflection spectra of cold giant planets with CPI-C, alongside transmission and emission spectra of the inner short-period gas giants from ground- and space-based facilities. Transmission and emission spectroscopy will provide precise measurements of atmospheric molecular abundances, C/O ratios, and thermal structures for the inner hot Jupiters~\citep{2023Natur.614..659R}, while reflection spectroscopy will reveal cloud properties and atmospheric compositions of the outer cold giants. By targeting dynamically linked planet pairs, this approach enables direct atmospheric comparisons across vastly different orbital and thermal environments. Such research will provide unique insights into the formation and migration history of hot Jupiters through systematic spectral comparisons with their cold giant companions.

Beyond following-up on specific discoveries from RV surveys, the broader scientific program of CPI-C will leverage synergies across an even wider range of detection techniques, especially the astrometry measurements. In astrometry, absolute proper-motion accelerations derived from the Hipparcos-Gaia cross-calibration (HGCA) efficiently identify stars hosting unseen companions; direct imaging can then confirm and characterize these objects~\citep{2021ApJS..254...42B}-e.g.,such as the prediction and subsequent imaging confirmation of HIP 99770 b~\citep{2023Sci...380..198C}. Conversely, once planetary positions are obtained through direct imaging, the three-dimensional orbital parameters and the planet’s mass can be determined with greater reliability. Throughout its mission, CPI-C will also observe planets in binary and multi-star systems, as well as trans-Neptunian objects within the solar system~\citep{Parker_2016,2025ApJ...991L..34P,2025PrA....43..100H}.

Finally, CPI-C will conduct a statistically analysis of both detections and non-detections within the complete survey sample. CPI-C will be capable of directly imaging mature exoplanets with effective temperatures below 800 K located within 5 AU of their host star in the target system. The direct imaging technique offers distinct advantages in detecting planetary systems with long orbital periods, thereby complementing indirect detection methods in terms of observational targets. It is anticipated to significantly expand the current exoplanet sample and enhance our understanding of the formation mechanisms of planetary systems~\citep{1996Icar..124...62P,2002ApJ...576..870P}. Long-period giant planets identified through this method will be particularly valuable for refining theoretical models of planet formation, including those based on core accretion, gravitational instability, and orbital migration. However, due to the limited availability of precise physical parameters for mature planets by imaging, existing evolutionary models still face notable limitations and substantial uncertainties in mass estimation. By comparing planetary systems at all-age stages, direct imaging with CPI-C will provide critical observational constraints for refining planetary evolution models. The expected results will enrich the number and diversity of detected exoplanets and provide important observational evidence to understand planetary formation and evolution mechanisms. The systematic study of the occurrence rates, orbital distributions, and atmospheric diversity of giant planets of all ages around nearby solar-type stars will create opportunities for comparative planetary science research. This work will ultimately lay a solid foundation for the future direct detection and spectroscopic analysis of ``Earth-mass'' planets in the habitable zone (HZ) around solar-type stars.
 
\subsection{CPI-C Scientific Instrument Introduction}
\begin{figure}
   \centering
   \includegraphics[width=13cm, angle=0]{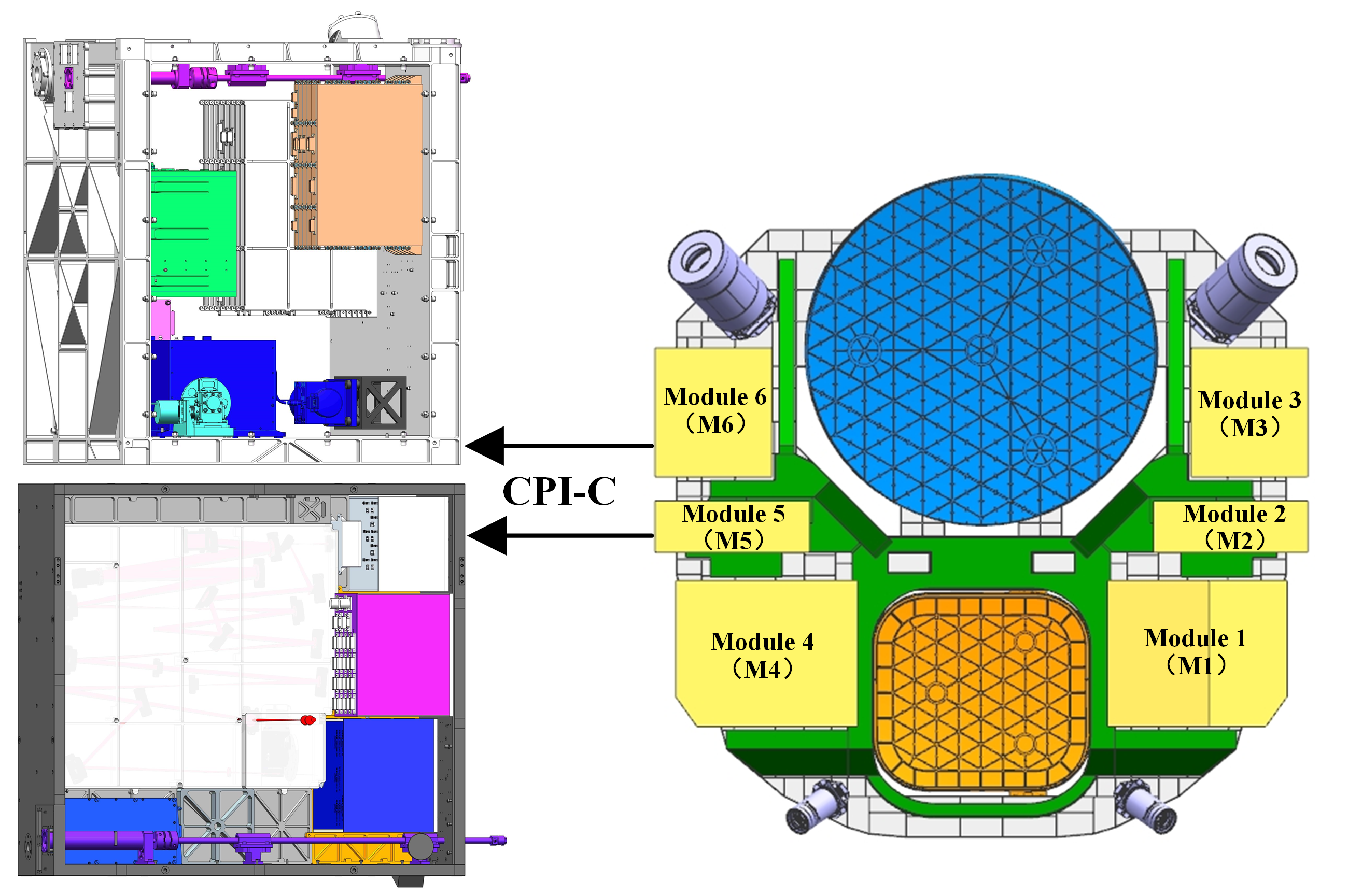}
   \caption{Schematic layout of the CPI-C instrument.} 
   \label{fig:Layout of CPI-C}
\end{figure}

The payload of CPI-C adopts a split layout and occupies two positions of Module5 (M5) and Module6 (M6) on the CSST. According to their functions, the two positions are respectively named as the Scientific Detection Unit (SDU) in M5 and the Control and Power Supply Unit (CPSU) in M6, as shown in Figure \ref{fig:Layout of CPI-C}. The SDU internally integrates the main optics of CPI-C, as well as a visible camera (VIS Cam), a Shack-Hartmann wavefront sensing camera (SH-WFS) and other electronic units that undertake core functions such as data acquisition and processing, and drive control of a deformable mirror and a tip-tilt mirror. The CPSU is mainly responsible for the communication between CPI-C and CSST, the overall temperature control function of the module, the internal power supply and distribution of the module, and the cooling control for the three cameras of CPI-C. Due to the payload space constraints in the SDU, the near-infrared imaging optics, including the near infrared camera (NIR Cam) and its filter wheel, are also placed in the CPSU.

The VIS Cam and the SH-WFS both deploy imaging sensors of the type electron-multiplying CCD (EMCCD), to achieve sufficient SNR for faint targets. The NIR Cam uses a scientific-grade InGaAs detector to guarantee its low-noise performance across the wavelength range of 0.9\textasciitilde1.6$\mu$m. 

To achieve the goal of high-contrast imaging at the order of $10^{−8}$, CPI-C employs a 31-step transmission filter for pupil apodization and a kilo-actuator deformable mirror (kilo-DM) to precisely control the phase aberrations from the imperfect optics. The system's measured contrast can reach an order of $10^{−6}$ after precise calibration of the CPI-C optics. The imaging contrast can be further improved by more than two orders of magnitude after the wavefront correction by the kilo-DM. Ultimately, a high-contrast image is obtained in a specific working area within the focal plane PSF. A series of high-contrast imaging and processing algorithms and techniques, including ADI/LOCI, IRS/O-IRS, and G-RDI~\citep{marois06,lafreniere07,ren12,2015ApJ...802...12D,2021MNRAS.502.2158R}, will be employed to further enhance the final observational contrast, which provides a sufficient signal-to-noise ratio (S/N) to support spectral analysis and studies of the physical properties of planets.

\section{Scientific Goals} \label{sec:secience}
\subsection{High-contrast imaging survey of cool planets around nearby stars}
A high-contrast imaging survey of nearby bright solar-type stars is one of the most important observational missions for CPI-C. To date, current ground-based exoplanet imagers have detected approximately 40 (planetary-mass companions with masses below 13 $M_\mathrm{J}$~\citep{2016PASP..128j2001B,2019AJ....158...13N}, which is the commonly adopted deuterium-burning limit distinguishing planets from brown dwarfs~\citep{2001RvMP...73..719B}. Among these planets, their masses are mostly above 10 $M_\mathrm{J}$, and they are predominantly very young, around 1--200 million years (Myr), and the orbital separation is far from their host stars, with a semi-major axes mostly ranging from tens to hundreds of AU. Representing the state-of-the-art of the ground-based instruments that dedicated for exoplanet imaging, such as the Gemini/GPI and the VLT/SPHERE, are all equipped with extreme adaptive optics and coronagraphs on 8-meter class telescopes. These instruments achieved the first light around year of 2014--2015 and have a routine observation since then. In the infrared bands, they have achieved an observational imaging contrast on the order of $10^{-6}$, at angular separation of 0.4--0.5 arcseconds, which represents the best observation contrast performance~\citep{2014PNAS..11112661M,2019A&A...631A.155B}.

Space-based telescopes also contribute to exoplanet imaging and disk observations. The Hubble Space Telescope (HST) has successfully imaged debris disks and conducted follow-up observations of known exoplanets, though limited by contrast ($\sim$ $10^{-5}$ at visible wavelengths) and inner working angle constraints~\citep{2008Sci...322.1345K,2011ApJ...741...55S}. The James Webb Space Telescope (JWST), launched in 2021, provides coronagraphic capabilities through its NIRCam (0.6–5 µm) and MIRI (5–28 µm) instruments~\citep{2007SPIE.6693E..0HK,2022A&A...667A.165B}. JWST primarily targets thermal emission from young, self-luminous giant planets in the near- to mid-infrared, achieving contrasts of $\sim 10^{-4.5}$ and $\sim 10^{-5}$ at IWA of 0.6$^{\prime\prime}$ and 1$^{\prime\prime}$, respectively~\citep{2023ApJ...951L..20C}. Such a contrast and IWA limit the primarily observation and discovery for young self-luminous planets in the mid-infrared. The upcoming Nancy Grace Roman Space Telescope will carry a coronagraph demonstrating starlight suppression techniques for future missions, with an anticipated contrast of $\sim$ $10^{-7}$ in the near-infrared~\citep{2020SPIE11443E..1UK} expected to deliver a contrast on the order of 10-8 with the support of advanced data reduction algorithms such as RDI..

CPI-C distinguishes itself from these facilities through several unique capabilities: (1) achieving deeper contrast ($10^{-8}$) in the visible to near-infrared bands (0.6--1.6 $\mu$m), crucial for detecting reflected light from cool, mature planets; (2) operating at smaller inner working angles (3–4 $\lambda$/D, or $\sim0.2^{\prime\prime}$ at 661 nm), enabling observations of planets at closer orbital separations (0.5–5 AU); (3) focusing specifically on the critical reflected-light regime where Neptune- to Jupiter-sized planets at several hundred Kelvin remain detectable, a parameter space largely inaccessible to current space- and ground-based facilities.

Between 2014 and 2021, we have successfully developed a high-contrast exoplanet imaging instrument, composed of a portable adaptive optics (PAO) system and a coronagraph optimized in near infrared wavelength. Interfaced with the international 4-meter class telescopes, including the 3.6-meter New Technology Telescope (NTT) in La Silla Observatory, and the Astrophysical Research Consortium (ARC) 3.5-meter telescope in Apache Point Observatory, we have successfully recovered the high-contrast image of the $\kappa$ And b ~\citep{2021RAA....21...82Z}. Among the observation from recent ground-based facilities, very few new exoplanets have been discovered, and no cool or mature exoplanets have been detected.

\begin{figure*} 
   \centering
   \includegraphics[width=13.0cm, angle=0]{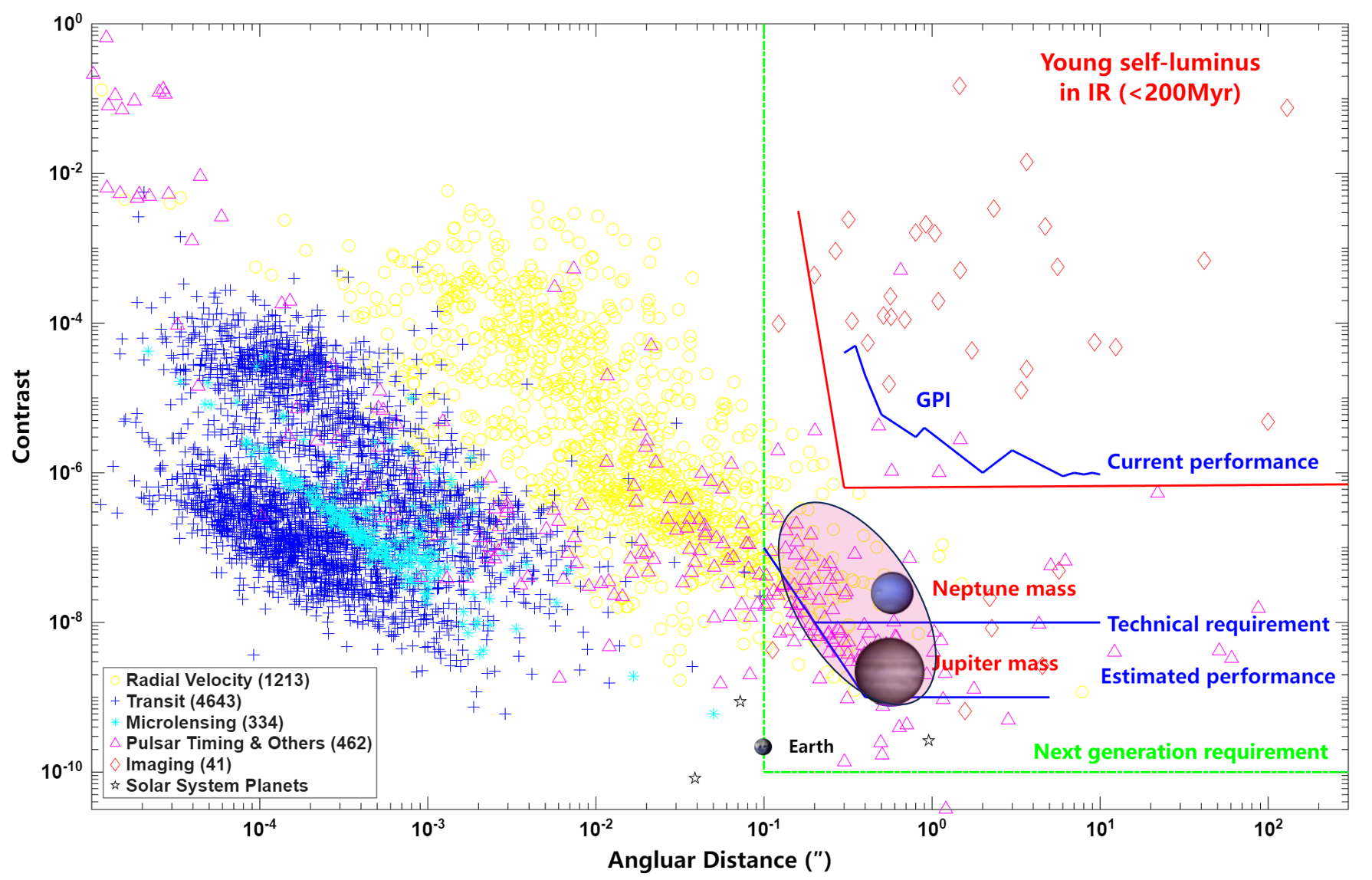}
   \caption{The detection capability of the direct imaging of cool planets for the CPI-C mission.} 
   \label{fig:cnt}
\end{figure*}

Therefore, CPI-C will fully take advantage of the space environment, combined with the extreme high-contrast technique \citep{2007Natur.446..771T,2016ApJ...832...84D}, will make a breakthrough of the current contrast limitation. Figure~\ref{fig:cnt} shows the detection capability of the direct imaging of cool planets for the CPI-C mission, comparing with current high-contrast imaging performance, indicating that CPI-C will be capable of discovering the cool Neptune-size to Jupiter-size exoplanets around nearby stars. The data source of the distribution of contrast and the angular separation of the discovered exoplanets is from the EU website (http://exoplanet.eu/). 

\begin{figure} 
   \centering
   \includegraphics[width=13cm, angle=0]{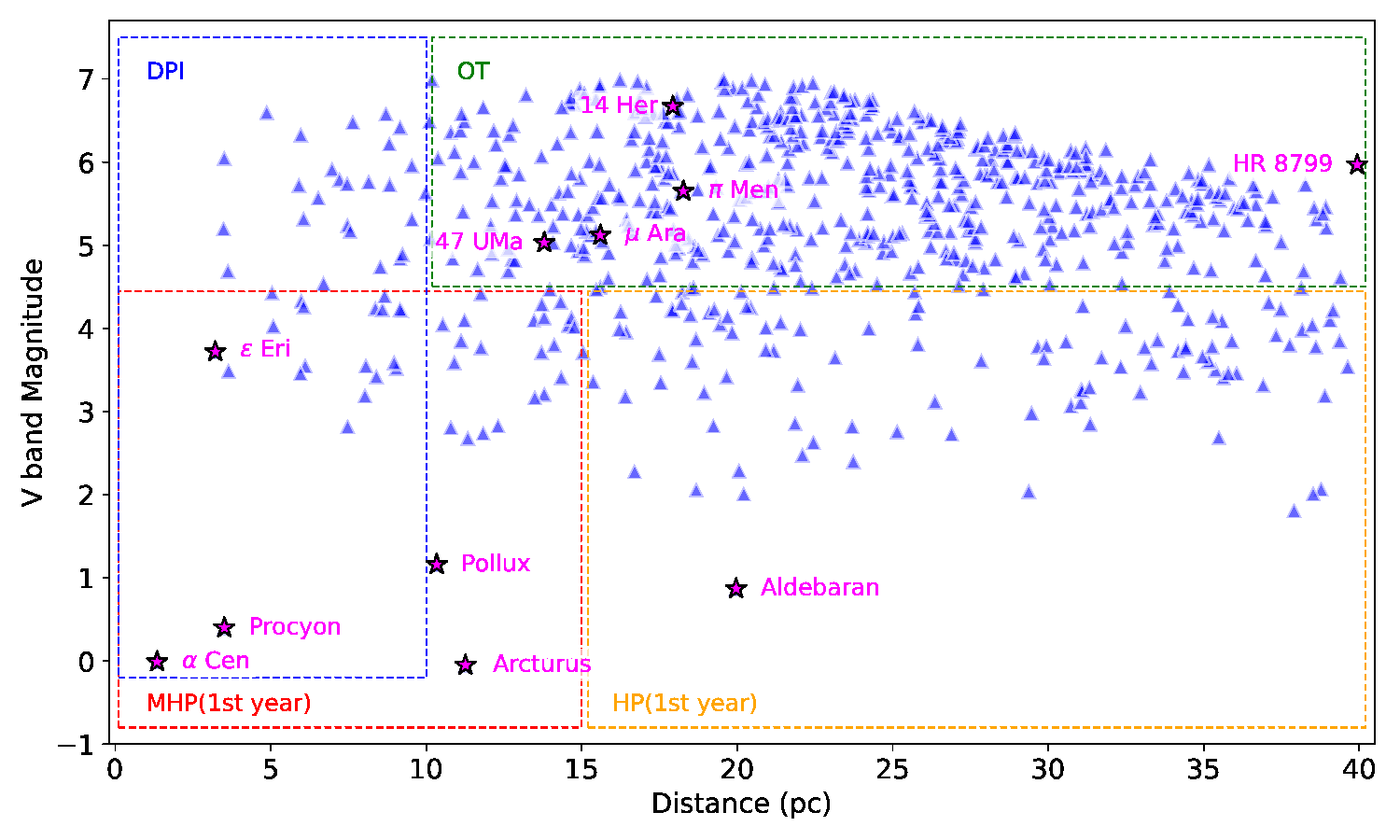}
   \caption{ Here we classify star candidates into the most high priority (MHP), high priority (HP), deep-imaging (DPI), and other targets (OT), respectively. Some of the early science targets have been marked with pentagrams.} 
   \label{fig:alltargets}
\end{figure}

We have selected approximately 700 nearby stars as the potenial observation candidates for CPI-C. These stars are selected within 40 pc and $V<7$ mag from the Hipparcos catalog, according to the detection capability of CPI-C \citep{2007A&A...474..653V}. For each star we summarize the spectral type, $V$ magnitude and equatorial coordinates for visibility and scheduling, and computed synthetic magnitudes in the CPI-C bands by convolving stellar spectra with the measured end-to-end throughput of each filter. We then assigned a survey priority based on a simple performance metric that scales with distance and the sensitivity, in order to fulfil a high scientific yield. As illustrated in Figure~\ref{fig:alltargets}, the star candidates for observation is classified into four sets, as the most-high-priority (MHP), high-priority (HP), deep-imaging (DPI), and other targets (OT), which covers the 10-year full mission timeline. The classification criteria are based on stellar distance and brightness: MHP (distance < 15 pc, V < 4.5 mag) and HP (15–40 pc, V < 4.5 mag) targets are bright nearby stars prioritized for early observations; DPI (< 10 pc, V up to 7.5 mag) includes very nearby but fainter stars suitable for deep exposures; OT comprises remaining targets for later mission phases. The observation targets will be selected from these sets accordingly. A typical target list is provided and shown in Appendix~\ref{sec:app}; readers  interested in the complete list are welcome to contact the authors directly.

In addition to exoplanets, the survey will include circumstellar disks, and selected Solar System small-body binaries as ancillary scientific targets, enabling comprehensive studies of planetary system architectures and disk–planet interactions (Bao et al., submitted). Each target will be observed multiple times over the survey to ensure robust planet detection and characterization. The first epoch serves as a discovery image to identify candidate point sources, followed by one or more confirmation epochs separated by months to a year to distinguish co-moving companions from background stars. For confirmed companions, additional visits are used to obtain high signal-to-noise photometry in all seven CPI-C bands and to measure accurate relative astrometry along the orbit.

\subsection{Follow-up Observation of Detected Exoplanets}
Cross-verification using multiple detection methods has always been an important and powerful approach in the discovery and characterization of exoplanets. The combination of the transit and the RV method is a good example. The transit method provides the planet-to-star radius ratio, while the RV method can measure the planet's minimum mass due to orbital inclination uncertainty. Combining the two methods allows for better determination of the planet's accurate mass and density, thereby confirming its internal structure. Currently, the radii and masses of dozens of exoplanets have been measured relatively accurately, due to the imporvement of the measurement precision of both techniques. Their internal structures can thus be preliminarily constrained by bulk density, providing crucial physical parameters and reference benchmarks for atmospheric characterization observations. However, the current capabilities of the two techniques favors the short-orbit or massive planets, in which the orbital periods is only tens of Earth days~\citep{2017A&A...603A..54L}.

Follow-up high-contrast imaging observations of planetary systems discovered via the RV method represents a very important cross-verification~\citep{2017A&A...603A..54L,2018ApJS..239...31B,2022BAAS...54e.226N}. Due to the contrast limitation of the ground-based observations, such an study could only be performed on brown dwarfs. A representative case demonstrating the combination of RV and direct imaging methods is shown below to analyze the mass of the brown dwarf GL 758 B \citet{bowler18}. This study integrated thirty years of radial velocity data and direct imaging data, measuring GL 758 B's orbital semi-major axis as 21.1$^{+2.7}_{-1.3}$ AU and orbital period as 96$^{+21}_{-9}$ years. This yielded a dynamical mass of 35--61 Jupiter masses. This mass estimate differs significantly from the 10--40 Jupiter masses predicted by different brown dwarf evolutionary models (COND, SM, \citealt{2001RvMP...73..719B}), suggesting that brown dwarf evolutionary models or age estimates require further refinement. 

\begin{figure*} 
  \centering
  \includegraphics[width=0.3\textwidth]{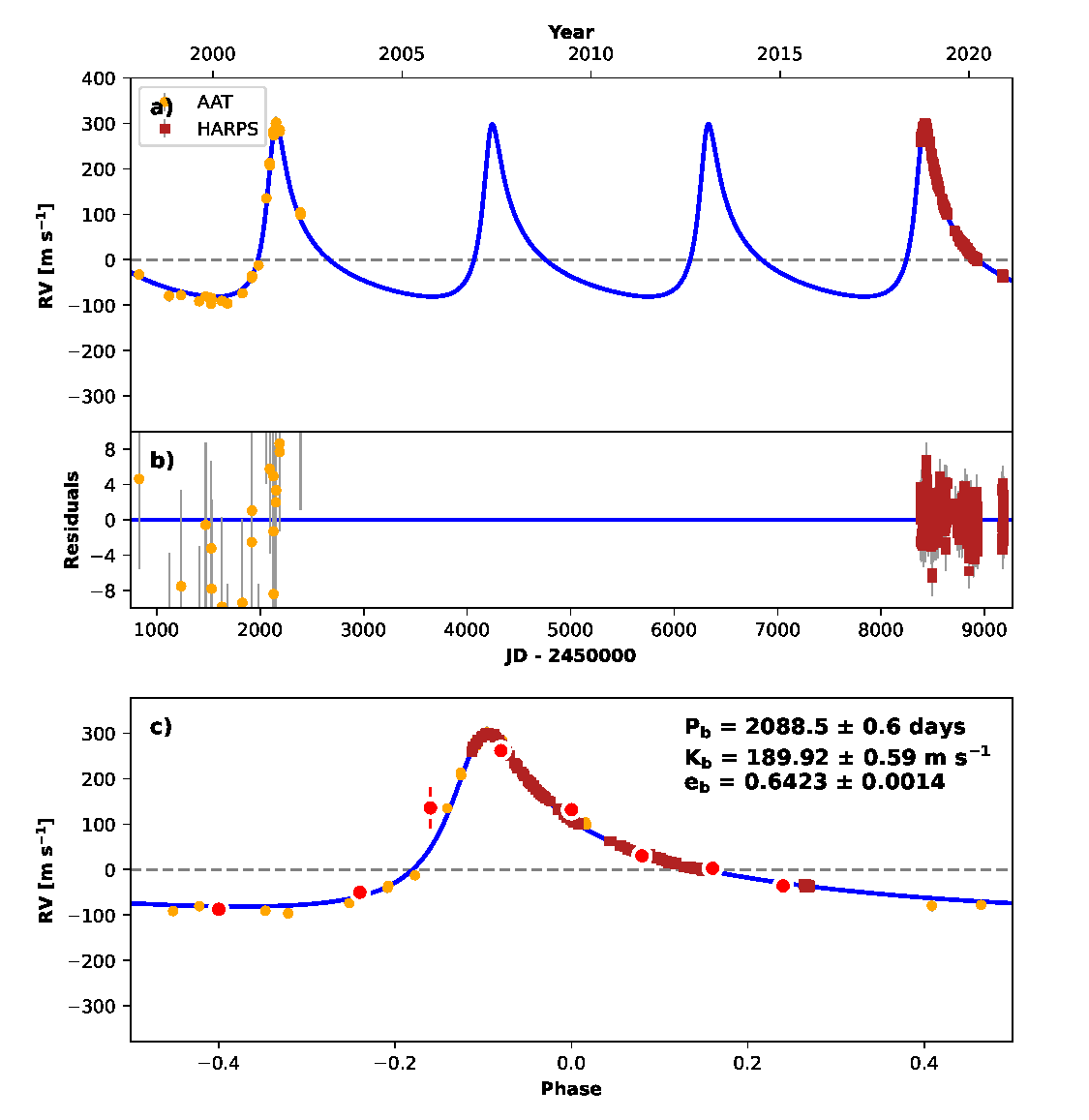}
  \includegraphics[width=0.3\textwidth]{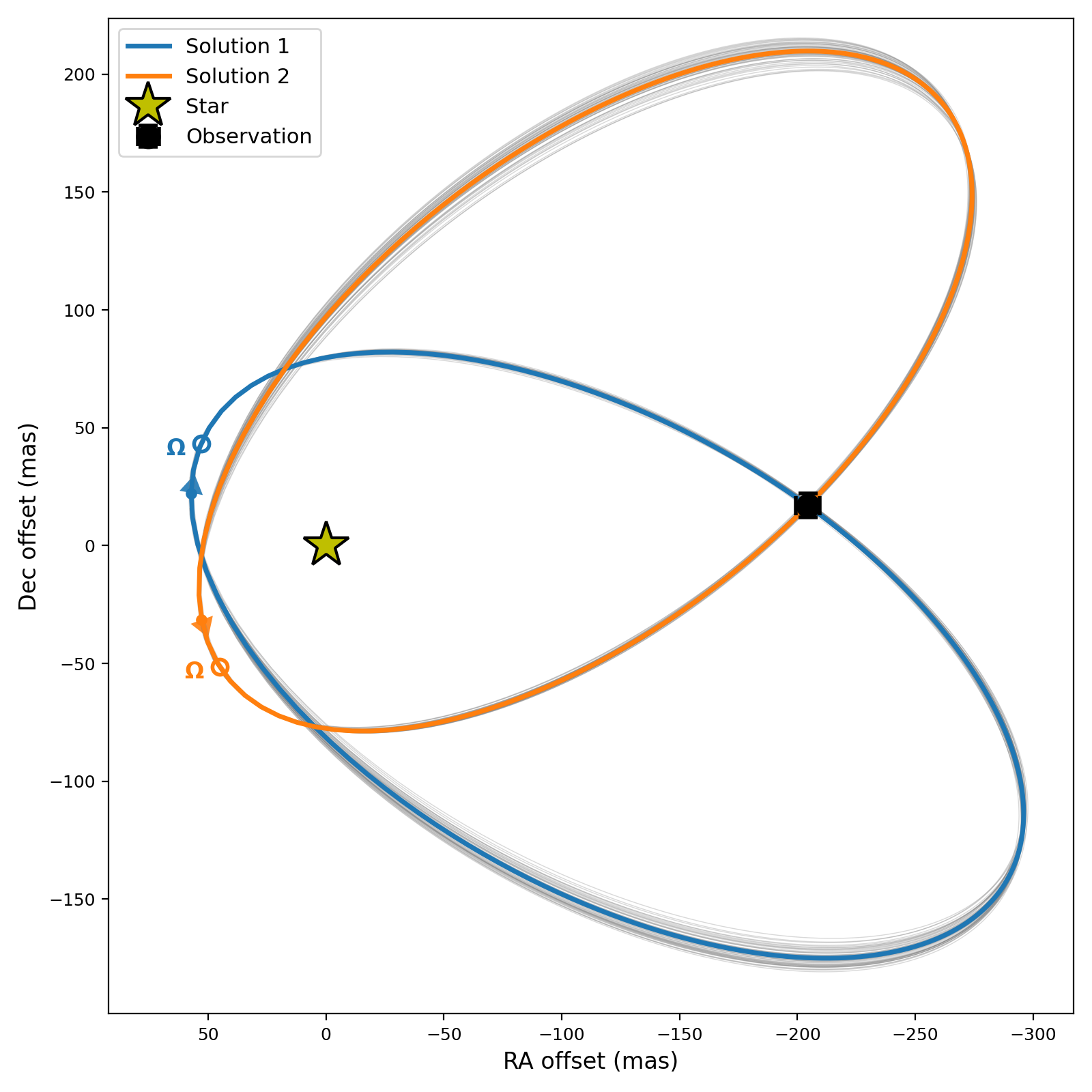}
  \includegraphics[width=0.3\textwidth]{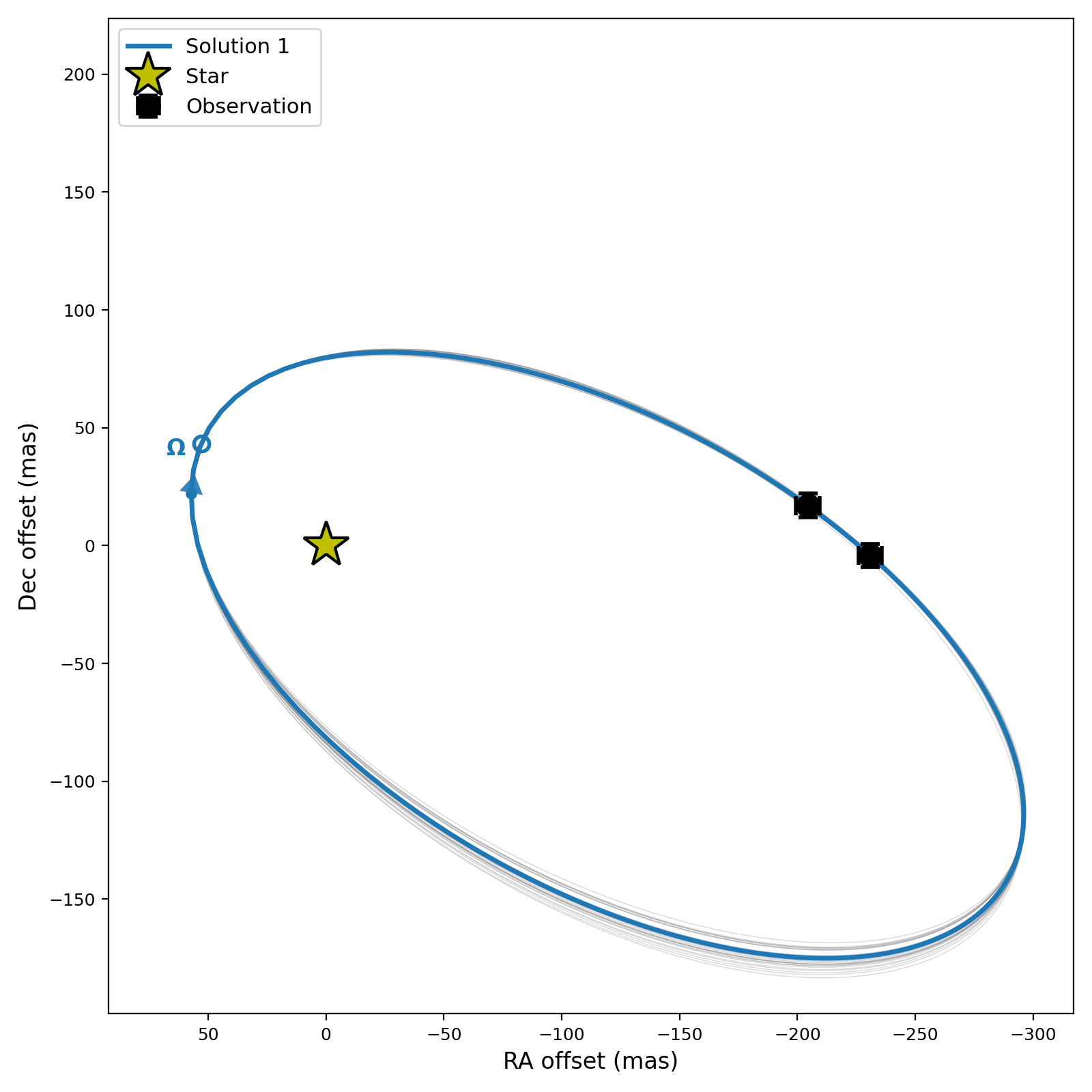}
  \caption{Left: RV ﬁtting results of HD 39091 ($\pi$ Men) b using AAT and HARPS data~\citep{2018A&A...619L..10G,2022AJ....163..223H}. Middle: With one hypothetical direct imaging measurement (black square), orbital fitting yields two distinct orbital solutions (blue and orange) that differ in their projected orientations but both pass through the observed position. The arrow indicates the position of the perigee, and the hollow circle next to $\Omega$ indicates the position of the ascending node. The two solutions are related by $i_₂ = \pi - i_₁$ (symmetric about $i = \pi / 2$), and their projected orbits are geometrically mirror-symmetric about the line connecting the star to the observed position. Right: With two hypothetical astrometric measurements (black squares), the degeneracy is broken, yielding a unique three-dimensional orbital solution (orange). This illustrates the scientific value of future direct imaging observations for RV targets.} 
  \label{fig:HD39091}
\end{figure*}

However, CPI-C will offer a high contrast of two orders better, thus enabling the following-up observations of several RV detected exoplanets, among which approximately twenty planets are in the detection area of CPI-C (see Table~\ref{tab:RV_targ}). With more astrometric information obtained through direct imaging, we can obtain the three-dimensional orbital information, which will further constrain the mass of these planetary systems, when combined with the RV data. To demonstrate the constraining power of future direct imaging observations, we performed orbital fitting for HD 39091 b using the existing RV measurements combined with hypothetical astrometric data. Figure \ref{fig:HD39091} illustrates the results: with one astrometric observation, the posterior distribution shows two distinct peaks in ($\Omega$, $i$), yielding two degenerate orbital solutions (middle panel); with two observations at different epochs, the posterior converges to a unique solution (right panel). This demonstrates that multi-epoch direct imaging is necessary to break the orbital degeneracy and uniquely constrain the three-dimensional orbit  of RV-detected targets.~\citep{bowler18}

\begin{table*}
\bc
\begin{minipage}[]{\textwidth}
\caption[]{Information of RV targets.\label{tab:RV_targ}}
\end{minipage}
\setlength{\tabcolsep}{3pt}
\small
\begin{tabular}{lcccccccc}
\hline\noalign{\smallskip}
Name & Parallax & $M_{s}$ & $P$ & $e$ & $\omega$ & $t_p$ & $M_p \sin i$ & Reference \\
 & (mas) & ($M_\odot$) & (days) &  & ($^\circ$) & (JD) & ($M_{\rm Jupiter}$) &  \\
\hline\noalign{\smallskip}
14 Her b    & 55.87  & 0.90    & 1767.56  & 0.37  & 22.3   & 2451368.0  & 4.58  & 1, 2, 3 \\
47 UMa b    & 72.01  & 1.03    & 1078.0   & 0.03  & 334.0  & 2451917.0  & 2.53  & 4 \\
47 UMa c    & 72.01  & 1.03    & 2391.0   & 0.10  & 295.0  & 2456441.0  & 0.54  & 4 \\
55 Cnc d    & 79.45  & 1.02    & 5574.2   & 0.03  & 254.0  & 2453490.0  & 3.84  & 5 \\
HD 114613 b & 48.87  & 1.27    & 4000.0   & 0.458 & 196.0  & 2455150.0  & 0.357 & 6 \\
HD 134987 c & 38.19  & 1.07    & 5000     & 0.12  & 195.0  & 2461100.0  & 0.82  & 7 \\
HD 142 b    & 38.19  & 1.10    & 349.7    & 0.17  & 327.0  & 2452683.0  & 1.25  & 8 \\
HD 142 c    & 38.19  & 1.10    & 6005.0   & 0.21  & 250.0  & 2455954.0  & 5.3   & 8 \\
HD 150706 b & 35.48  & 0.94    & 5894.0   & 0.38  & 132.0  & 2458179.0  & 2.71  & 9 \\
HD 154345 b & 54.74  & 0.88    & 3538.0   & 0.26  & 346.0  & 2454701.0  & 1.0   & 9 \\
HD 162004 b & 44.05  & 1.19    & 3117.0   & 0.40  & 64.0   & 2449344.0  & 1.53  & 10 \\
HD 190360 b & 62.49  & 1.04    & 2867.9   & 0.343 & 14.7   & 2459271.0  & 1.495 & 11, 12 \\
HD 217107 c & 49.78  & 1.02    & 5059.3   & 0.40  & 204.0  & 2450921.0  & 4.09  & 13 \\
$\pi$ Men b  & 54.68  & 1.09    & 2088.33  & 0.65  & 331.15 & 2456306.5  & 9.82  & 14, 15 \\
HD 62509 b  & 96.54  & 1.47    & 589.64   & 0.02  & 354.58 & 2447739.02 & 2.9   & 16 \\
HD 87883 b  & 54.67  & 0.82    & 2754.0   & 0.53  & 191.0  & 2451139.0  & 12.1  & 17 \\
HD 92987 b  & 22.98  & 1.08    & 10354.6  & 0.21  & 195.1  & 2457889    & 16.88 & 18 \\
$\epsilon$ Eri b & 310.58 & 0.83 & 2671.0   & 0.26  & 130.6  & 2444411.5  & 0.77  & 19, 20 \\
$\mu$ Arae e     & 64.09  & 1.08 & 4205.8   & 0.10  & 57.6   & 2452955.2  & 1.814 & 21 \\
\noalign{\smallskip}\hline
\end{tabular}
\ec
\tablecomments{0.96\textwidth}{%
Refs: 1~--~\citet{2007ApJ...654..625W},
2~--~\citet{2021ApJ...922L..43B},
3~--~\citet{2023AJ....166...27B},
4~--~\citet{2010MNRAS.403..731G},
5~--~\citet{2014MNRAS.441..442N},
6~--~\citet{2019AJ....157..149L},
7~--~\citet{2010MNRAS.403.1703J},
8~--~\citet{2012ApJ...753..169W},
9~--~\citet{2012A&A...545A..55B},
10~--~\citet{2016ApJ...818...34E},
11~--~\citet{2005ApJ...632..638V},
12~--~\citet{2015A&A...581A..38C},
13~--~\citet{2009ApJ...693.1084W},
14~--~\citet{2018A&A...619L..10G},
15~--~\citet{2022AJ....163..223H},
16~--~\citet{2008JKAS...41...59H},
17~--~\citet{2009ApJ...703.1545F},
18~--~\citet{2019A&A...625A..71R},
19~--~\citet{2021AJ....162..181L},
20~--~\citet{2019AJ....157...33M},
21~--~\citet{2007A&A...462..769P}.%
}
\end{table*}
 
In addition to the RV method, high-contrast imaging observations from CPI-C can also be combined with astrometric method of exoplanet detection. Precision astrometric data from space telescopes like Gaia and the under-construction CHES Mission~\citep{2022RAA....22g2003J} can identify proper motion acceleration of stars due to gravitational perturbations from planets, thereby indirectly inferring the existence of planet. When planet candidates are identified through astrometry, CPI-C can be used for follow-up observations. This strategy significantly increases the efficiency of direct imaging~{\citep{2021SPIE11823E..0FM, 2024arXiv240102039P, 2025arXiv250916761B}} . A successful example is HIP 99770 b~\citep{2023Sci...380..198C}, whose existence was first inferred from its proper motion acceleration in the Hipparcos-Gaia Catalog of Accelerating Stars (HGCA) and later confirmed through direct imaging by the SCExAO instrument~\citep{2021ApJS..254...42B}. With its high contrast capabilities, CPI-C will be able to directly image and confirm more of these astrometric exoplanet candidates, including those with lower masses.

Combining direct imaging, RV, and astrometry yields a complete three-dimensional orbital solution even when only a fraction of the orbit is observed~\citep{2019AJ....158..140B}. This synergy enables dynamical mass measurements independent of stellar age or evolutionary models, providing robust constraints on planetary system architectures and formation pathways.

\subsection{High-Contrast Imaging of the Circumstellar Disks}
A protoplanetary disk is a disk-shaped structure composed primarily of gas and dust that surrounds young stars—either those that have recently formed or are still forming. These disks can extend from a few stellar radii to hundreds of astronomical units (AU) and typically persist for several million years. It is widely accepted that planets, particularly Jupiter-like giants, form within these disks. During the planet formation process, material evolves from micron-sized dust particles and gas, gradually aggregating and growing into planetary-scale bodies. However, the precise mechanisms underlying this process remain unclear within the scientific community. The most effective observational approach to studying planet formation is to directly search for and detect planets that are currently forming within these disks. Such protoplanets can provide direct evidence regarding the location, timescale, and local environmental conditions of planet formation. This kind of information is nearly impossible to obtain through any other means. Nevertheless, directly observing planet formation within protoplanetary disks presents significant observational challenges. Conventional planet detection techniques, such as radial velocity and transit methods, require a stable and quiescent host star. However, young stars that are only a few million years old are typically highly active and undergoing accretion processes, making them unsuitable targets for these methods. While direct imaging offers a promising alternative for detecting forming planets, it must overcome the challenge of interference caused by the brightness and structure of the surrounding protoplanetary disk.

Over the last decade, a new method for finding and studying planets forming within protoplanetary disks has been developed. In this method, the location and properties of planets are inferred from the structures created in the disk by their gravitational perturbations~{\citep{2015ApJ...809L...5D, 2015ApJ...809...93D}}. Acting as gravitational perturbers within the disk, planets gravitationally interact with the disk material. This process can create large-scale structures such as spiral-arm density waves and gaps, and can also trigger instabilities like the Rossby Wave Instability, forming vortices. The size of these structures can reach the same order of magnitude as the planet's orbital semi-major axis. In the last two decades, a series of high-resolution observational facilities at optical/near-infrared and millimeter/centimeter wavelengths have been developed, such as Subaru/HiCIAO, Gemini/GPI, VLT/SPHERE, ALMA, and JVLA. They have conducted numerous imaging observations of disks, finding many structures suspected to be caused by perturbations from forming planets. By studying these structures and comparing them with theoretical simulations, constraints can be placed on the planet's properties, particularly mass and orbit~{\citep{2015ApJ...815L..21F, 2017ApJ...835...38D, 2017ApJ...835..146D}}. As an example, through hydrodynamic fits to ring, gap and spiral substructures~{\citep{2023ASPC..534..423B}}, compiles constraints on the masses and semi-major axes of candidate protoplanets in disks.

The CPI-C can conduct direct imaging observations of protoplanetary disks. At optical and near-infrared wavelengths, the primary signal from the disk is starlight scattered by (sub-)micron-sized dust. Structures created by planetary perturbations can be identified and characterized in these scattered light images. By studying these structures, CPI-C holds promise for inferring the presence of planets.

Simultaneously, CPI-C will also be used to measure the distribution of exozodiacal dust (EzD) around nearby stars, which is crucial for future imaging detection of Earth-like planets. Analysis by \citet{absil11} indicates that abundant EzD around main-sequence stars is a major noise source for future direct imaging detection of exo-Earths. Future missions aiming for imaging and spectroscopic characterization of Earth-like planets will require longer exposures to improve the signal-to-noise ratio of planetary images. However, exposure time increases dramatically with the intensity of the EzD-scattered background light, significantly raising the requirements for telescope light-gathering power, meaning larger aperture telescopes and longer integration times are needed. Currently, EzD detection, whether using spectro-photometric methods (e.g., instruments like MIPS and IRS on Spitzer) or infrared interferometry (e.g., Keck/KIN, VLTI/VINCI), has a sensitivity limited to around 1000 times the Solar Zodiacal Cloud strength (SZC, defined as $10^{-7}$ times the solar brightness). Future ground-based observations will also struggle to reach 100 times SZC. This sensitivity is insufficient for quantitatively analyzing the noise contribution from EzD-scattered light, making it difficult to assess its impact on future imaging searches for exo-Earths \citep{2010SPIE.7734E..0LA}.

The sensitivity of CPI-C to exozodiacal dust and its impact on the mission performance requires careful consideration for target selection. Recent measurements from the HOSTS (Hunt for Observable Signatures of Terrestrial Systems) survey using the Large Binocular Telescope Interferometer have provided critical constraints on the exozodiacal dust levels around nearby stars\citep{2020AJ....159..177E}. The survey found that a strong correlation  between the detection of exozodiacal dust and the presence 
of cold debris disks, with $\sim 78\%$ of stars with known cold dust also showing detectable warm dust in their habitable zones. For future missions targeting Earth-like planets at contrast levels of $\sim 10^{-10}$ in visible wavelengths, EzD represents a major source of background noise that can significantly increase integration times\citep{2010SPIE.7734E..0LA}. However, CPI-C's contrast requirement of $10^{-8}$ for detecting cool Jupiter-like and Neptune-like planets is less demanding, and the typical EzD levels should not fundamentally limit CPI-C's capability to achieve its primary science objectives.

\subsection{Synergy between Reflection Spectra of Cold Jupiter-like Planets and Transmission/Emission Spectra of Short-Period Giant Planets in same Systems}

This project will, for the first time, enable direct atmospheric comparisons between short-period giant planets and their co-existing cold giant planets within the same planetary systems. By combining ground-based and space-based transmission and emission spectra of close-in transiting gas giants with reflection spectra of cold Jupiters located several AU from their host stars, obtained via CPI-C’s unprecedented high-contrast imaging ($10^{-8}$) and multi-band photometry. Transmission and emission spectroscopy will yield precise constraints on molecular abundances, C/O ratios, and thermal structures for inner hot Jupiters(for instance, as demonstrated in the transmission spectrum of WASP-39b)~\citep{2023Natur.614..659R}. Meanwhile, reflection spectroscopy obtained by CPI-C will provide insights into the cloud properties and atmospheric compositions of outer cold giant planets (synthetic spectra for such objects are presented in Figure \ref{fig:sim_model} of the CSST White Paper). The combination of these complementary diagnostics will help constrain the formation locations and migration pathways of hot Jupiters, while also leveraging the potentially primordial atmospheres of both types of planets to trace the early chemical conditions and dynamical evolution of their systems. This study will focus on systems in which inner and outer giant planets are dynamically linked, thus enabling meaningful comparative planetology across divergent orbital and thermal regimes.

The transit timing variation (TTV) technique has proven to be a powerful method for large-scale exoplanet discovery. In recent research, Kepler-725-c, a super-Earth in the habitable zone of a late G-type dwarf star, was discovered for the first time using the TTV method \citep{2025NatAs...9.1184S}. Short-period exoplanets, which are readily detected via the transit method, are likely to co-exist with cold giant planets on orbits of several AU \citep{2021ARA&A..59..291Z}. Cold giant exoplanets that are in or near mean-motion resonances (MMRs) with inner transiting planets can induce significant TTVs in the latter ones \citep{2022MNRAS.512.3113B,2025AJ....169..342W}. This makes the TTV technique an ideal tool for identifying cold giant exoplanets with co-exist hot planets, which are particularly suitable candidates for the proposed synergy study.

A compelling scientific case lies in the first-ever simultaneous observation of reflection and transmission/emission spectra within the same planetary systems. Transmission and emission spectra of short-period transiting gas giants can be obtained through ground-based observations, whereas CPI-C is capable of acquiring reflection spectra of long-period planets. This unprecedented capability of CPI-C will enable, for the first time, direct atmospheric comparisons between hot Jupiters and their co-existing cold planetary companions, thereby providing a unique opportunity to gain insights into the formation and migration history of hot Jupiters. Such a study will facilitate system-level spectral comparisons that were previously unattainable.

\subsection{High-contrast Imaging of Exoplanets in multiple-Stars system and Trans-Neptune Objects in Solar system}
The HST has demonstrated its observational capability through the discovery of the moon S/2015(136472) 1 orbiting Makemake\citep{2025arXiv250905880B}, despite the significant challenges associated with its nearly edge-on orbital orientation. This achievement highlights the potential to detect faint companions around bright Trans-Neptunian Objects (TNOs), a critical factor in determining system masses and evaluating formation models, such as those involving giant impacts. The moon’s exceptionally low relative brightness (approximately 7.8 magnitudes fainter than Makemake) provides support for the ``dark moon hypothesis", suggesting that the dark material inferred from global thermal measurements is preferentially located on the satellite’s surface rather than on Makemake itself. Observations from the JWST have further advanced our understanding by transforming Makemake from a seemingly static body into a dynamically active world\citep{2025ApJ...991L..34P}. These observations confirmed the presence of a complex surface composition dominated by hydrocarbon ices and achieved the first detection of gaseous methane, indicative of ongoing surface or atmospheric activity. Additionally, a new, precise measurement of the deuterium-to-hydrogen (D/H) ratio offers crucial constraints on the origin and evolution of volatile materials within the system. The upcoming CPI-C mission will build upon these findings by enabling the detection of even fainter satellites, thereby improving our characterization of TNO physical and dynamical properties. Through such observations, CPI-C will serve as a bridge between planetary science and astrophysics, contributing to a more unified understanding of planet formation processes across diverse spatial scales. 

\subsection{ Sample Statistics of the Direct Imaging Planets}
Current two prominent theories of giant exoplanet formation are the Core Accretion and the disk gravitational instability mechanism, respectively. The Core Accretion model envisions the giant planets will form when planetesimal accumulation produces a solid core sufficiently to attract a massive atmosphere~\citep{2002ApJ...567L.149B,2015ApJ...800...82P}. The giant planets within 3 AU that detected by RV or transit approaches, have shown a consistent with the core accretion theory. However, the detection of long-period giant planet via direct imaging technique poses challenges to the Core Accretion model. The first observations of forming protoplanetary systems at distant orbits beyond 50 AU, such as HD 100546 b and LkCa 15 b and c, provide new observational evidence for theoretical models of planet formation. Some of the directly detected planets orbit far beyond the snow line (>30 AU), which is difficult to explain with the Core Accretion model and instead lends more support to the Gravitational Instability model~\citep{1996Icar..124...62P,2002ApJ...576..870P,2015ApJ...814L..27C,2015Natur.527..342S}. Expanding the sample of distant planets and further constraining the distributions of semi-major axes, masses, and mass ratios relative to the host star for these wide-orbit giant planets will be crucial for understanding the planet formation mechanisms and identifying key factors in the process~\citep{2016ARA&A..54..271K,2018haex.bookE.143M}. CPI-C will firstly image and characterization the mature exoplanets that is located within 5 AU distant from the host star, with a pretty low effective temperatures down to a few hundred K. The expected results will provide an important complementry to the curerent ground-based observations samples (>10AU, ~1000K).

Currently, due to a lack of physical information of mature planets by direct imaging, existing evolution models have certain shortcomings. The masses of directly imaged exoplanets are primarily determined by comparison with current planetary evolution models. For example, the masses of the HR 8799 planetary system were mainly determined by fitting the relationship between planetary age and luminosity \citep{2007ApJ...655..541M}. However, most planets detected by imaging are relatively young (a few hundred Myr), and their physical properties depend on the initial conditions during their formation, introducing uncertainties in mass measurements. For instance, \citet{2008Sci...322.1348M} calculated the mass of exoplanet HR 8799 b to be between 5--11 $M_\mathrm{J}$ (Jupiter masses) based on a cooling model, with the other three planets c, d, and e between 7--13 $M_\mathrm{J}$. In contrast, \citet{2011ApJ...729..128C} using a planetary evolution model combined with dynamical stability constraints, found a mass of 6--7 $M_\mathrm{J}$ for HR 8799 b and 7--10 $M_\mathrm{J}$ for c, d, and e. More recently, dynamical mass measurements combining Gaia–Hipparcos astrometry multi-epoch imaging constrain HR 8799 e to $9.6^{+1.9}_{-1.8} M\mathrm{J}$ \citep{2021ApJ...915L..16B}, providing an important anchor for evolution models. This shows that exoplanet mass measurements are dependent on the employed model and have a large uncertainty. Therefore, the CPI-C will conduct imaging and spectroscopic analysis of mature exoplanets, enabling the systematic study of planets at different evolutionary stages (including relatively stable mature stages and young stages shortly after formation). Finally, by performing a statistical analysis of the detections and non-detections of planets within the whole survey samples, CPI-C will provide insights into the occurrence rates, orbital distributions, and atmospheric diversity of all-age giant planets around nearby solar type stars. The potential results will help to understand the formation and evolution mechanisms of the planets.

\subsection{Numerical Simulation and Expected Scientific Yield} \label{sec:simulation}

To evaluate the feasibility of CPI-C’s core science objectives, we performed end-to-end mock observations of exoplanets. The simulations incorporate realistic instrumental effects, including diffraction suppression, wavefront errors, speckle noise, and photon/background noise, followed by post-processing with optimized speckle suppression algorithms~\citep{2025arXiv251108886Z,2025arXiv251109862Z}.

\begin{figure*} 
  \centering
  \includegraphics[width=7.5cm]{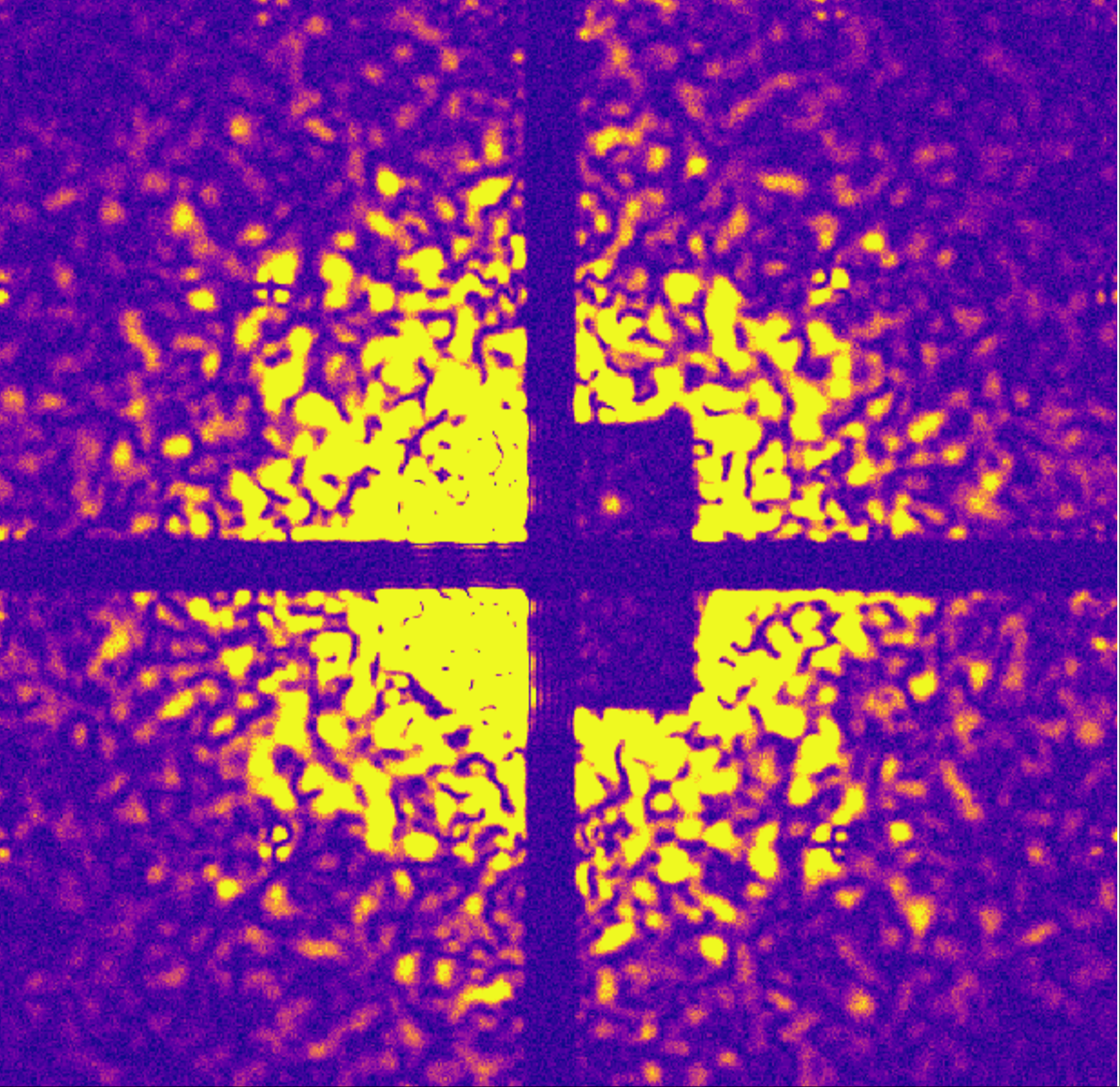}
  \caption{A simulated image of CPI-C observation, and an artificial planet can be seen in dark hole regions.} 
  \label{fig:sim_image}
\end{figure*}

As an illustrative case, we simulated observations of a K1V-type star with a V-band magnitude of 0. A hypothetical Jupiter-sized planet was placed at an angular separation of 0.65$^{\prime\prime}$ from the host star, with a phase angle of 90° and a position angle of 315°. The planetary parameters assume a radius of 1 $R_{\mathrm{J}}$, medium metallicity, and low sedimentation efficiency. 
Figure \ref{fig:sim_image} presents the simulated image of CPI-C observation, and an artificial planet can be seen in a dark hole region.

\begin{figure*}[h!]
  \centering
  \includegraphics[width=0.59\textwidth]{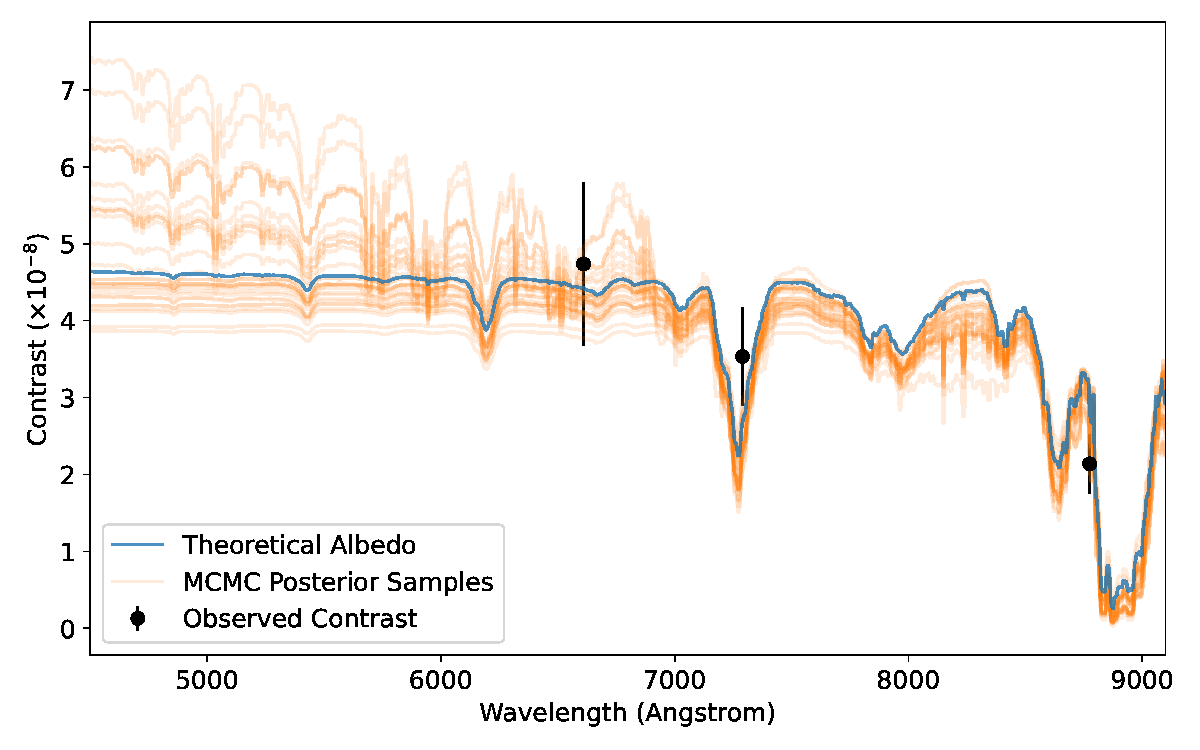}
  \includegraphics[width=0.35\textwidth]{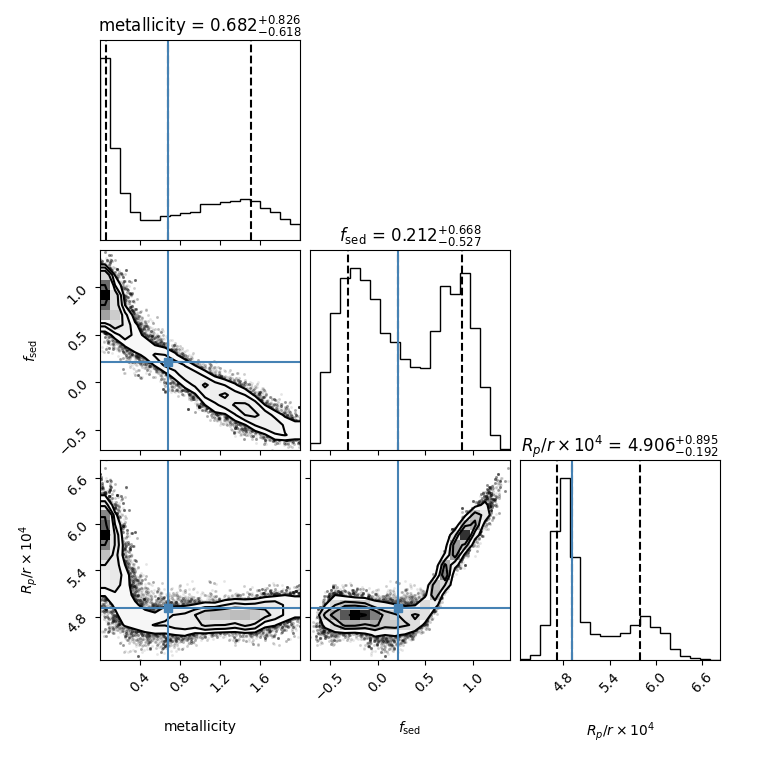}
  \caption{\textit{Left}: Simulated reflected-light spectrum of a Jupiter-like planet. The blue curve shows the theoretical albedo spectrum, while black points with error bars represent mock CPI-C observations in optical bands. The orange curves correspond to spectral realizations drawn from the posterior distribution of the MCMC retrieval, illustrating the range of models consistent with the data. \textit{Right}: Posterior distributions from MCMC fitting.} 
  \label{fig:sim_model}
\end{figure*}

We applied the \texttt{emcee} Markov Chain Monte Carlo (MCMC) sampler~\citep{2013PASP..125..306F} to fit the mock CPI-C observations, and the results are presented in Figure~\ref{fig:sim_model}. It is important to note that Figure~\ref{fig:sim_model} demonstrates atmospheric retrieval for cool, mature planets observed in reflected light using CPI-C's optical bands. In this regime, the retrieved parameters include atmospheric metallicity ($M$), sedimentation efficiency ($f_{\rm sed}$), and the planet radius-to-separation ratio ($R/d$). The posterior samples reproduce the underlying reflected-light spectrum, confirming that the retrieval framework can successfully recover planetary atmospheric features from simulated data.

The achievability of CPI-C’s science objectives is further supported by both instrumental tests and data analysis. Laboratory measurements and tests have demonstrated stable performance across multiple bands, with accurate centroiding and high signal-to-noise observations. This confirms that CPI-C’s sensitivity comfortably meets the requirements for the planned survey of nearby stars. In addition, spectral fitting of mock observations reveals that the instrument is capable of distinguishing different atmospheric compositions (see Figure \ref{fig:sim_model}), highlighting its unique advantage in probing the reflected-light spectra of giant planets. The integration of retrieval algorithms with high-quality test data establishes a complete workflow from raw image acquisition to atmospheric model inversion. This foundation not only validates CPI-C’s immediate feasibility but also provides valuable experience for applying more sophisticated retrieval frameworks to real observations in the future.

\section{Scientific Requirements}
\begin{figure*}[h!]
  \centering
  \includegraphics[width=10cm]{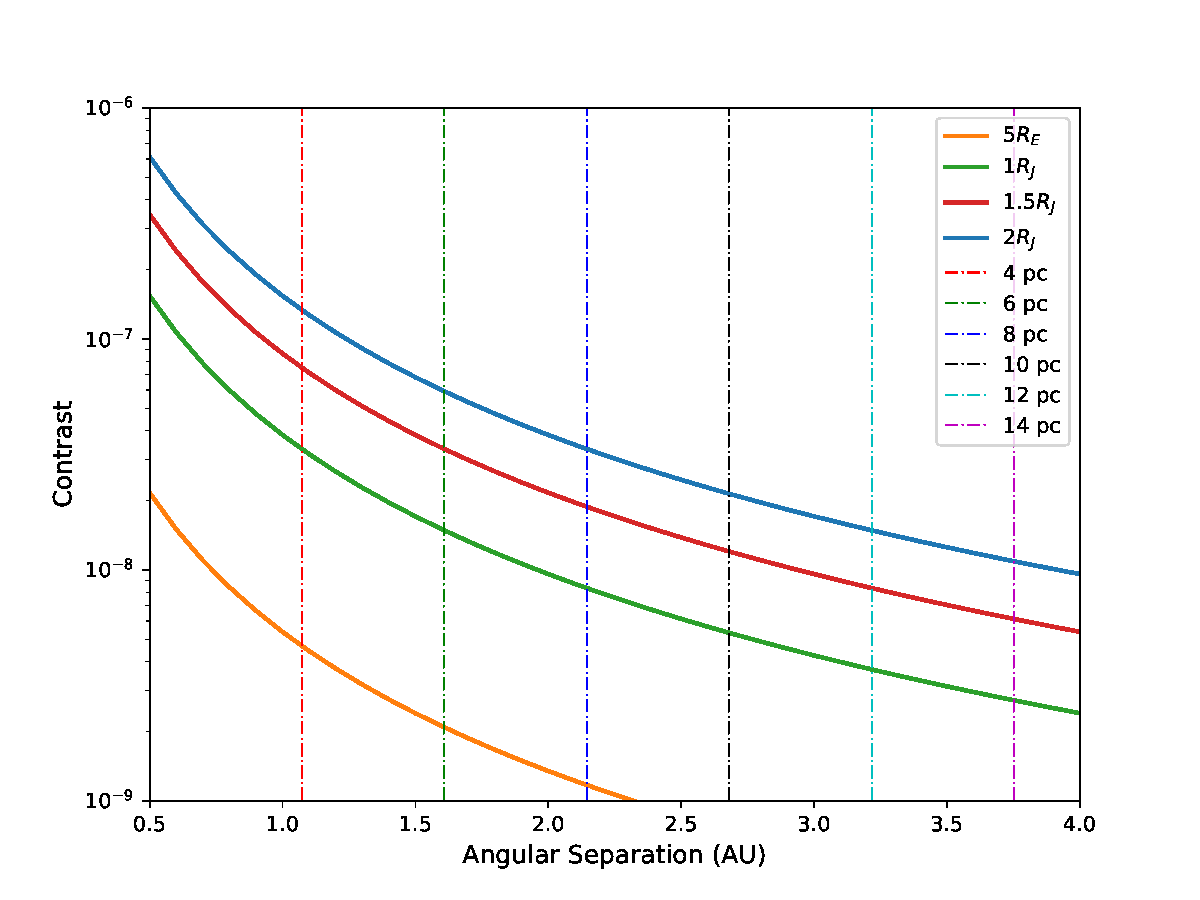}
  \caption{Minimum contrast required for CPI-C to detect exoplanets of different sizes (1, 1.5, 2 Jupiter radii and 5 Earth radii) at various angular separations from their host star with different distances. The vertical dashed lines represent the inner working angle limits at different stellar distances.} 
  \label{fig:contrast_require}
\end{figure*}

CPI-C will enable direct imaging observations of mature Jupiter-like/Neptune-like planets with effective temperatures of several hundred Kelvin orbiting solar type stars, obtaining high-quality images of such planets. In the visible band, the primary energy from these planets originates from starlight reflected through their atmospheres, requiring an imaging contrast of $10^{-7}$ to $10^{-8}$. The contrast requirement decreases with smaller separation from the host star and larger planet size, which is indicated in Figure \ref{fig:contrast_require}. The contrast ratio of the reflected planet light to the primary starlight and asscociated sensitivity calculation can refer to the previous papers~\citep{1990Icar...87..484B,2012OptEn..51a1002L,2016ApJ...832...84D}.

To detect new planetary candidates, CPI-C must operate from visible to near-infrared wavelengths and will be equipped with corresponding broadband filters to cover spectral lines of characteristic elements  for life. Photometry across multiple bands will yield fitted spectra of these planetary atmospheres, thereby constraining key physical properties of the planets. In Table \ref{tecreq}, we list the baseline technical specifications required for CPI-C. The instrument's performance should exceed the requirements outlined in this table. Based on the instrument design and existing test results, we provide the estimated performance of CPI-C during formal observations in Table \ref{parameters} .

\begin{table}[h!]
\centering
\caption{Technical Specifications of CPI-C.\label{tecreq}}
\renewcommand{\arraystretch}{1.5} 
\begin{tabular}{|m{7.5cm}|m{5.5cm}|}
\hline
\textbf{Technical Specification} & \textbf{Requirement} \\
\hline
Observation Wavelength Range & 0.6$\mu$m -- 1.6$\mu$m \\
\hline
Imaging Dynamic Contrast & $\le 10^{-8}$ @ 0.60$\sim$0.90$\mu$m \\ & $\le 10^{-4}$ @ 1.165$\sim$1.340$\mu$m \\
\hline
Inner Working Angle (IWA) & $\le 0.55''$ @ 0.633$\mu$m  \\ & $\le 1^{\prime\prime}$ @ 1.25$\mu$m \\
\hline
Outer Working Angle (OWA) & $\ge 0.8''$ @ 0.633$\mu$m \\ & $\ge 1.5^{\prime\prime}$ @ 1.25$\mu$m \\
\hline
Visible Band Detection Sensitivity & SNR $\ge 5$ @ 0.642$\mu$m$\sim$0.681$\mu$m for a 20th magnitude point source \footnote{30s exposure by the visible camera, main optics efficiency of 0.78, and background brightness better than 22 mag/arcsec$^2$.}\\
\hline
Near-Infrared Band Detection Sensitivity & SNR $\ge 5$ @ 1.165$\mu$m$\sim$1.340$\mu$m for a 15th magnitude point source \footnote{30s exposure by the NIR camera, main optics efficiency of 0.84, and background brightness better than 22 mag/arcsec$^2$.}\\
\hline
Field of View (FOV) & Visible Channel: $\ge \pm 6^{\prime\prime}$ \\ & Near-Infrared Channel: $\ge \pm 5^{\prime\prime}$ \\
\hline

\hline
\end{tabular}
\end{table}



\begin{table}[h]
\bc
\begin{minipage}[]{150mm}
\caption[]{Estimated Performance of CPI-C instrument.\label{parameters}}
\end{minipage}
\setlength{\tabcolsep}{3pt}
\small
\begin{tabular}{cc}
\hline\noalign{\smallskip}
Parameter Name & Parameter Value or Range \\
\hline\noalign{\smallskip}
Satellite & orbit $\sim$400 km, lifetime $\sim$10 years \\
Telescope & \SI{2}{m} primary, off-axis TMA, focal length = 28 m \\
Observation Wavelength Range & 0.58--1.6 $\mu$m \\
Inner working angle (IWA) & \SI{3}{\lambda/D}; \SI{0.21}{\arcsecond}@\SI{661}{\nm} (adjustable according to different scientific goals) \\
Outer working angle (OWA) & \SI{14}{\lambda/D}; \SI{0.95}{\arcsecond}@\SI{661}{\nm} (adjustable according to different scientific goals) \\
Width of the cross-shape focal mask & \SI{0.4}{\arcsecond} \\
\hline
\multicolumn{2}{c}{\textbf{Visible Band}} \\
\hline
Dark zone contrast & better than \num{1E-08} after data reduction with optimization algorithms \\
Sensitivity & Mag 20 @ 0.642--0.681 $\mu$m (SNR$\ge$5, 30 s exposure) \\
Focal plane scale & \SI{1.62e-2}{\arcsecond/pix} \\
Camera effective pixels & \num{1024} $\times$ \num{1024} \\
Read noise ($1\times$ EM gain) & \SI{160}{e^{-}/pix} \\
Clock induced charge (CIC) & \SI{0.2}{e^{-}/pix} \\
Dark current & \SI{1.0e-3}{e^{-}/pix/s} \\
Conversion gain ($1\times$ EM gain) & \SI{59}{e^{-}/ADU} \\
\hline
\multicolumn{2}{c}{\textbf{Near Infrared Band}} \\
\hline
Dark zone contrast & \num{1E-06} after data reduction with optimization algorithms \\
Sensitivity & Mag 15 @ 1.165--1.340 $\mu$m (SNR$\ge$5, 30 s exposure) \\
Focal plane scale & \SI{2.49e-2}{\arcsecond/pix} \\
Camera effective pixels & \num{640} $\times$ \num{512} \\
Read noise & \SI{50}{e^{-}/pix} \\
Dark current & \SI{10}{e^{-}/pix/s} \\
Conversion gain & \SI{7}{e^{-}/ADU} \\
\noalign{\smallskip}\hline
\end{tabular}
\ec
\tablecomments{0.86\textwidth}{Parameters are estimated for CPI-C baseline design. IWA and OWA values may vary depending on specific scientific goals.}
\end{table}

\section{Instruments}
\subsection{Optics Layout}
To obtain an imaging contrast better than $10^{-8}$, CPI-C must be able to suppress the diffracted light and correct the quasi-static phase aberrations induced by the telescope and the optics of CPI-C itself. For such requirements, CPI-C optical system consists of the three channels, such as Internal calibration laser channel, Wavefront sensing channel, and Scientific imaging channel. The scientific imaging channel consists of a 31-step-transmission pupil apodized filter (PM), a kilo-actuator deformable mirror (kilo-DM), a visible-light camera and a near-infrared camera.

\begin{figure} 
   \centering
   \includegraphics[width=12.0cm, angle=0]{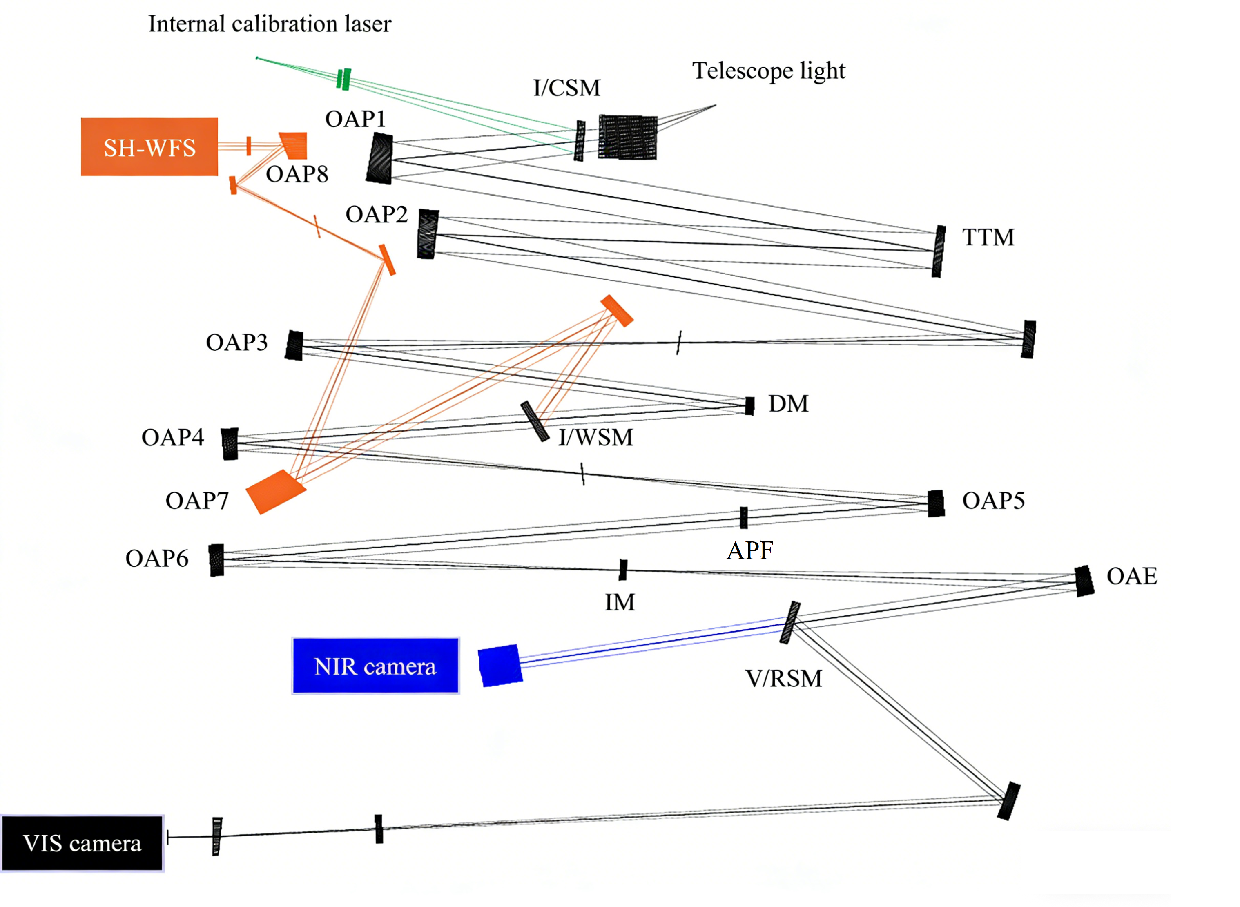}
   \caption{CPI-C optics layout.} 
   \label{fig:optics layout}
\end{figure}

CPI-C adopts a multi-pupil design for its optical layout, which has four conjugate pupils in sequence: the tip-tilt mirror (TTM), the kilo-DM, the pupil apodized filter mirror (PM) and the microlens array (MLA). The optical layout is shown in Figure\ref{fig:optics layout}. The light from the CSST telescope (F/14) is input to the CPI-C optical system, first reflected by the fold mirror M1, then transmitted to the image/calibration splitting mirror (I/CSM) which allows transmission of the telescope light and reflection of an internal laser source. The laser source, which is required to be conjugated with the telescope focus, provides the reference wavefront for both the low-order wavefront sensing by CPI-C’s Shack-Hartmann wavefront sensor and the high-order sensing at its focal plane.

The incident light is subsequently collimated by the OAP1 mirror (Off-Axis Paraboloid 1), which establishes the first pupil plane on the TTM. Furthermore, the exit pupil of the telescope must be conjugated with the TTM to ensure proper optical alignment. The light is recollimated by the OAP2 and OAP3 mirrors, which creates the second pupil plane with a size of 10 mm on the kilo-DM. After the kilo-DM, the light is split into two parts by a beam splitter (I/WSM, image/wavefront splitting mirror) and conducted into the wavefront sensing channel and scientific imaging channel, respectively.

\begin{figure} 
   \centering
   \includegraphics[width=8.5cm, angle=0]{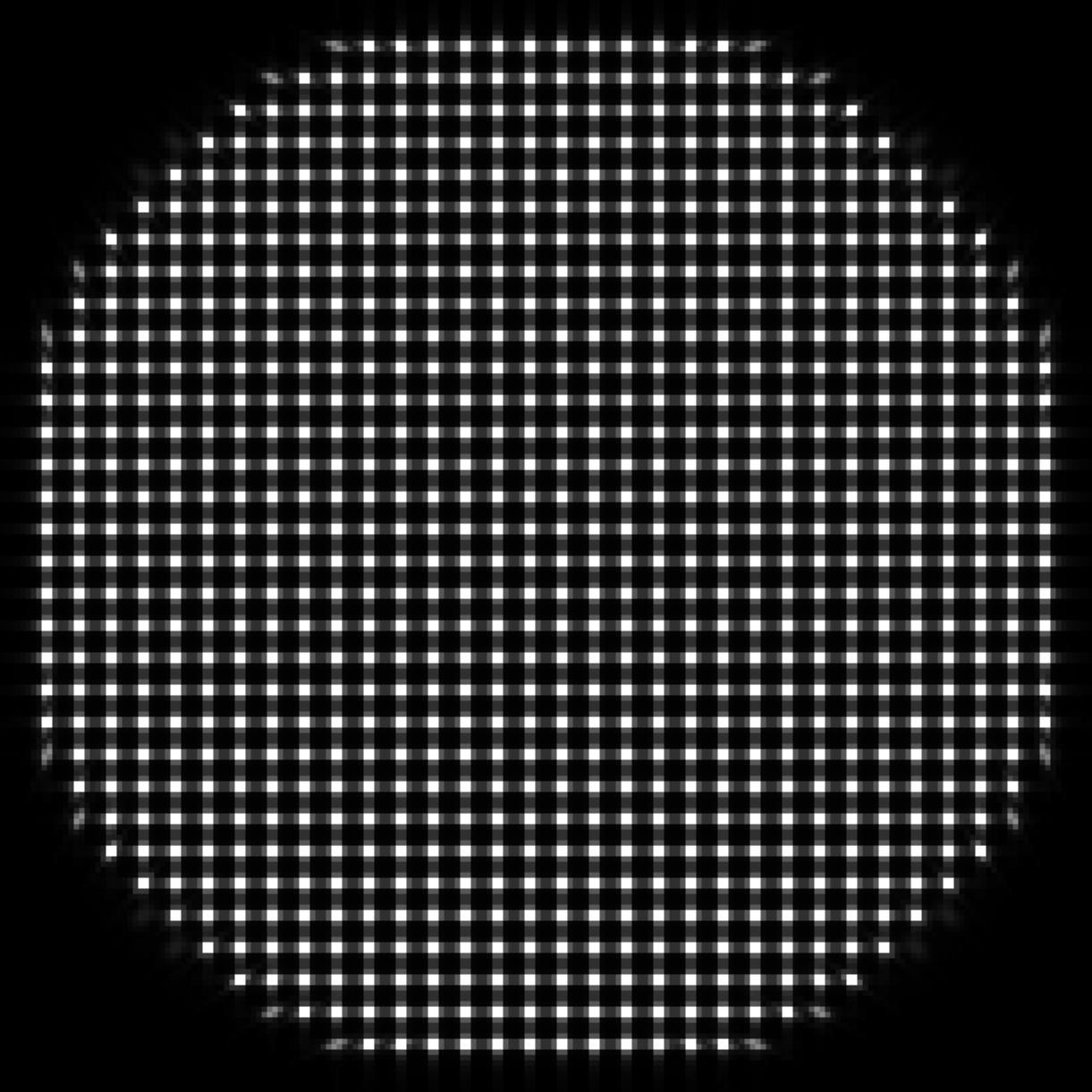}
   \caption{The simulation of the SH-WFS image.} 
   \label{fig:SH-WFS}
\end{figure}

In the wavefront sensing channel, the reflected light from the I/WSM propagates through the OAP7 and OAP8 mirrors, resulting in a reduction of the pupil plane diameter from 10 mm to 4.8 mm. At the exit pupil position, the microlens array of CPI-C’s low-order Shack-Hartmann wavefront sensor (SH-WFS) is installed to enable real-time acquisition of low-order wavefront information. The imaging detector of the wavefront sensor has been comprehensively characterized, with established calibration procedures and algorithms for non-uniformity correction\citep{Dou2025}. A simulated image of the SH-WFS output is presented in Figure \ref{fig:SH-WFS}.CPI-C is designed with two independent correction closed-loops. One of them is that the kilo-DM corrects the first 66 orders of low-order aberrations, excluding tilt and piston errors, based on the wavefront detected and reconstructed by the SH-WFS. The automatic alignment of a star PSF precisely in the center FOV from any of its initial positions in other areas of the imaging FOV is also enabled, through the closed-loop of the SH-WFS and the TTM.

\begin{figure*} 
  \centering
  \includegraphics[width=0.46\textwidth]{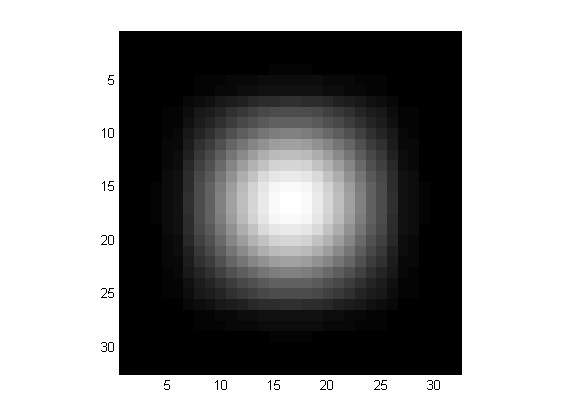}
  \includegraphics[width=0.46\textwidth]{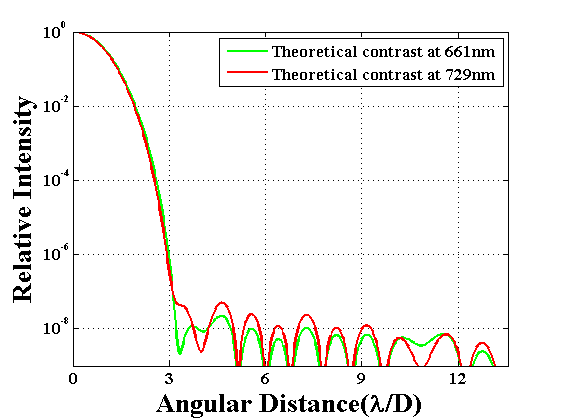}
  \caption{Left: A 31-step transmission pupil apodized filter. Right: the theoretical imaging contrast in the Visible Band.} 
  \label{fig:PupilAndCNTVis}
\end{figure*}

\begin{figure} 
   \centering
   \includegraphics[width=12.0cm, angle=0]{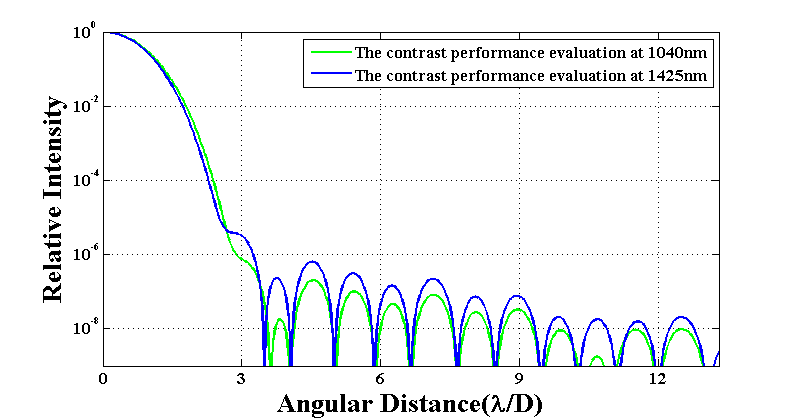}
   \caption{ The theoretical imaging contrast in the Infrared Band} 
   \label{fig:contrastNIR}
\end{figure}

To suppress the aperture diffraction noise, a 31-step-transmission pupil apodized filter (PM) is installed at the third pupil plane in the optical layout of CPI-C. The PM has been optimized to deliver its best performance in the visible bands. It also achieves relatively high contrast in the near infrared bands. The transmittance distribution and the theoretical imaging contrast are shown in Figure~\ref{fig:PupilAndCNTVis} and Figure~\ref{fig:contrastNIR}, respectively. An image-plane mask (IM) is installed at the intermediate focal plane before the OAP mirror in the scientific imaging channel to reflect most of the energy from the Airy disk and its vicinity in starlight out of CPI-C’s optical system. This design prevents the blooming effect the EMCCD sensor of the VIS Cam, which actually improves the measured contrast. Subsequently, the light is split into two paths at the wavelength of 920nm by a dichroic mirror (V/RSM, visible/infrared splitting mirror), which enables scientific imaging in the visible and near-infrared bands with respective fields of view of $±6''$ and $±5''$. A filter wheel is placed in front of each scientific camera to facilitate switching between different observation bands.

\subsection{Performance Evaluation}
To obtain better than $10^{-8}$ image contrast, CPI-C firstly needs to record the reference wavefront, and calibrate TTM and DM influence function by using the internal laser source. Then, switch to the telescope light. The SH-WFS detects the low/middle-order aberrations introduced by the telescope, and corrects the above aberrations by controlling the DM and TTM.

\begin{figure} 
   \centering
   \includegraphics[width=12.0cm, angle=0]{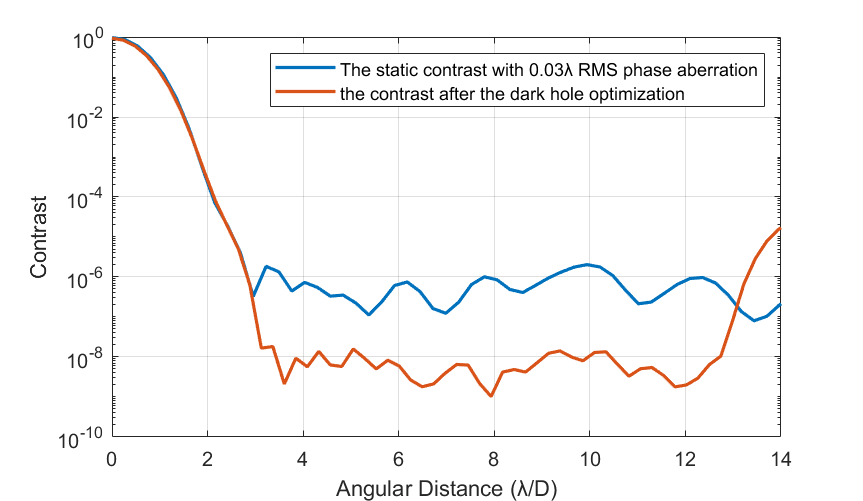}
   \caption{The static contrast of $10^{-6}$ at IWA of $3\lambda/D$ with 0.03$\lambda$ RMS phase aberration, and the contrast after the dark hole optimization.} 
   \label{fig:contrast}
\end{figure}

After that, CPI-C needs to correct the speckle noise introduced by the optical components of telescope and CPI-C self. Thus, a closed-loop correction system with the DM and scientific camera, is formed to correct the above speckle noise. The DM will provide an optimal phase to create two high contrast dark-holes in a target region.

The simulation of CPI-C reveals that the achievement of high contrast of $10^{-8}$  relies on the precise control of wavefront error. This work includes the correction towards absolute phase aberration and the contrast gain provided by the deformable mirror. In theory, the total absolute phase aberration of the coronagraph itself and the telescope should be suppressed under 0.03 RMS, to obtain the static contrast of $10^{-6}$ at $3\lambda/D$. And the kilo DM could provide another 100 times gain of contrast by applying the dark-hole optimization in Figure \ref{fig:contrast}. Finally, the imaging contrast better than $10^{-8}$ is possible with additional post-processing methods.

\subsection{Wide-band filters Design}

To fully exploit its high-contrast capability, CPI-C adopts a compact seven-band photometric system covering 600--1600\,nm. The optical bands (F661, F729, F877) target the red-optical continuum and CH$_4$ absorption, providing both a robust detection channel and leverage on cloud opacity. The four near-infrared bands (F1040, F1265, F1425, F1532) open the $900$--$1100$\,nm Y window, probe the broad $1.4\,\mu$m H$_2$O/CH$_4$ feature, and are bracketed by J/H windows to anchor the continuum. This division of labor is central to CPI-C's retrieval strategy, allowing reflected-light and thermal-emission diagnostics to be combined in a single observing sequence (Zhu et al., in preparation).

While Section~\ref{sec:simulation} demonstrated CPI-C's capability to constrain reflected-light properties (metallicity, cloud structure) using the three optical bands, the addition of four near-infrared bands enables complementary thermal-emission diagnostics. The NIR filters can provide constraints on fundamental physical parameters—effective temperature ($T_{\rm eff}$), surface gravity ($g$), and radius ($R$)—that are difficult to constrain from optical photometry alone.

Although most mirrors and windows in the optical train have $>95\%$ transmission, the apodized pupil mask and focal-plane mask together reduce total throughput to $\sim$20--22\% in both channels, and the polarization modulation element contributes only $\sim$10\% transmission. These losses strongly motivate moderately wide passbands: excessively narrow filters would yield inadequate photon counts, while overly broad filters would dilute spectral features. The adopted widths (Table~\ref{tab:cpic_filters}) balance signal-to-noise ratio (S/N) and feature contrast, ensuring detectability of cool planets.

\begin{table}[htbp]
\centering
\caption{Wavelength design of CPI-C filters.\label{tab:cpic_filters}}
\begin{tabular}{cccccc}
\hline
Camera (VIS/NIR) & Filter & Central Wavelength & Bandwidth & cut-on & cut-off \\
                 &        & (nm)               & (nm)      & (nm)                    & (nm) \\
\hline
VIS & F661  & 661.0  & 66.0  & 615.0  & 707.0  \\
VIS & F729  & 729.0  & 52.0  & 690.0  & 768.0  \\
VIS & F877  & 877.5  & 35.0  & 847.0  & 908.0  \\
NIR & F1040 & 1040.0 & 160.0 & 947.0  & 1138.0 \\
NIR & F1265 & 1265.0 & 250.0 & 1122.0 & 1408.0 \\
NIR & F1425 & 1425.0 & 150.0 & 1332.0 & 1518.0 \\
NIR & F1532 & 1532.5 & 95.0  & 1467.0 & 1598.0 \\
\hline
\end{tabular}

\vspace{2mm}
\parbox{0.9\linewidth}{\small
Bandwidths are defined as the wavelength intervals between the points where the in-band transmission drops to $90\%$ of the peak on the blue and red sides.
}
\end{table}

\begin{figure}
\centering
\includegraphics[width=12.0cm]{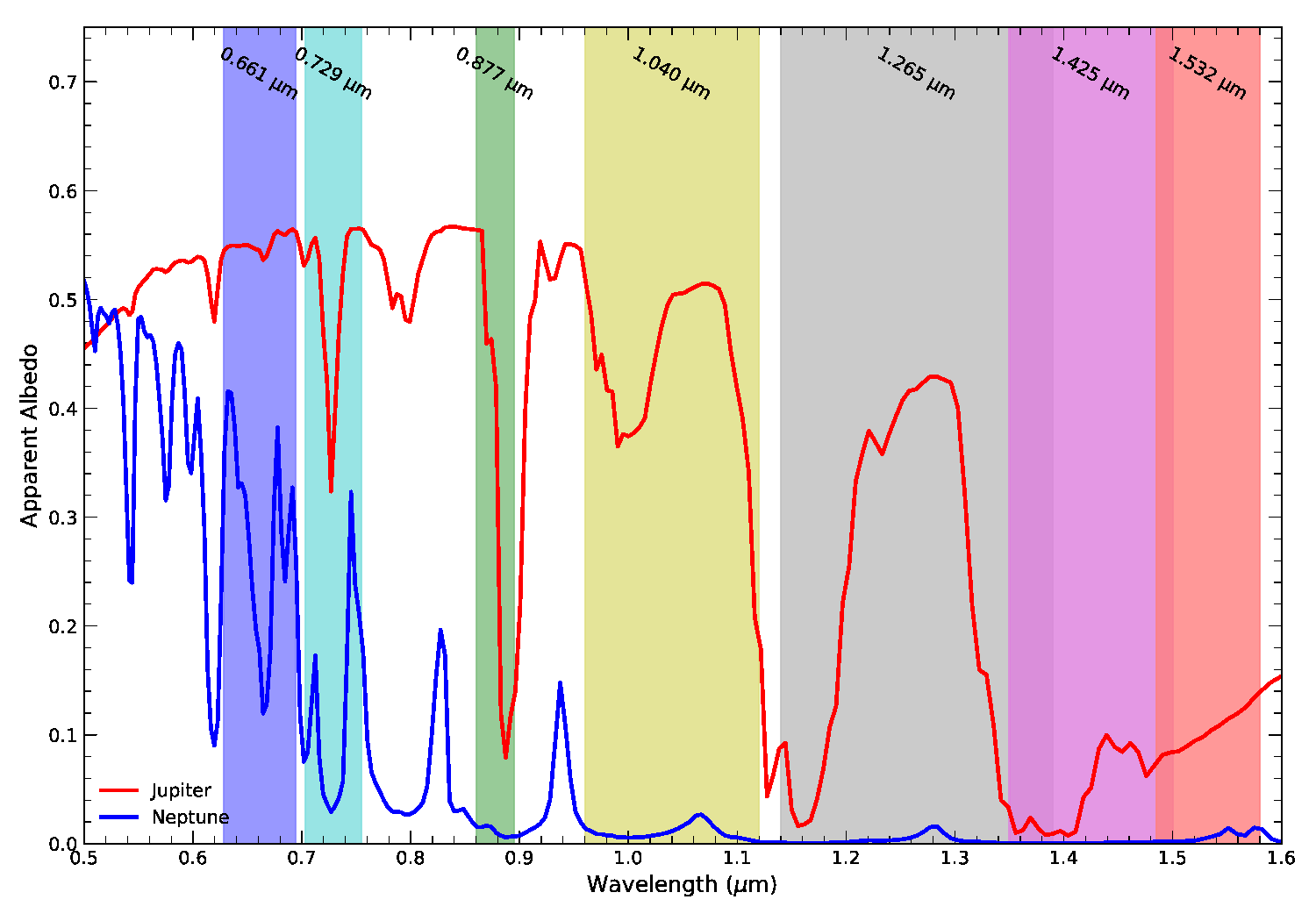}
\caption{Apparent geometric albedo spectra of Jupiter (red) and Neptune (blue) over 0.5--1.6\,$\mu$m, with the adopted CPI-C bandpasses overlaid as shaded regions. 
The optical filters (F661, F729, F877) straddle methane absorption and continuum windows in the red-optical regime, while the four NIR filters (F1040, F1265, F1425, F1532) probe the Y-band continuum, the strong 1.4\,$\mu$m H$_2$O/CH$_4$ complex, and continuum regions on either side. }
\label{fig:bandpass_albedo}
\end{figure}

Figure~\ref{fig:bandpass_albedo} illustrates the spectral placement of the seven CPI-C bandpasses on representative reflected-light spectra of Jupiter and Neptune~\citep{2018JQSRT.217...86V}. The three optical filters are positioned to isolate the deep CH$_4$ absorptions near 730\,nm and 890\,nm from adjacent continuum windows, providing sensitivity to methane abundance and cloud top pressure. The four near-infrared bands extend the coverage into the Y window (1.0–1.1,$\mu$m) and sample the broad 1.4,$\mu$m H$_2$O/CH$4$ feature with flanking continuum reference bands (1.26 and 1.53,$\mu$m). This design ensures that the retrieved spectra can disentangle molecular absorption from continuum level, thereby constraining $T_{\mathrm{eff}}$, $\log g$, atmospheric composition, and planetary radius when combined with thermal-emission measurements.

\subsection{Mechanics, configuration and thermal control}
\begin{figure}[t]
   \centering
   \includegraphics[width=12.0cm, angle=0]{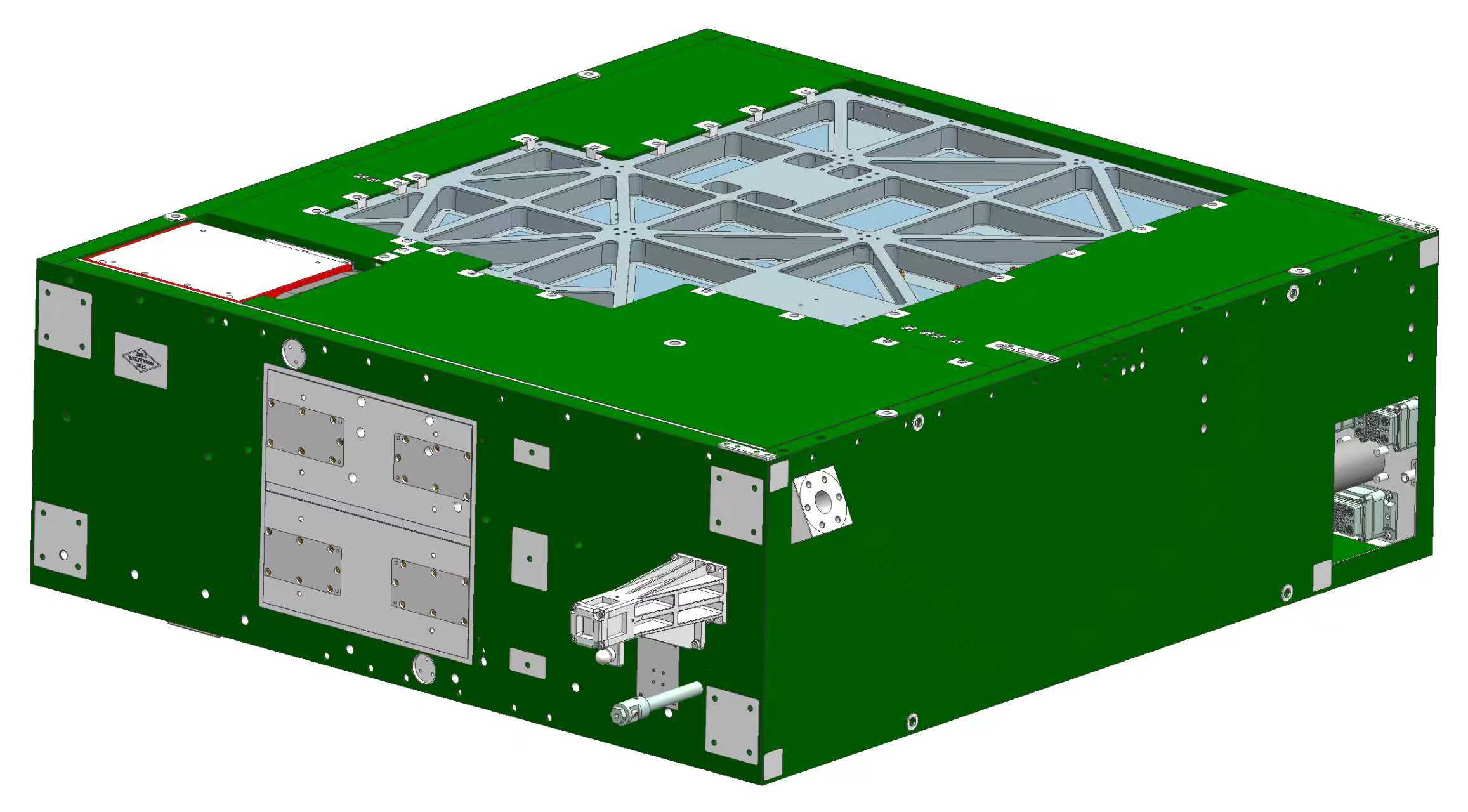}
   \caption{Schematic diagram of M5 structure.} 
   \label{fig:Schematic diagram of M5 structure}
\end{figure}

\begin{figure}[h]
   \centering
   \includegraphics[width=12.0cm, angle=0]{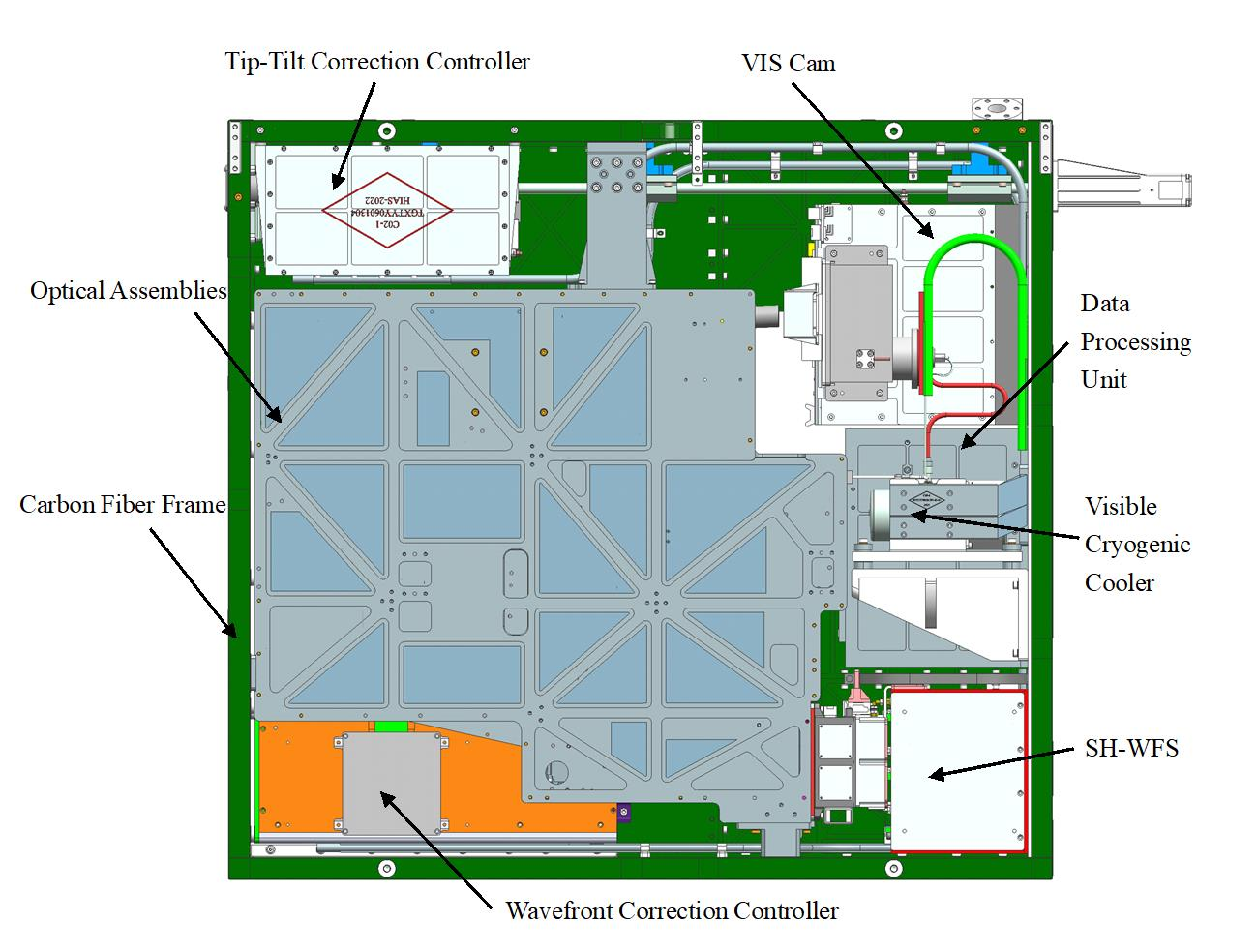}
   \caption{Internal Layout Diagram of M5.} 
   \label{fig:Internal Layout Diagram of Module 5}
\end{figure}

M5 (as shown in Figure \ref{fig:Schematic diagram of M5 structure}), serving as the SDU, is securely connected to the telescope via three titanium alloy interface points (A, B, and C)~\citep{https://doi.org/10.1155/vib/5575921}. It primarily consists of a carbon fiber frame, optical assemblies, a visible-light imaging camera, a visible filter wheel, a visible cryogenic cooler, a wavefront sensor camera, an aberration correction system (including a wavefront correction controller, a tip-tilt correction controller, a tip-tilt mirror, and an alignment laser source), and a data processing unit(as shown in Figure \ref{fig:Internal Layout Diagram of Module 5}). The module features a box-type structure, in which the carbon fiber frame forms an internal cavity. Each functional unit is distributed along the left and right sides of the cavity, with the rear side fixed to the carbon fiber frame. The optical assemblies are positioned in the upper section of the cavity and are mounted to the carbon fiber frame via substrate mounting bases 1 to 3. Incident light from the telescope enters through the optical substrate’s entrance aperture and is redirected by the M1 mirror into the internal optical system. On the left side of the module, heat is transferred to the evaporator mounting surface on the right via heat pipes. The right-side units dissipate heat primarily through the mounting surface, with some units employing heat pipes for additional thermal management. The S6 cable assembly within M5 is connected to the connectors of each functional unit and is routed along the mid-crossbeam of the frame for secure fixation.

\begin{figure} 
   \centering
   \includegraphics[width=12.0cm, angle=0]{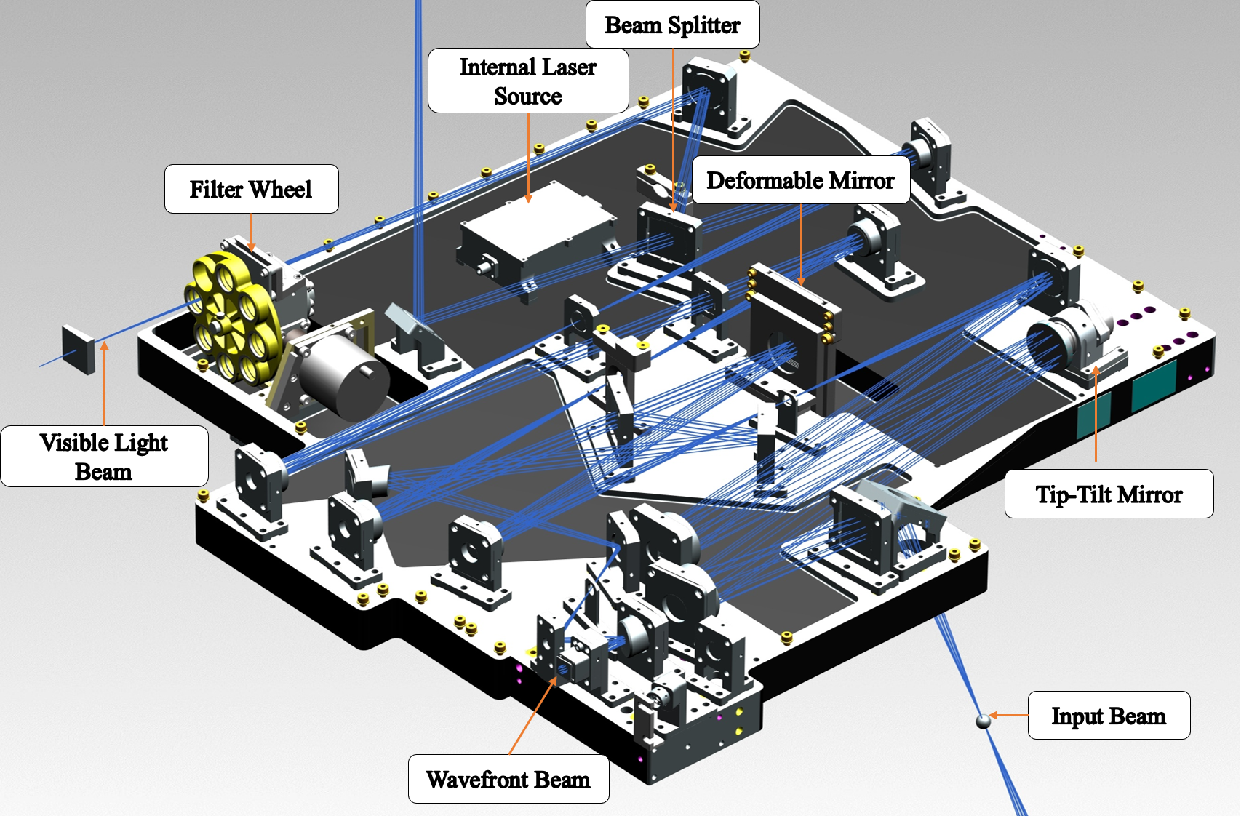}
   \caption{The optical system of CPI-C.} 
   \label{fig:The optical system of CPI-C}
\end{figure}

The optical system of CPI-C is primarily housed within the optical assembly, with all optical elements mounted on the optical substrate. A pick-off flat mirror located at the output of the telescope serves to relay the off-axis output beam to the coronagraph. The optical system (as shown in Figure \ref{fig:The optical system of CPI-C}) is composed of eight off-axis paraboloidal (OAP) mirrors. These optical relays create three different pupil planes and three focal planes along the optical path. The Fast Steering Mirror and the Deformable Mirror are also located in pupil planes to manipulate the wavefront exactly in its pupil. The focal points insert occulting masks. Two Beam splitters split the optical path into three beams: one is visible light observation, another is infrared observation, and the third is wavelength sensing.

\begin{figure} 
   \centering
   \includegraphics[width=12.0cm, angle=0]{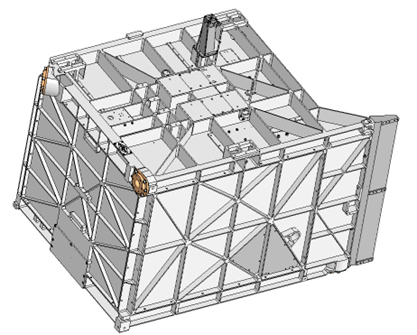}
   \caption{Schematic diagram of M6 structure.} 
   \label{fig:Schematic diagram of Module 6 structure}
\end{figure}
\begin{figure} 
   \centering
   \includegraphics[width=12.0cm, angle=0]{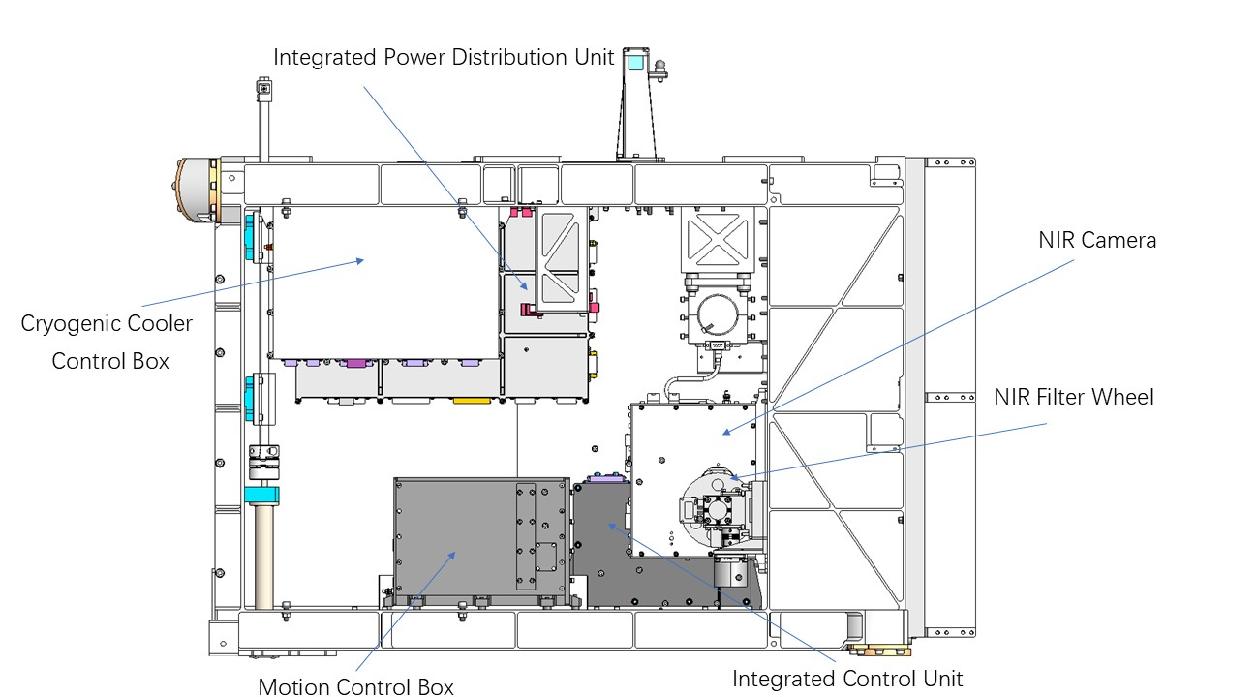}
   \caption{Internal Layout Diagram of M6.} 
   \label{fig:Internal Layout Diagram of Module 6}
\end{figure}
M6, serving as the CPSU, is securely connected to the telescope via four titanium alloy interface points (A, B, C, D). It primarily consists of an aluminum alloy frame, an integrated control unit, an integrated power distribution unit, a near-infrared camera, a refrigerator control box, a motion control box, and a filter wheel. The module features a box-type structure, in which the aluminum alloy frame forms an internal cavity. Each unit is fixed to the aluminum alloy frame with screws. Light from M5 enters the interior of M6 through the entrance aperture and finally enters the inside of the camera.
On the lower side of the module, heat is transferred to the evaporator mounting surface on the upper via heat pipes. The upper-side units dissipate heat primarily through the mounting surface, with some units employing heat pipes for additional thermal management. The S6 cable assembly within M6 is connected to the connectors of each functional unit and is routed along the frame for secure fixation. Figure \ref{fig:Schematic diagram of Module 6 structure} shows the structure of M6. Figure \ref{fig:Internal Layout Diagram of Module 6} shows the internal layout of M6.

The thermal design of the CPI-C utilizes both passive and active controls to achieve the desired operating temperature and thermal stability while minimizing the heater power required and the mass. The CPI-C consists of the fold mirror assembly, optics, optical bench, support structure, thermal cover, cameras, and electronics. The support structure, thermal cover, cameras, and electronics are all individually wrapped with multilayer insulation (MLI). The thermal hardware includes MLI, heaters, radiators, thermal straps, temperature sensors, and grooved heat pipes.

\begin{figure} 
   \centering
   \includegraphics[width=12.0cm, angle=0]{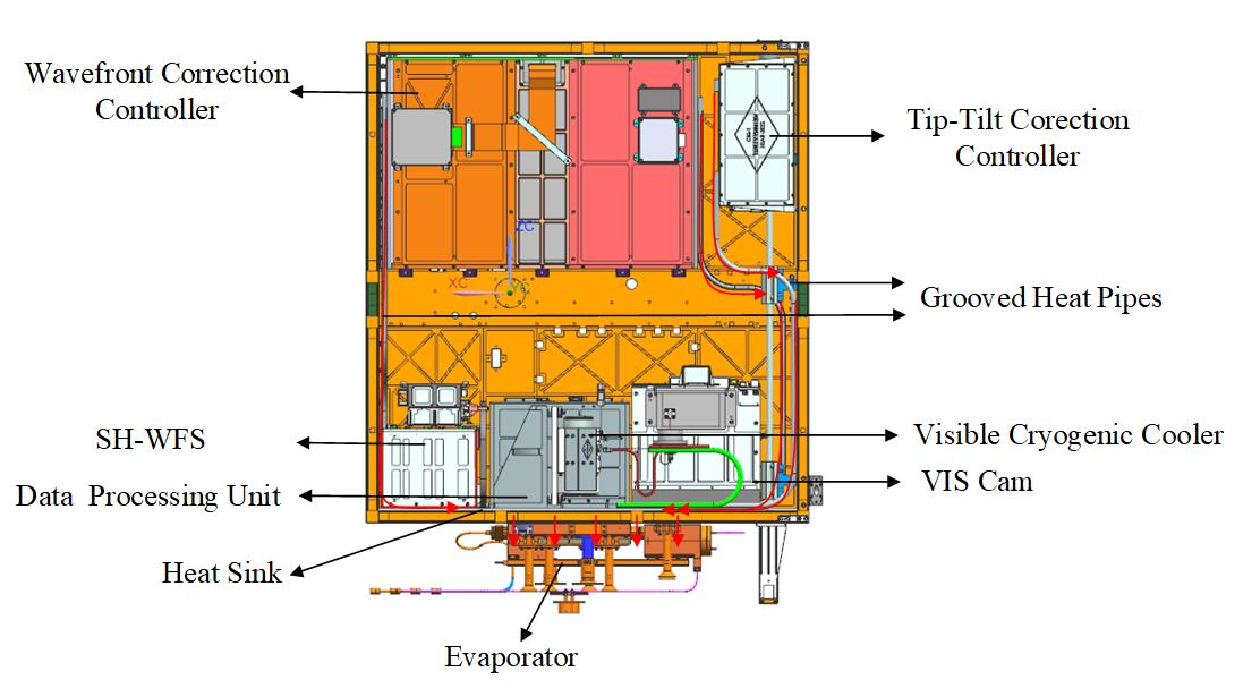}
   \centering
   \caption{M5 heat dissipation channel.} 
   \label{fig:Module 5 heat dissipation channel}
\end{figure}
\begin{figure} 
   \centering
   \includegraphics[width=12.0cm, angle=0]{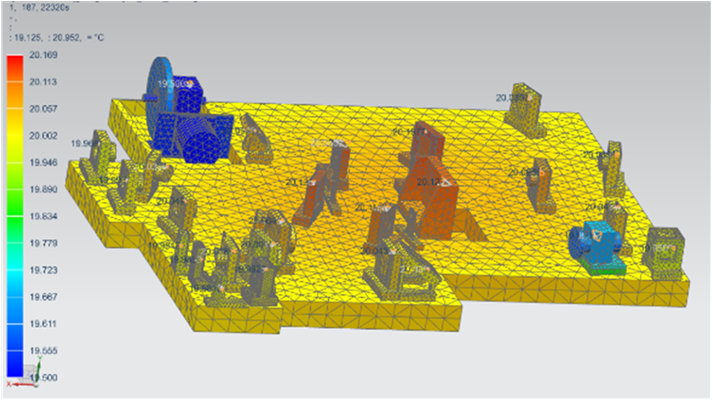}
   \centering
   \caption{The thermal math model generated temperature maps of the coronagraph optical bench.} 
   \label{fig:The thermal math model generated temperature maps of the coronagraph optical bench}
\end{figure}

The loop heat pipe (LHP) provided by CSST serves as the main heat dissipation channel for the heat generated by the electronic products inside the CPI-C. Since the heat-generating devices are distributed at multiple locations in the internal space and all have a certain distance from the LHP's evaporator, the thermal control system adopts a scheme of the common heat dissipation plate and grooved heat pipe bridging to establish thermal conduction between each heat source and the evaporator. In addition, some single units are installed on the common heat dissipation plate through direct heat conduction for heat dissipation. Figure \ref{fig:Module 5 heat dissipation channel} shows the heat dissipation channel of M5, including the connection between heat-generating devices, grooved heat pipes, common heat dissipation plate, and the evaporator of the loop heat pipe (LHP).

\begin{figure} 
   \centering
   \includegraphics[width=12.0cm, angle=0]{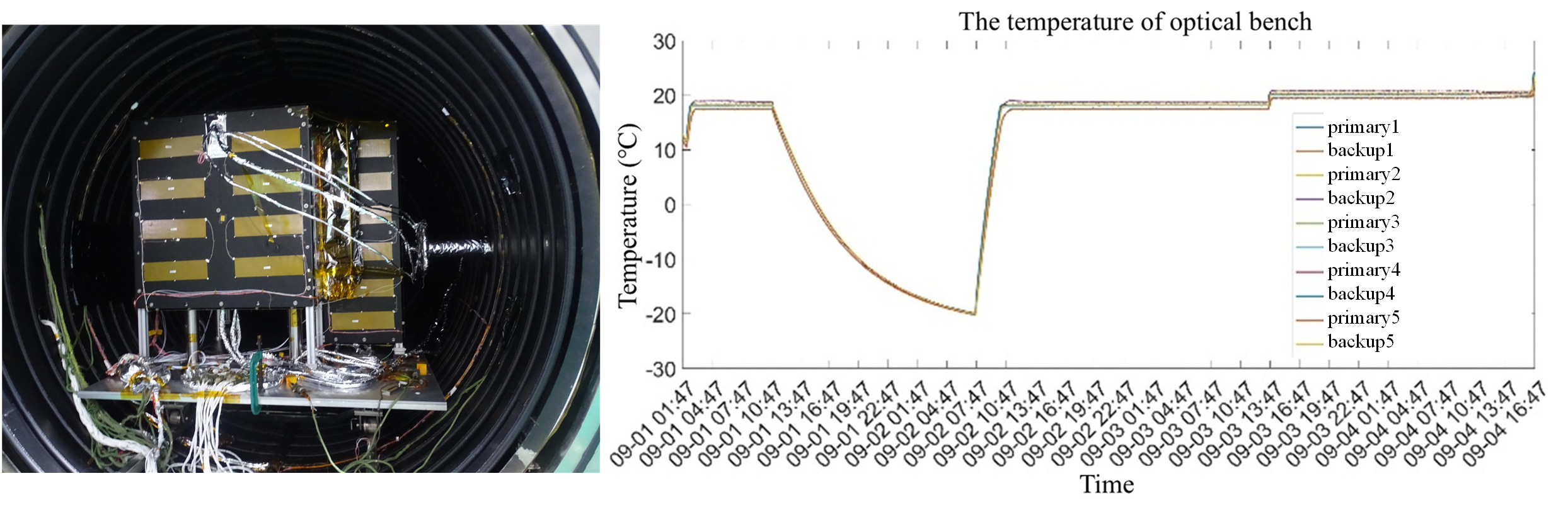}
   \centering
   \caption{CPI-C's thermal balance test layout diagram and temperature curves of the CPI-C’s optical bench in thermal balance test.} 
   \label{fig:CPI-C's thermal balance test layout diagram and temperature distribution map of the coronagraph optical bench in thermal balance test}
\end{figure}
By conducting thermal balance test to simulate the on-orbit operating modes of the CPI-C, including storage condition, low-temperature condition, high-temperature condition, and high-temperature alternating condition, the thermal control design of M5 and M6 is verified to meet the operating temperature requirements. The optical bench and thermal cover are isolated from the CPI-C support structure, and their temperature is controlled within the range of 20 ± 0.5 °C, which is a critical requirement for system performance. A thermal math model (TMM) of the CPI-C was developed to support structural thermal optical analysis and trade studies. The TMM simulates all the major components and structures of the CPI-C.  Figure \ref{fig:The thermal math model generated temperature maps of the coronagraph optical bench} shows the TMM-generated temperature maps of the coronagraph optical bench. Figure \ref{fig:CPI-C's thermal balance test layout diagram and temperature distribution map of the coronagraph optical bench in thermal balance test}  shows thermal balance test layout diagram and temperature curves of the CPI-C's optical bench in thermal balance test.

\subsection{Micro-vibration control \& test}

When the CPI-C is in operation, the cryocooler near the camera can generate a stable sinusoidal disturbance in the form of micro-vibrations. Once transmitted to the optical system, these disturbances can substantially degrade its imaging quality. Consequently, effective control of micro-vibrations is crucial for fully exploiting the performance potential of the coronagraph.

The cryocooler operates within a frequency range of 92-102 Hz. Accordingly, the vibration isolation system must provide effective attenuation within this range and its harmonics. Additionally, as with all space missions, the structural impact of the intense vibrations experienced during rocket launch must also be taken into account.

\begin{figure} [b]
   \centering
   \includegraphics[width=12.0cm, angle=0]{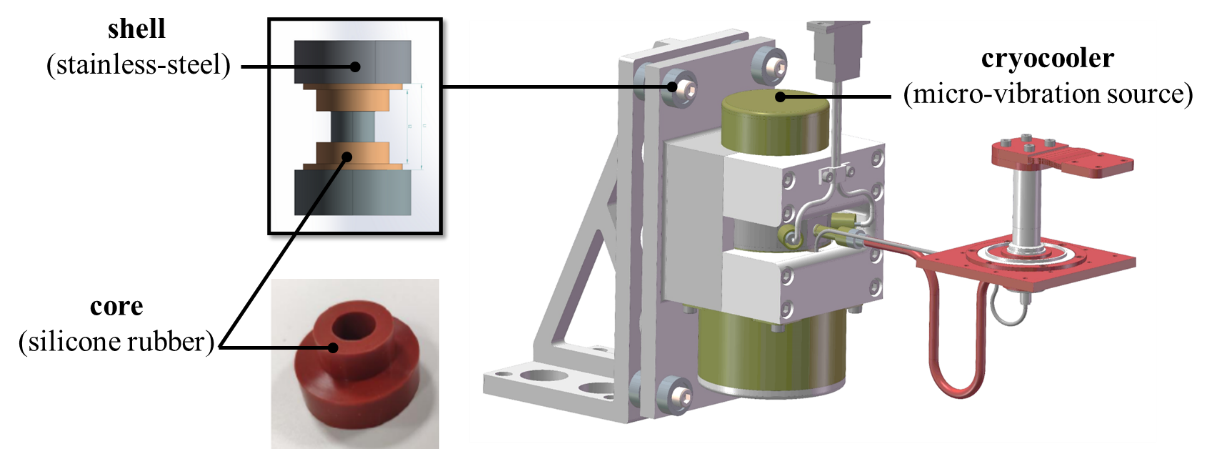}
   \caption{Micro-vibration isolation system} 
   \label{fig:mvibration}
\end{figure}

To meet these requirements, we proposed a variable-stiffness vibration isolator, as illustrated in Figure \ref{fig:mvibration}. It comprises a silicone rubber core enclosed within a stainless-steel shell. The stainless-steel shell serves to adjust the compression of the silicone rubber and provides physical limit protection when the vibration amplitude becomes excessive. As the compression of the silicone rubber changes, its stiffness exhibits a nonlinear response: at low compression levels, the stiffness is minimal, under on-orbit conditions, this low-stiffness behavior enhances micro-vibration isolation. During the launch phase, the intense vibrations induce greater compression of the silicone rubber, causing the isolator’s stiffness to increase correspondingly, thereby protecting the cryocooler from damage.

\begin{figure}
   \centering
   \includegraphics[width=15.0cm, angle=0]{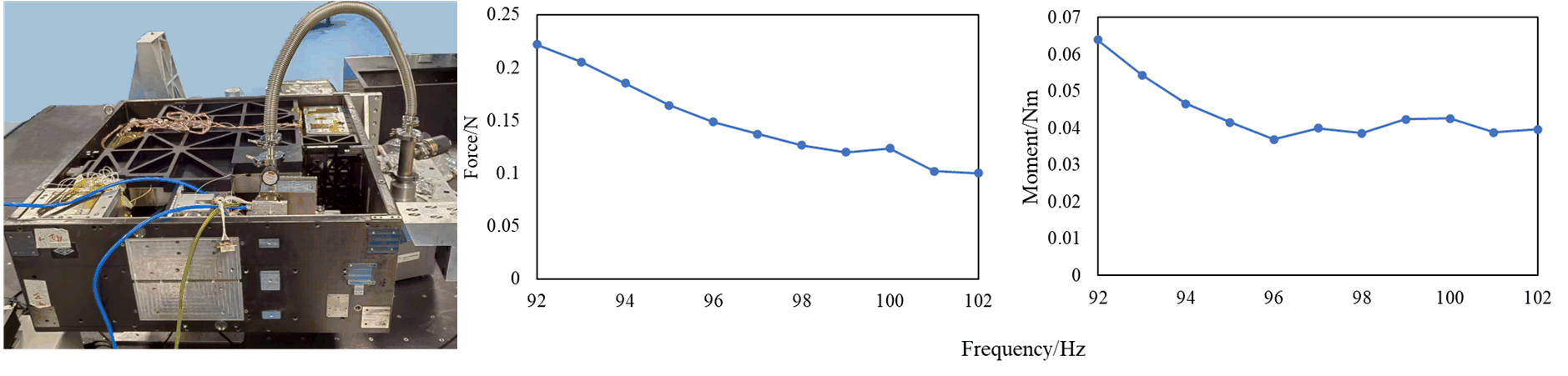}
   \caption{Left: CPI-C module micro-vibration test site. Right: Micro-vibration test results on the CPI-C module.} 
   \label{fig:mvresult}
\end{figure}

After completing the design of the vibration isolation system, a micro-vibration test was carried out on the entire CPI-C module to verify that the vibration isolation system could control the module’s micro-vibration output within the target range. As illustrated in Figure \ref{fig:mvresult} (Left), the module was mounted on a test fixture, with force sensors installed at the interface to measure the three-dimensional output forces and torques. Recently, we have proposed a noval six-dimensional micro-vibration measurement platform, which can offfer enhanced stiffness and expands the measurable frequency bandwidth, thus finally to improve the measurement precision. Details on the development of the micro-vibration measurement platform can be found in a recent paper~\citep{2025Meas..25117206C}. These measurements were then synthesized into two vectors representing the resultant force and resultant torque. The test results, shown in Figure \ref{fig:mvresult} (Right), indicate that the maximum disturbance force was 0.22 N and the maximum disturbance torque was 0.064 Nm, both of which meet the design requirements of 0.4 N and 0.1 Nm, respectively. In recent vibration AIT specific test, M5 in CPI-C on CSST has delivered a ~0.0053 arcseonds ($3\sigma$) vibration, when integrated with the setallite platform.

\subsection{Electronics}

\begin{figure} [b]
   \centering
   \includegraphics[width=12.0cm, angle=0]{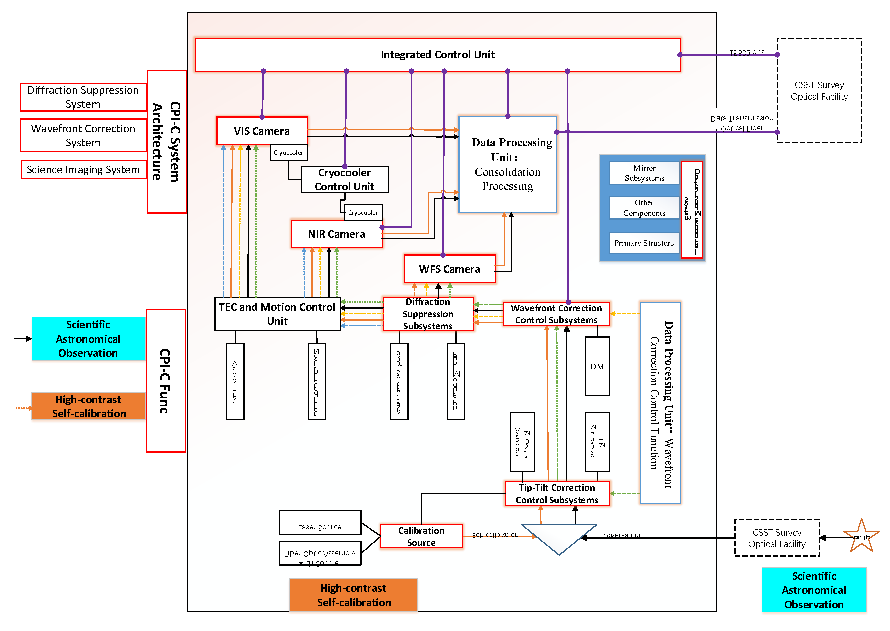}
   \caption{Block diagram of the CPI-C system} 
   \label{fig:function_block}
\end{figure}

Figure \ref{fig:function_block} shows the block diagram of the CPI-C system and its electronics. The CPI-C aims to detect faint exoplanets around bright star. Diffraction occurs when starlight passes through the telescope aperture, and static wavefront errors from the system's optical surfaces generate speckle noise. The diffraction pattern and speckle noise can overwhelm the signal of a planet. Diffracted light in the system's Point Spread Function (PSF) image is suppressed by modulating the energy distribution across the telescope pupil.
Speckle noise is mitigated by controlling the quasi-static wavefront errors  using a DM. After effectively suppressing diffracted photon noise and speckle noise, high-contrast imaging can be achieved in specific regions of the PSF.

The integrated control unit of the CPI-C receives command injections (e.g., data instructions and calibration files) from the telescope via the 1553B bus. It then distributes control commands to internal subsystems (e.g., data processing unit and visible imaging camera) through its internal control bus. Telemetry data collected locally and from internal subsystems are aggregated by the integrated control unit and transmitted back to the CSST Survey Optical Facility via the 1553B bus.
Telemetry data collected by the integrated control unit itself, combined with telemetry of other subsystems gathered via the internal bus, is summarized and transmitted back to the telescope via the 1553B bus.

Starlight and calibration light sequentially pass through CPI-C's tip-tilt correction, wavefront error correction, and diffraction suppression systems before reaching the visible imaging camera, near-infrared imaging camera, and wavefront sensing camera, respectively. Images from these three cameras undergo consolidation processing in the data processing unit and are then transmitted to the compression/storage module of the CSST Survey Optical Facility via a data transmission optical fiber.

\subsection{Software}

\begin{figure} 
   \centering
   \includegraphics[width=12.0cm, angle=0]{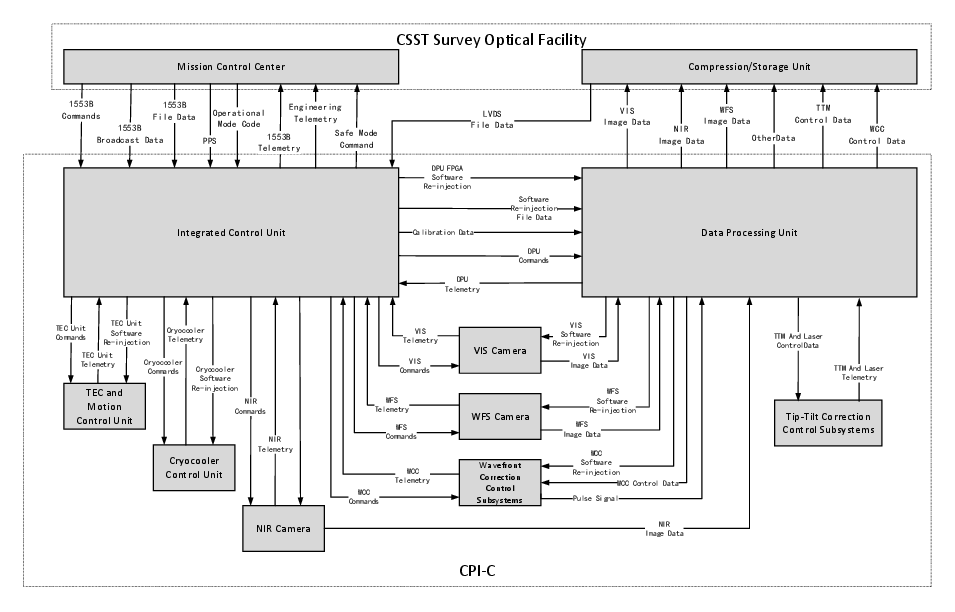}
   \caption{Information flow of CPI-C.} 
   \label{fig:software}
\end{figure}

The onboard software system of the CPI-C comprises 11 software modules including:
\begin{itemize}
    \item Integrated Control Software
    \item Visible Camera Signal Processing Software
    \item Near-Infrared Camera Signal Processing Software
    \item Wavefront Sensing Camera Signal Processing Software
    \item Consolidation Processing Software
    \item Image Processing Software
\end{itemize}

These software components are distributed across electronic units such as the integrated control unit, data processing unit, and visible imaging camera. Figure \ref{fig:software} illustrates information flows between external interfaces and internal electronic units of the CPI-C module. The CPI-C receives command injections and data files from the telescope, while returning digital telemetry and engineering telemetry reflecting the instrument's overall status. The CPI-C data processing unit transmits consolidated scientific data (including camera images) to the compression/storage module of the 
telescope.

\section{Proposed Mission Configuration and Profile}
\subsection{Mission Requirements and Analysis}

CPI-C employs a staring mode for observations. Upon pointing at the target star, it creates a high-contrast imaging dark region around the target and performs continuous exposure and readout for a maximum duration of 1200 seconds. The exposure time for individual frames can be adjusted between 0.01 seconds and 300 seconds. For deep observations requiring long integration times, multiple continuous exposure sequences are performed, often at different telescope roll angles. Between sequences, CPI-C conducts on-orbit geometric calibration to ensure that the planetary target remains optimally positioned within the high-contrast dark zone. This calibration accounts for small drift in telescope pointing and instrument alignment, compensating for thermal variations and mechanical settling over the observation period.

To achieve better calibration of system aberrations and avoid field obstruction, CPI-C can utilize a field rotation observation mode in orbit. Upon completing the readout of a set of exposure images, the telescope is rotated by a specific angle. By leveraging the positional changes of the exoplanet target and the variations in quasi-static aberrations at different rotation angles, this mode partially removes quasi-static aberrations, thereby enhancing the target's signal-to-noise ratio.

The approach of multiple sequences with intermediate calibration offers several advantages over a single ultra-long exposure: (1) it mitigates the accumulation of systematic errors from quasi-static speckles that may evolve on timescales of tens of minutes; (2) it enables rejection of cosmic-ray contaminated frames without losing the entire dataset; (3) it maintains the planet signal within the optimized dark-zone region where contrast performance is highest, preventing signal loss due to drift outside the working angle. The individual exposure sequences are subsequently co-added during post-processing, with alignment based on stellar centroiding and speckle pattern 
registration, yielding signal-to-noise ratios that scale approximately as √N for N 
independent sequences.

The survey follows a three-stage workflow to detect and validate planetary candidates:

\textbf{Stage 1: Candidate Detection}~
CPI-C using two high-contrast dark zones to suppress starlight and identify faint companions or circumstellar disks. Since planetary orbits may shift candidates outside the dark zones over time, multiple epochs of observation are required for each target.

\textbf{Stage 2: Candidate Validation}~
Candidates are validated through proper motion and parallax analysis. Co-moving companions share the same astrometric motion as their host stars, but background objects are in fixed positions.

\textbf{Stage 3: Orbital and Atmospheric Characterization}~
Validated candidates undergo: 1) Astrometric Monitoring: Multi-epoch positional measurements constrain orbital parameters (semi-major axis, eccentricity, inclination). 2) Multi-Band Photometry: Flux measurements in F661, F729, and F877 bands provide spectral energy distributions (SEDs). By fitting these SEDs to theoretical models, planetary radii, albedo, and atmospheric composition (e.g., methane abundance) are derived.

\subsection{Mission Profile}
Within the 10-year operational lifetime of CSST, CPI-C is allocated an estimated 10\% of the total time, equating to approximately one year. CPI-C can conduct joint observations with the Integral Field Spectrograph (IFS) and the Multi-Channel Imager (MCI), which are instruments located at the same plane of telescope's field of view. 

Over the one-year operational time, CPI-C's primary scientific objective is to conduct a high-contrast imaging survey of approximately 700 nearby stars to detect planets orbiting them. The instrument also supports multiple other scientific goals, including high-contrast infrared observations of exoplanets; observations of planetary disks; observations of planets orbiting binary or multi-star systems; observations of asteroids and the trans-neptune objects in the solar system. 

Beyond scientific observations, CPI-C requires calibration time, estimated at 10\% of its total allocated hours. Planned calibration activities include acquiring bias frames, dark frames, background frames, EM gain calibration frames, and flat-field frames. Furthermore, calibration of the instrument's real-time wavefront sensing system is essential. Additionally, when significant variations in system aberrations occur, CPI-C must also perform high-contrast imaging optimizations to re-establish the optimized dark region achieving a contrast of $10^{-8}$.

During a given observation window, targets may be at different stages of this detection process. Establishing a suitable target prioritization evaluation methodology ensures that targets at various detection stages and of differing magnitudes are rationally scheduled, thereby maximizing scientific productivity. 

CPI-C is subject to the following observational constraints during operations: (1) The angle between the line of sight and the solar vector must be no less than 65$^\circ$ (revised from 50$^\circ$); (2) The angle between the line of sight and the lunar vector must be no less than 40$^\circ$; (3) The angles between the line of sight and the Earth's bright limb/dark limb must be no less than 70$^\circ$ and 30$^\circ$, respectively; (4) No observational exposures are performed when traversing the South Atlantic Anomaly (SAA) region; (5) During sunlit orbital phases, the normal vector of the solar panels must maintain parallelism with the solar direction; (6) Orbital maneuvers must comply with satellite Control Moment Gyroscope (CMG) limitations, for example, allowing two 180-degree maneuvers or five 90-degree maneuvers per orbit.

\begin{figure} 
   \centering
   \includegraphics[width=12.0cm, angle=0]{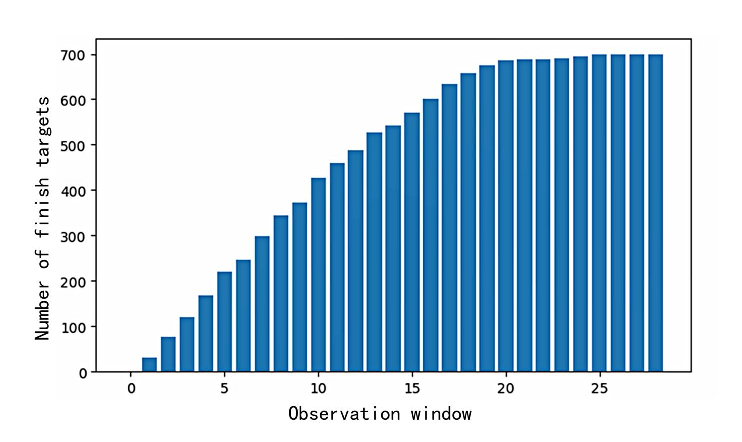}
   \caption{The cumulative number of targets observed within each time window was obtained using dynamic programming-based scheduling.} 
   \label{fig:windows}
\end{figure}

We employed a dynamic programming method to schedule the imaging survey of the 700 nearby stellar targets. The observation windows were initially planned based on the first batch of scientific observation windows, totaling 25 windows and accumulating 4320 hours of effective observation time. The results are as follows: The total exposure time required to complete one full observation cycle for all 700 targets is approximately 65 days. Figure \ref{fig:windows} shows the cumulative number of targets observed within each window, indicating that approximately 20 observation windows are needed to complete one full survey observation cycle.

\section{data reduction}
When images captured by CPI-C are transmitted to the ground, they enter the CSST's science data processing system. The data processing pipeline automatically processes the Level 0 data into Level 1 data. In the resulting Level 1 data, instrumental effects such as camera bias, dark current, flat-field response, and electron-multiplying (EM) gain are removed, and cosmic rays are flagged and corrected.

Within the high-contrast imaging dark zone of the Level 1 image data, speckle noise still exists, which is one of the most significant limiting factors for the direct imaging of exoplanets. When CPI-C images a target star, the high-contrast dark zones contain speckle noise caused by residual wavefront correction errors and quasi-static aberrations within the system. This speckle noise degrades the signal-to-noise ratio of planetary photometry. Developing the most effective methods to suppress this speckle noise will be a key area for future scientific research.

\begin{figure} 
   \centering
   \includegraphics[width=12cm, angle=0]{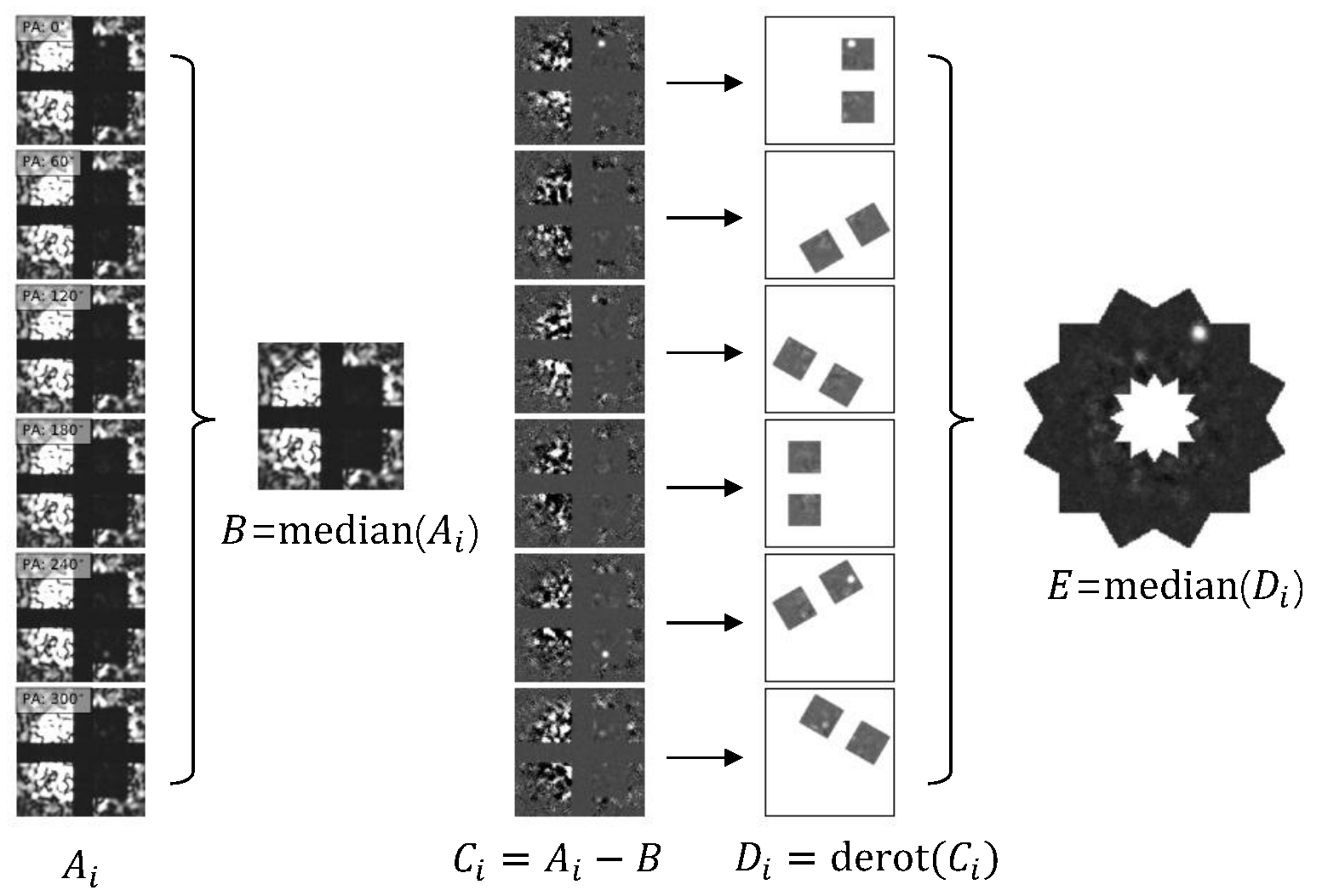}
   \caption{Schematic diagram of the ADI-based speckle removal algorithm for CPI-C data. Step 1: Capture images of the star at 6 position angles. Step 2: Generate a reference image by taking the median of the captured images. Step 3: Subtract the reference image from the original images to remove speckles. Step 4: Rotate and align the planets, then extract the high-contrast dark region. Step 5: Stack the dark region images to obtain the final processed image.} 
   \label{fig:ADI}
\end{figure}

Influenced by phase aberrations from the telescope and the instrument optics, speckle noise exhibits unique characteristics. Speckle noise does not follow a random distribution but has a certain lifetime. This means that simply increasing exposure time or co-adding multiple images does not effectively mitigate the impact of speckle noise, requiring more advanced processing techniques. The most commonly used high-contrast data processing technique in ground-based observations is the Angular Differential Imaging (ADI) \citep{marois06} method. For ground-based observation, ADI  takes the advantage of the field rotation in altazimuth alt-azimuth telescopes during tracking, which keeps the stellar speckles relatively stationary while causing planetary signals to rotate around the star.  One can fully use the field rotation to construct a reference PSF for subtraction to suppress the specekles while keep the planet signal stayed, the subtracted  images are derotated and aligned to produce a final planetary image with sufficient S/N. Building upon ADI, the Locally Optimized Combination of Images (LOCI) \citep{lafreniere07}  algorithm employs optimization techniques to construct an optimized reference image for each small region of the science image, rather than generating a single reference PSF for the entire frame.

Differentiating between high-contrast imaging techniques for ground-based and space-based telescopes is essential. The ADI method relies on field rotation induced by Earth's rotation and is inherently applicable to ground-based alt-azimuth telescopes. In contrast, space telescopes such as the CSST or RST cannot utilize this natural rotation. Instead, a similar observational effect can be achieved by physically rolling the spacecraft, a method often referred to as the roll subtraction ~\citep{2007Natur.446..771T}. 
In Figure \ref{fig:ADI}, we present a schematic diagram indicate the principles of processing CPI-C data using the ADI method. During observation, images are captured at 6 position angles. By constructing a reference frame and performing subtraction, the speckles can be removed. These images are then rotated to align the planets and stacked and produce the final image. Using ADI, one can not only remove speckles but also provides a complete 360-degree field of view around the star.
This maneuver generates the necessary field rotation for speckle suppression, enabling the use of angular differential imaging algorithms such as the Optimized Image Rotation and Subtraction (O-IRS) technique developed by Dou et al. or LOCI, which have been successfully demonstrated in previous space-based observations.
The development of speckle suppression algorithms has progressed along two parallel paths. For ground-based telescopes equipped with extreme adaptive optics (AO), Ren et al. identified the second-order components of the residual wavefront as the dominant speckle noise source and demonstrated that their symmetry properties could be exploited through image rotation ~\citep{ren12}. Dou et al. later advanced this concept into the O-IRS technique, which incorporates optimization to achieve a higher signal-to-noise ratio ~\citep{2015ApJ...802...12D}.  Importantly, the quasi-static wavefront characteristics of a space coronagraph are analogous to the post-AO residual wavefront in ground-based systems. Therefore, techniques like O-IRS that leverage image rotation and optimization are also expected to be highly applicable to CPI-C data.
Alongside these angular differential methods, Reference Star Differential Imaging (RDI) serves as a powerful and versatile approach used by both ground-based and space-based observatories. For scenarios with minimal or no field rotation, an advanced form of this technique is particularly effective. Ren \& Chen  developed a global optimization-based reference star differential imaging algorithm (G-RDI)~\citep{2021MNRAS.502.2158R}, which is especially well-suited for the stable  PSF of space-based instruments like the CSST and is capable of approaching the photon noise limit. Thus, for CPI-C data reduction, we anticipate that both roll-subtraction-assisted algorithms (e.g., O-IRS) and the advanced G-RDI method will be highly beneficial.

\section{summary}
As one of the most important scientific instruments on CSST, CPI-C is composed of SDU in M5 and CPSU in M6, rescpectively. CPI-C employs a step-transmission apodization technique to suppress the diffraction noises and a precise phase aberration control technique to eliminate the speckle noises, which will deliver a contrast on the order better than $10^{−8}$ , with the support of the specialized data reduction technique. CPI-C is allocated approximately 10\% of CSST's total observation time, over its 10-year mission timeline, with the support of ``joint observation mode''. CPI-C's primary mission is to conduct the high-contrast direct imaging survey of nearby solar-type stars, to firstly detect and characterize the``cool'' exoplanets, ranging in size from Neptunes to Jupiters, at an orbital separation within 5 AU. We will also conduct the follow-up observations of planets detected by the RV or astrometric method to further constrain the mass, and to perform the high-contrast imaging of circumstellar disks to study the planet formation. The Key scientific goals include not only the discovery of the cool planets but also the further characterization through multi-band photometry to further constrain its physical properties, including the mass, radius, effective temperature and even the atmospheric composition. Finally, we will provide a systematic statistical analysis of the CPI-C's whole target samples, including both the detections and non-detections. The observation results will further enrich the number and variety of exoplanets, providing important evidence to understand the formation and evolution mechanisms of planets, and create possibilities for the comparative planetary science research. Such a research will lay a solid foundation for the future direct detection and spectroscopy analysis of the “Earth-mass” planets in the HZ of a solar type star, which will help to finally answer "Are we alone in the universe?", one the most fundamental scientific questions for human-beings. 

\appendix

\section{CPI-C target source selection (partial)} \label{sec:app}

In this section, we provide a partial list of the stellar targets considered for CPI-C observations (Table~\ref{tab:hip}). 
The sources represent potential candidates for high-contrast imaging due to their proximity and brightness, which are favorable for exoplanet detection. 
For brevity, only a subset of the targets is presented here. 
The full list of sources, including additional stellar parameters and completeness notes, is available from the corresponding authors upon request.

\begin{table}[ht]
\bc
\begin{minipage}[]{100mm}
\caption[]{Basic information of selected stars.\label{tab:hip}}
\end{minipage}
\setlength{\tabcolsep}{3pt}
\small
\begin{tabular}{cccc}
\hline\noalign{\smallskip}
Hipparcos identifier & parallax(mas) & $V$ (mag) & spectral type \\
\hline\noalign{\smallskip}
69673  & 88.85  & $-0.05$ & K2IIIp \\
8102   & 274.17 & 3.49    & G8V \\
99240  & 163.73 & 3.55    & G5IV-Vvar \\
16537  & 310.75 & 3.72    & K2V \\
88601  & 196.62 & 4.03    & K0V\\
22449  & 124.60 & 3.19    & F6V \\
84405  & 167.08 & 4.33    & K2III \\
19849  & 198.24 & 4.43    & K1V \\
2021   & 133.78 & 2.82    & G2IV \\
108870 & 275.76 & 4.69    & K5V \\
104214 & 287.13 & 5.20    & K5V \\
3821   & 167.99 & 3.46    & G0V\\
37279  & 285.93 & 0.40    & F5IV-V \\
37826  & 96.74  & 1.16    & K0IIIvar \\
21421  & 50.09  & 0.87    & K5III \\
71681  & 742.12 & 1.35    & K1V \\
71683  & 742.12 & $-0.01$ & G2V \\
17378  & 110.58 & 3.52    & K0IV \\
15510  & 165.02 & 4.26    & G8V \\
27072  & 111.49 & 3.59    & F7V \\
104217 & 285.42 & 6.05    & K7V \\
81693  & 92.63  & 2.81    & F9IV \\
86974  & 119.05 & 3.42    & G5IV \\
61941  & 84.53  & 2.74    & F0V \\
89937  & 124.11 & 3.55    & F7Vvar \\
67927  & 88.17  & 2.68    & G0IV \\
68933  & 53.52  & 2.06    & K0IIIb \\
72659  & 149.26 & 4.54    & G8V+K4V \\
116727 & 72.50  & 3.21    & K1IV \\
9884   & 49.48  & 2.01    & K2III \\
77952  & 81.24  & 2.83    & F2III \\
99461  & 165.24 & 5.32    & K2V \\
61317  & 119.46 & 4.24    & G0V \\
57757  & 91.74  & 3.59    & F8V \\
1599   & 116.38 & 4.23    & F9V \\
746    & 59.89  & 2.28    & F2III-IV \\
105858 & 108.50 & 4.21    & F6V \\
64394  & 109.23 & 4.23    & G0V \\
82396  & 49.85  & 2.29    & K2IIIb \\
27913  & 115.43 & 4.39    & G0V \\
\noalign{\smallskip}\hline
\end{tabular}
\ec
\tablecomments{0.95\textwidth}{Only part of the table is listed here for illustration. The full dataset can be obtained by contacting the corresponding authors.}
\end{table}

\begin{acknowledgements}
We thank the anonymous referee for valuable comments that have significantly improved the manuscript. This research was supported by the China Manned Space Project (grants nos.CMS-CSST-201906, CMS-CSST-2025-A18, CMS-CSST-2021-A11 and CMS-CSST-2021-B04) and the National Natural Science Foundation of China (grants nos.11827804, U2031210, 11433007).
\end{acknowledgements}

\bibliographystyle{raa}
\bibliography{bibtex}

@article{absil11,
  title = {Performance Study of Ground-Based Infrared Bracewell Interferometers . Application to the Detection of Exozodiacal Dust Disks with GENIE},
  author = {Absil, O. and {den Hartog}, R. and Gondoin, P. and Fabry, P. and Wilhelm, R. and Gitton, P. and Puech, F.},
  year = {2011},
  month = mar,
  journal = {\aap},
  volume = {527},
  pages = {C4},
  issn = {0004-6361},
  doi = {10.1051/0004-6361/20053516e},
  urldate = {2025-09-04}
}

@article{bowler18,
  title = {Orbit and Dynamical Mass of the Late-T Dwarf GL 758 B},
  author = {Bowler, Brendan P. and Dupuy, Trent J. and Endl, Michael and Cochran, William D. and MacQueen, Phillip J. and Fulton, Benjamin J. and Petigura, Erik A. and Howard, Andrew W. and Hirsch, Lea and Kratter, Kaitlin M. and Crepp, Justin R. and Biller, Beth A. and Johnson, Marshall C. and Wittenmyer, Robert A.},
  year = {2018},
  month = apr,
  journal = {\aj},
  volume = {155},
  pages = {159},
  issn = {0004-6256},
  doi = {10.3847/1538-3881/aab2a6},
  urldate = {2025-09-04}
}

@article{ren12,
  title = {SPECKLE NOISE SUBTRACTION AND SUPPRESSION WITH ADAPTIVE OPTICS CORONAGRAPHIC IMAGING},
  author = {Ren, Deqing and Dou, Jiangpei and Zhang, Xi and Zhu, Yongtian},
  year = {2012},
  month = jul,
  journal = {\apj},
  volume = {753},
  number = {2},
  pages = {99},
  issn = {0004-637X, 1538-4357},
  doi = {10.1088/0004-637X/753/2/99},
  urldate = {2022-04-29},
  langid = {english}
}

@article{marois06,
  title = {Angular Differential Imaging: A Powerful High-Contrast Imaging Technique},
  shorttitle = {Angular Differential Imaging},
  author = {Marois, Christian and Lafreniere, David and Doyon, Rene and Macintosh, Bruce and Nadeau, Daniel},
  year = {2006},
  month = apr,
  journal = {\apj},
  volume = {641},
  number = {1},
  pages = {556--564},
  issn = {0004-637X, 1538-4357},
  doi = {10.1086/500401},
  urldate = {2025-04-10},
  langid = {english}
}

@article{lafreniere07,
  title = {A New Algorithm for Point-Spread Function Subtraction in High-Contrast Imaging: A Demonstration with Angular Differential Imaging},
  shorttitle = {A New Algorithm for Point-Spread Function Subtraction in High-Contrast Imaging},
  author = {Lafreniere, David and Marois, Christian and Doyon, Rene and Nadeau, Daniel and Artigau, Etienne},
  year = {2007},
  month = may,
  journal = {\apj},
  volume = {660},
  number = {1},
  pages = {770--780},
  issn = {0004-637X, 1538-4357},
  doi = {10.1086/513180},
  urldate = {2025-04-10},
  langid = {english}
}

@ARTICLE{2021MNRAS.502.2158R,
       author = {{Ren}, Deqing and {Chen}, Yili},
        title = "{Global optimization-based reference star differential imaging for high-contrast exoplanet imaging survey}",
      journal = {\mnras},
         year = 2021,
        month = apr,
       volume = {502},
       number = {2},
        pages = {2158-2171},
          doi = {10.1093/mnras/stab022},
       adsurl = {https://ui.adsabs.harvard.edu/abs/2021MNRAS.502.2158R}
}

@inproceedings{dou14,
  title = {A Coronagraph Based on Two Spatial Light Modulators for Active Amplitude Apodizing and Phase Corrections},
  booktitle = {Proc.SPIE},
  author = {Dou, Jiangpei and Ren, Deqing and Zhang, Xi and Zhu, Yongtian and Zhao, Gang and Wu, Zhen and Chen, Rui and Liu, Chengchao and Yang, Feng and Yang, Chao},
  year = {2014},
  month = aug,
  volume = {9147},
  pages = {91478O},
  doi = {10.1117/12.2055327},
  urldate = {2025-09-04}
}

@article{lacy19,
  title = {Characterization of Exoplanet Atmospheres with the Optical Coronagraph on WFIRST},
  author = {Lacy, B. and Shlivko, D. and Burrows, A.},
  year = {2019},
  month = mar,
  journal = {\aj},
  volume = {157},
  pages = {132},
  issn = {0004-6256},
  doi = {10.3847/1538-3881/ab0415},
  urldate = {2025-09-04}
}

@article{liu20,
  title = {A Tale of Planet Formation: From Dust to Planets},
  shorttitle = {A Tale of Planet Formation},
  author = {Liu, Beibei and Ji, Jianghui},
  year = {2020},
  month = oct,
  journal = {RAA},
  volume = {20},
  pages = {164},
  issn = {1674-4527},
  doi = {10.1088/1674-4527/20/10/164},
  urldate = {2025-09-04}
}

@article{ren07,
  title = {A Coronagraph Based on Stepped-Transmission Filters},
  author = {Ren, Deqing and Zhu, Yongtian},
  year = {2007},
  month = sep,
  journal = {\pasp},
  volume = {119},
  pages = {1063--1068},
  issn = {0004-6280},
  doi = {10.1086/522015},
  urldate = {2025-09-04}
}

@article{ren10,
  title = {A Transmission-Filter Coronagraph: Design and Test},
  shorttitle = {A Transmission-Filter Coronagraph},
  author = {Ren, Deqing and Dou, Jiangpei and Zhu, Yongtian},
  year = {2010},
  month = may,
  journal = {\pasp},
  volume = {122},
  number = {891},
  pages = {590},
  issn = {0004-6280},
  doi = {10.1086/652958},
  urldate = {2025-09-04},
  langid = {english}
}

@article{2021ARA&A..59..291Z,
  title = {Exoplanet Statistics and Theoretical Implications},
  author = {Zhu, Wei and Dong, Subo},
  year = {2021},
  month = sep,
  journal = {Annual Review of A\&A},
  volume = {59},
  pages = {291--336},
  issn = {0066-4146},
  doi = {10.1146/annurev-astro-112420-020055},
  urldate = {2025-09-04}
}

@ARTICLE{1996Icar..124...62P,
       author = {{Pollack}, James B. and {Hubickyj}, Olenka and {Bodenheimer}, Peter and {Lissauer}, Jack J. and {Podolak}, Morris and {Greenzweig}, Yuval},
        title = "{Formation of the Giant Planets by Concurrent Accretion of Solids and Gas}",
      journal = {\icarus},
         year = 1996,
        month = nov,
       volume = {124},
       number = {1},
        pages = {62-85},
          doi = {10.1006/icar.1996.0190},
       adsurl = {https://ui.adsabs.harvard.edu/abs/1996Icar..124...62P}
}

@ARTICLE{2002ApJ...576..870P,
       author = {{Padoan}, Paolo and {Nordlund}, {\r{A}}ke},
        title = "{The Stellar Initial Mass Function from Turbulent Fragmentation}",
      journal = {\apj},
         year = 2002,
        month = sep,
       volume = {576},
       number = {2},
        pages = {870-879},
          doi = {10.1086/341790},
archivePrefix = {arXiv},
       eprint = {astro-ph/0011465},
 primaryClass = {astro-ph},
       adsurl = {https://ui.adsabs.harvard.edu/abs/2002ApJ...576..870P}
}

@ARTICLE{2008Sci...322.1348M,
       author = {{Marois}, Christian and {Macintosh}, Bruce and {Barman}, Travis and {Zuckerman}, B. and {Song}, Inseok and {Patience}, Jennifer and {Lafreni{\`e}re}, David and {Doyon}, Ren{\'e}},
        title = "{Direct Imaging of Multiple Planets Orbiting the Star HR 8799}",
      journal = {Science},
         year = 2008,
        month = nov,
       volume = {322},
       number = {5906},
        pages = {1348},
          doi = {10.1126/science.1166585},
archivePrefix = {arXiv},
       eprint = {0811.2606},
 primaryClass = {astro-ph},
       adsurl = {https://ui.adsabs.harvard.edu/abs/2008Sci...322.1348M}
}

@ARTICLE{2015ApJ...802...12D,
       author = {{Dou}, Jiangpei and {Ren}, Deqing and {Zhao}, Gang and {Zhang}, Xi and {Chen}, Rui and {Zhu}, Yongtian},
        title = "{A High-contrast Imaging Algorithm: Optimized Image Rotation and Subtraction}",
      journal = {\apj},
         year = 2015,
        month = mar,
       volume = {802},
       number = {1},
          eid = {12},
        pages = {12},
          doi = {10.1088/0004-637X/802/1/12},
archivePrefix = {arXiv},
       eprint = {1501.03893},
 primaryClass = {astro-ph.IM},
       adsurl = {https://ui.adsabs.harvard.edu/abs/2015ApJ...802...12D}
}

@ARTICLE{2006ApJS..167...81G,
       author = {{Guyon}, O. and {Pluzhnik}, E.~A. and {Kuchner}, M.~J. and {Collins}, B. and {Ridgway}, S.~T.},
        title = "{Theoretical Limits on Extrasolar Terrestrial Planet Detection with Coronagraphs}",
      journal = {\apjs},
         year = 2006,
        month = nov,
       volume = {167},
       number = {1},
        pages = {81-99},
          doi = {10.1086/507630},
archivePrefix = {arXiv},
       eprint = {astro-ph/0608506},
 primaryClass = {astro-ph},
       adsurl = {https://ui.adsabs.harvard.edu/abs/2006ApJS..167...81G}
}

@ARTICLE{2010RAA....10..189D,
       author = {{Dou}, Jiang-Pei and {Ren}, De-Qing and {Zhu}, Yong-Tian},
        title = "{High-contrast coronagraph for ground-based imaging of Jupiter-like planets}",
      journal = {RAA},
         year = 2010,
        month = feb,
       volume = {10},
       number = {2},
        pages = {189-198},
          doi = {10.1088/1674-4527/10/2/010},
archivePrefix = {arXiv},
       eprint = {0910.5355},
 primaryClass = {astro-ph.IM},
       adsurl = {https://ui.adsabs.harvard.edu/abs/2010RAA....10..189D}
}

@ARTICLE{2021RAA....21...82Z,
       author = {{Zhu}, Yong-Tian and {Dou}, Jiang-Pei and {Zhang}, Xi and {Zhao}, Gang and {Guo}, Jing and {Infante}, Leopoldo},
        title = "{Portable adaptive optics for exoplanet imaging}",
      journal = {RAA},
         year = 2021,
        month = may,
       volume = {21},
       number = {4},
          eid = {082},
        pages = {082},
          doi = {10.1088/1674-4527/21/4/82},
       adsurl = {https://ui.adsabs.harvard.edu/abs/2021RAA....21...82Z}
}

@ARTICLE{1965Natur.207..568L,
       author = {{Lovelock}, J.~E.},
        title = "{A Physical Basis for Life Detection Experiments}",
      journal = {\nat},
         year = 1965,
        month = aug,
       volume = {207},
       number = {4997},
        pages = {568-570},
          doi = {10.1038/207568a0},
       adsurl = {https://ui.adsabs.harvard.edu/abs/1965Natur.207..568L}
}

@ARTICLE{1975RSPSB.189..167L,
       author = {{Lovelock}, J.~E.},
        title = "{Thermodynamics and the Recognition of Alien Biospheres}",
      journal = {P ROY SOC A-MATH PHY Series B},
         year = 1975,
        month = may,
       volume = {189},
       number = {1095},
        pages = {167-180},
          doi = {10.1098/rspb.1975.0051},
       adsurl = {https://ui.adsabs.harvard.edu/abs/1975RSPSB.189..167L}
}

@ARTICLE{2011PASP..123..341R,
       author = {{Ren}, Deqing and {Zhu}, Yongtian},
        title = "{A Coronagraph Using a Liquid Crystal Array and a Deformable Mirror for Active Apodizing and Phase Corrections}",
      journal = {\pasp},
         year = 2011,
        month = mar,
       volume = {123},
       number = {901},
        pages = {341},
          doi = {10.1086/659127},
       adsurl = {https://ui.adsabs.harvard.edu/abs/2011PASP..123..341R}
}

@Article{Shen20,
title = {High-contrast Imaging Design and Numerical Simulation of Exoplanets Detection in Full-working Area},
journal = {ATI},
volume = {17},
number = {1},
pages = {68-75},
year = {2020},
issn = {2097-3675},
url = {http://www.ati.ac.cn/en/article/id/2379},
author = {{Shen}, Yuliang and {Dou}, Jiangpei}
}

@ARTICLE{2011RAA....11..198D,
       author = {{Dou}, Jiang-Pei and {Ren}, De-Qing and {Zhu}, Yong-Tian},
        title = "{An iterative wavefront sensing algorithm for high-contrast imaging systems}",
      journal = {RAA},
         year = 2011,
        month = feb,
       volume = {11},
       number = {2},
        pages = {198-204},
          doi = {10.1088/1674-4527/11/2/008},
archivePrefix = {arXiv},
       eprint = {1009.2561},
 primaryClass = {astro-ph.IM},
       adsurl = {https://ui.adsabs.harvard.edu/abs/2011RAA....11..198D}
}

@ARTICLE{2012PASP..124..247R,
       author = {{Ren}, Deqing and {Dong}, Bing and {Zhu}, Yongtian and {Christian}, Damian J.},
        title = "{Correction of Non-Common-Path Error for Extreme Adaptive Optics}",
      journal = {\pasp},
         year = 2012,
        month = mar,
       volume = {124},
       number = {913},
        pages = {247},
          doi = {10.1086/664947},
       adsurl = {https://ui.adsabs.harvard.edu/abs/2012PASP..124..247R}
}

@ARTICLE{2016ApJ...832...84D,
       author = {{Dou}, Jiangpei and {Ren}, Deqing},
        title = "{Phase Quantization Study of Spatial Light Modulator for Extreme High-contrast Imaging}",
      journal = {\apj},
         year = 2016,
        month = nov,
       volume = {832},
       number = {1},
          eid = {84},
        pages = {84},
          doi = {10.3847/0004-637X/832/1/84},
archivePrefix = {arXiv},
       eprint = {1609.04870},
 primaryClass = {astro-ph.IM},
       adsurl = {https://ui.adsabs.harvard.edu/abs/2016ApJ...832...84D}
}

@ARTICLE{2019OptEn..58a4102R,
       author = {{Ren}, Deqing and {Wang}, Gang and {Zhang}, Xi},
        title = "{Approach for deformable mirror wavefront error correction}",
      journal = {OE},
         year = 2019,
        month = jan,
       volume = {58},
          eid = {014102},
        pages = {014102},
          doi = {10.1117/1.OE.58.1.014102},
       adsurl = {https://ui.adsabs.harvard.edu/abs/2019OptEn..58a4102R}
}

@ARTICLE{2020PASJ...72...30R,
       author = {{Ren}, Deqing and {Wang}, Gang},
        title = "{A low-cost and duplicable portable solar adaptive optics system based on LabVIEW hybrid programming}",
      journal = {\pasj},
         year = 2020,
        month = apr,
       volume = {72},
       number = {2},
          eid = {30},
        pages = {30},
          doi = {10.1093/pasj/psaa006},
       adsurl = {https://ui.adsabs.harvard.edu/abs/2020PASJ...72...30R}
}

@Article{zhang22,
title = {Staticaberration correction technique for adaptive optics system based on focal-plane copy approach},
journal = {Chinese Optics},
volume = {15},
number = {3},
pages = {545-551},
year = {2022},
issn = {2097-1842},
doi = {10.37188/CO.2021-0182},
url = {https://www.chineseoptics.net.cn/cn/article/doi/10.37188/CO.2021-0182},
author  = {Zhang, Tian-Yu and Wang, Gang and Zhang, Xi and Dou, Jiang-Pei},
}

@ARTICLE{2001RvMP...73..719B,
       author = {{Burrows}, Adam and {Hubbard}, W.~B. and {Lunine}, J.~I. and {Liebert}, James},
        title = "{The theory of brown dwarfs and extrasolar giant planets}",
      journal = {RMP},
         year = 2001,
        month = jul,
       volume = {73},
       number = {3},
        pages = {719-765},
          doi = {10.1103/RevModPhys.73.719},
archivePrefix = {arXiv},
       eprint = {astro-ph/0103383},
 primaryClass = {astro-ph},
       adsurl = {https://ui.adsabs.harvard.edu/abs/2001RvMP...73..719B}
}

@ARTICLE{2015ApJ...814L..27C,
       author = {{Currie}, Thayne and {Cloutier}, Ryan and {Brittain}, Sean and {Grady}, Carol and {Burrows}, Adam and {Muto}, Takayuki and {Kenyon}, Scott J. and {Kuchner}, Marc J.},
        title = "{Resolving the HD 100546 Protoplanetary System with the Gemini Planet Imager: Evidence for Multiple Forming, Accreting Planets}",
      journal = {\apjl},
         year = 2015,
        month = dec,
       volume = {814},
       number = {2},
          eid = {L27},
        pages = {L27},
          doi = {10.1088/2041-8205/814/2/L27},
archivePrefix = {arXiv},
       eprint = {1511.02526},
 primaryClass = {astro-ph.EP},
       adsurl = {https://ui.adsabs.harvard.edu/abs/2015ApJ...814L..27C}
}

@ARTICLE{2015Natur.527..342S,
       author = {{Sallum}, S. and {Follette}, K.~B. and {Eisner}, J.~A. and {Close}, L.~M. and {Hinz}, P. and {Kratter}, K. and {Males}, J. and {Skemer}, A. and {Macintosh}, B. and {Tuthill}, P. and {Bailey}, V. and {Defr{\`e}re}, D. and {Morzinski}, K. and {Rodigas}, T. and {Spalding}, E. and {Vaz}, A. and {Weinberger}, A.~J.},
        title = "{Accreting protoplanets in the LkCa 15 transition disk}",
      journal = {\nat},
         year = 2015,
        month = nov,
       volume = {527},
       number = {7578},
        pages = {342-344},
          doi = {10.1038/nature15761},
archivePrefix = {arXiv},
       eprint = {1511.07456},
 primaryClass = {astro-ph.EP},
       adsurl = {https://ui.adsabs.harvard.edu/abs/2015Natur.527..342S}
}

@ARTICLE{2007ApJ...655..541M,
       author = {{Marley}, Mark S. and {Fortney}, Jonathan J. and {Hubickyj}, Olenka and {Bodenheimer}, Peter and {Lissauer}, Jack J.},
        title = "{On the Luminosity of Young Jupiters}",
      journal = {\apj},
         year = 2007,
        month = jan,
       volume = {655},
       number = {1},
        pages = {541-549},
          doi = {10.1086/509759},
archivePrefix = {arXiv},
       eprint = {astro-ph/0609739},
 primaryClass = {astro-ph},
       adsurl = {https://ui.adsabs.harvard.edu/abs/2007ApJ...655..541M}
}

@ARTICLE{2011ApJ...729..128C,
       author = {{Currie}, Thayne and {Burrows}, Adam and {Itoh}, Yoichi and {Matsumura}, Soko and {Fukagawa}, Misato and {Apai}, Daniel and {Madhusudhan}, Nikku and {Hinz}, Philip M. and {Rodigas}, T.~J. and {Kasper}, Markus and {Pyo}, T. -S. and {Ogino}, Satoshi},
        title = "{A Combined Subaru/VLT/MMT 1-5 {\ensuremath{\mu}}m Study of Planets Orbiting HR 8799: Implications for Atmospheric Properties, Masses, and Formation}",
      journal = {\apj},
         year = 2011,
        month = mar,
       volume = {729},
       number = {2},
          eid = {128},
        pages = {128},
          doi = {10.1088/0004-637X/729/2/128},
archivePrefix = {arXiv},
       eprint = {1101.1973},
 primaryClass = {astro-ph.EP},
       adsurl = {https://ui.adsabs.harvard.edu/abs/2011ApJ...729..128C}
}

@ARTICLE{2015Sci...350...64M,
       author = {{Macintosh}, B. and {Graham}, J.~R. and {Barman}, T. and {De Rosa}, R.~J. and {Konopacky}, Q. and {Marley}, M.~S. and {Marois}, C. and {Nielsen}, E.~L. and {Pueyo}, L. and {Rajan}, A. and {Rameau}, J. and {Saumon}, D. and {Wang}, J.~J. and {Patience}, J. and {Ammons}, M. and {Arriaga}, P. and {Artigau}, E. and {Beckwith}, S. and {Brewster}, J. and {Bruzzone}, S. and {Bulger}, J. and {Burningham}, B. and {Burrows}, A.~S. and {Chen}, C. and {Chiang}, E. and {Chilcote}, J.~K. and {Dawson}, R.~I. and {Dong}, R. and {Doyon}, R. and {Draper}, Z.~H. and {Duch{\^e}ne}, G. and {Esposito}, T.~M. and {Fabrycky}, D. and {Fitzgerald}, M.~P. and {Follette}, K.~B. and {Fortney}, J.~J. and {Gerard}, B. and {Goodsell}, S. and {Greenbaum}, A.~Z. and {Hibon}, P. and {Hinkley}, S. and {Cotten}, T.~H. and {Hung}, L. -W. and {Ingraham}, P. and {Johnson-Groh}, M. and {Kalas}, P. and {Lafreniere}, D. and {Larkin}, J.~E. and {Lee}, J. and {Line}, M. and {Long}, D. and {Maire}, J. and {Marchis}, F. and {Matthews}, B.~C. and {Max}, C.~E. and {Metchev}, S. and {Millar-Blanchaer}, M.~A. and {Mittal}, T. and {Morley}, C.~V. and {Morzinski}, K.~M. and {Murray-Clay}, R. and {Oppenheimer}, R. and {Palmer}, D.~W. and {Patel}, R. and {Perrin}, M.~D. and {Poyneer}, L.~A. and {Rafikov}, R.~R. and {Rantakyr{\"o}}, F.~T. and {Rice}, E.~L. and {Rojo}, P. and {Rudy}, A.~R. and {Ruffio}, J. -B. and {Ruiz}, M.~T. and {Sadakuni}, N. and {Saddlemyer}, L. and {Salama}, M. and {Savransky}, D. and {Schneider}, A.~C. and {Sivaramakrishnan}, A. and {Song}, I. and {Soummer}, R. and {Thomas}, S. and {Vasisht}, G. and {Wallace}, J.~K. and {Ward-Duong}, K. and {Wiktorowicz}, S.~J. and {Wolff}, S.~G. and {Zuckerman}, B.},
        title = "{Discovery and spectroscopy of the young jovian planet 51 Eri b with the Gemini Planet Imager}",
      journal = {Science},
         year = 2015,
        month = oct,
       volume = {350},
       number = {6256},
        pages = {64-67},
          doi = {10.1126/science.aac5891},
archivePrefix = {arXiv},
       eprint = {1508.03084},
 primaryClass = {astro-ph.EP},
       adsurl = {https://ui.adsabs.harvard.edu/abs/2015Sci...350...64M}
}

@ARTICLE{2013PASP..125..306F,
       author = {{Foreman-Mackey}, Daniel and {Hogg}, David W. and {Lang}, Dustin and {Goodman}, Jonathan},
        title = "{emcee: The MCMC Hammer}",
      journal = {\pasp},
         year = 2013,
        month = mar,
       volume = {125},
       number = {925},
        pages = {306},
          doi = {10.1086/670067},
archivePrefix = {arXiv},
       eprint = {1202.3665},
 primaryClass = {astro-ph.IM},
       adsurl = {https://ui.adsabs.harvard.edu/abs/2013PASP..125..306F}
}

@ARTICLE{2007ApJ...658..598K,
       author = {{Kaltenegger}, Lisa and {Traub}, Wesley A. and {Jucks}, Kenneth W.},
        title = "{Spectral Evolution of an Earth-like Planet}",
      journal = {\apj},
         year = 2007,
        month = mar,
       volume = {658},
       number = {1},
        pages = {598-616},
          doi = {10.1086/510996},
archivePrefix = {arXiv},
       eprint = {astro-ph/0609398},
 primaryClass = {astro-ph},
       adsurl = {https://ui.adsabs.harvard.edu/abs/2007ApJ...658..598K}
}

@ARTICLE{2007Natur.446..771T,
       author = {{Trauger}, John T. and {Traub}, Wesley A.},
        title = "{A laboratory demonstration of the capability to image an Earth-like extrasolar planet}",
      journal = {\nat},
         year = 2007,
        month = apr,
       volume = {446},
       number = {7137},
        pages = {771-773},
          doi = {10.1038/nature05729},
       adsurl = {https://ui.adsabs.harvard.edu/abs/2007Natur.446..771T}
}

@ARTICLE{2019AJ....158..140B,
       author = {{Brandt}, Timothy D. and {Dupuy}, Trent J. and {Bowler}, Brendan P.},
        title = "{Precise Dynamical Masses of Directly Imaged Companions from Relative Astrometry, Radial Velocities, and Hipparcos-Gaia DR2 Accelerations}",
      journal = {\aj},
         year = 2019,
        month = oct,
       volume = {158},
       number = {4},
          eid = {140},
        pages = {140},
          doi = {10.3847/1538-3881/ab04a8},
archivePrefix = {arXiv},
       eprint = {1811.07285},
 primaryClass = {astro-ph.SR},
       adsurl = {https://ui.adsabs.harvard.edu/abs/2019AJ....158..140B}
}

@ARTICLE{2021ApJS..254...42B,
       author = {{Brandt}, Timothy D.},
        title = "{The Hipparcos-Gaia Catalog of Accelerations: Gaia EDR3 Edition}",
      journal = {\apjs},
         year = 2021,
        month = jun,
       volume = {254},
       number = {2},
          eid = {42},
        pages = {42},
          doi = {10.3847/1538-4365/abf93c},
archivePrefix = {arXiv},
       eprint = {2105.11662},
 primaryClass = {astro-ph.GA},
       adsurl = {https://ui.adsabs.harvard.edu/abs/2021ApJS..254...42B}
}

@ARTICLE{2023Sci...380..198C,
       author = {{Currie}, Thayne and {Brandt}, G. Mirek and {Brandt}, Timothy D. and {Lacy}, Brianna and {Burrows}, Adam and {Guyon}, Olivier and {Tamura}, Motohide and {Liu}, Ranger Y. and {Sagynbayeva}, Sabina and {Tobin}, Taylor and {Chilcote}, Jeffrey and {Groff}, Tyler and {Marois}, Christian and {Thompson}, William and {Murphy}, Simon J. and {Kuzuhara}, Masayuki and {Lawson}, Kellen and {Lozi}, Julien and {Deo}, Vincent and {Vievard}, Sebastien and {Skaf}, Nour and {Uyama}, Taichi and {Jovanovic}, Nemanja and {Martinache}, Frantz and {Kasdin}, N. Jeremy and {Kudo}, Tomoyuki and {McElwain}, Michael and {Janson}, Markus and {Wisniewski}, John and {Hodapp}, Klaus and {Nishikawa}, Jun and {He{\l}miniak}, Krzysztof and {Kwon}, Jungmi and {Hayashi}, Masahiko},
        title = "{Direct imaging and astrometric detection of a gas giant planet orbiting an accelerating star}",
      journal = {Science},
         year = 2023,
        month = apr,
       volume = {380},
       number = {6641},
        pages = {198-203},
          doi = {10.1126/science.abo6192},
archivePrefix = {arXiv},
       eprint = {2212.00034},
 primaryClass = {astro-ph.EP},
       adsurl = {https://ui.adsabs.harvard.edu/abs/2023Sci...380..198C}
}

@ARTICLE{2017A&A...603A..54L,
       author = {{Lannier}, J. and {Lagrange}, A.~M. and {Bonavita}, M. and {Borgniet}, S. and {Delorme}, P. and {Meunier}, N. and {Desidera}, S. and {Messina}, S. and {Chauvin}, G. and {Keppler}, M.},
        title = "{Combining direct imaging and radial velocity data towards a full exploration of the giant planet population. I. Method and first results}",
      journal = {\aap},
         year = 2017,
        month = jul,
       volume = {603},
          eid = {A54},
        pages = {A54},
          doi = {10.1051/0004-6361/201628677},
archivePrefix = {arXiv},
       eprint = {1704.07432},
 primaryClass = {astro-ph.EP},
       adsurl = {https://ui.adsabs.harvard.edu/abs/2017A&A...603A..54L}
}

@INPROCEEDINGS{2022BAAS...54e.226N,
       author = {{Newman}, Patrick D. and {Plavchan}, Peter and {Burt}, Jennifer and {Teske}, Johanna and {Mamajeck}, Eric E. and {Leifer}, Stephanie and {Blackwood}, Gary and {Morgan}, Rhonda},
        title = "{Simulations for Planning Next-Generation Exoplanet Radial Velocity Surveys}",
    booktitle = {Bulletin of the AAS},
         year = 2022,
       volume = {54},
        month = jun,
          eid = {102.226},
        pages = {102.226},
       adsurl = {https://ui.adsabs.harvard.edu/abs/2022BAAS...54e.226N}
}

@ARTICLE{2016PASP..128j2001B,
       author = {{Bowler}, Brendan P.},
        title = "{Imaging Extrasolar Giant Planets}",
      journal = {\pasp},
         year = 2016,
        month = oct,
       volume = {128},
       number = {968},
        pages = {102001},
          doi = {10.1088/1538-3873/128/968/102001},
archivePrefix = {arXiv},
       eprint = {1605.02731},
 primaryClass = {astro-ph.EP},
       adsurl = {https://ui.adsabs.harvard.edu/abs/2016PASP..128j2001B}
}

@ARTICLE{2019AJ....158...13N,
       author = {{Nielsen}, Eric L. and {De Rosa}, Robert J. and {Macintosh}, Bruce and {Wang}, Jason J. and {Ruffio}, Jean-Baptiste and {Chiang}, Eugene and {Marley}, Mark S. and {Saumon}, Didier and {Savransky}, Dmitry and {Ammons}, S. Mark and {Bailey}, Vanessa P. and {Barman}, Travis and {Blain}, C{\'e}lia and {Bulger}, Joanna and {Burrows}, Adam and {Chilcote}, Jeffrey and {Cotten}, Tara and {Czekala}, Ian and {Doyon}, Rene and {Duch{\^e}ne}, Gaspard and {Esposito}, Thomas M. and {Fabrycky}, Daniel and {Fitzgerald}, Michael P. and {Follette}, Katherine B. and {Fortney}, Jonathan J. and {Gerard}, Benjamin L. and {Goodsell}, Stephen J. and {Graham}, James R. and {Greenbaum}, Alexandra Z. and {Hibon}, Pascale and {Hinkley}, Sasha and {Hirsch}, Lea A. and {Hom}, Justin and {Hung}, Li-Wei and {Dawson}, Rebekah Ilene and {Ingraham}, Patrick and {Kalas}, Paul and {Konopacky}, Quinn and {Larkin}, James E. and {Lee}, Eve J. and {Lin}, Jonathan W. and {Maire}, J{\'e}r{\^o}me and {Marchis}, Franck and {Marois}, Christian and {Metchev}, Stanimir and {Millar-Blanchaer}, Maxwell A. and {Morzinski}, Katie M. and {Oppenheimer}, Rebecca and {Palmer}, David and {Patience}, Jennifer and {Perrin}, Marshall and {Poyneer}, Lisa and {Pueyo}, Laurent and {Rafikov}, Roman R. and {Rajan}, Abhijith and {Rameau}, Julien and {Rantakyr{\"o}}, Fredrik T. and {Ren}, Bin and {Schneider}, Adam C. and {Sivaramakrishnan}, Anand and {Song}, Inseok and {Soummer}, Remi and {Tallis}, Melisa and {Thomas}, Sandrine and {Ward-Duong}, Kimberly and {Wolff}, Schuyler},
        title = "{The Gemini Planet Imager Exoplanet Survey: Giant Planet and Brown Dwarf Demographics from 10 to 100 au}",
      journal = {\aj},
         year = 2019,
        month = jul,
       volume = {158},
       number = {1},
          eid = {13},
        pages = {13},
          doi = {10.3847/1538-3881/ab16e9},
archivePrefix = {arXiv},
       eprint = {1904.05358},
 primaryClass = {astro-ph.EP},
       adsurl = {https://ui.adsabs.harvard.edu/abs/2019AJ....158...13N}
}

@ARTICLE{2014PNAS..11112661M,
       author = {{Macintosh}, Bruce and {Graham}, James R. and {Ingraham}, Patrick and {Konopacky}, Quinn and {Marois}, Christian and {Perrin}, Marshall and {Poyneer}, Lisa and {Bauman}, Brian and {Barman}, Travis and {Burrows}, Adam S. and {Cardwell}, Andrew and {Chilcote}, Jeffrey and {De Rosa}, Robert J. and {Dillon}, Daren and {Doyon}, Rene and {Dunn}, Jennifer and {Erikson}, Darren and {Fitzgerald}, Michael P. and {Gavel}, Donald and {Goodsell}, Stephen and {Hartung}, Markus and {Hibon}, Pascale and {Kalas}, Paul and {Larkin}, James and {Maire}, Jerome and {Marchis}, Franck and {Marley}, Mark S. and {McBride}, James and {Millar-Blanchaer}, Max and {Morzinski}, Katie and {Norton}, Andrew and {Oppenheimer}, B.~R. and {Palmer}, David and {Patience}, Jennifer and {Pueyo}, Laurent and {Rantakyro}, Fredrik and {Sadakuni}, Naru and {Saddlemyer}, Leslie and {Savransky}, Dmitry and {Serio}, Andrew and {Soummer}, Remi and {Sivaramakrishnan}, Anand and {Song}, Inseok and {Thomas}, Sandrine and {Wallace}, J. Kent and {Wiktorowicz}, Sloane and {Wolff}, Schuyler},
        title = "{First light of the Gemini Planet Imager}",
      journal = {PNAS},
         year = 2014,
        month = sep,
       volume = {111},
       number = {35},
        pages = {12661-12666},
          doi = {10.1073/pnas.1304215111},
archivePrefix = {arXiv},
       eprint = {1403.7520},
 primaryClass = {astro-ph.EP},
       adsurl = {https://ui.adsabs.harvard.edu/abs/2014PNAS..11112661M}
}

@ARTICLE{2019A&A...631A.155B,
       author = {{Beuzit}, J. -L. and {Vigan}, A. and {Mouillet}, D. and {Dohlen}, K. and {Gratton}, R. and {Boccaletti}, A. and {Sauvage}, J. -F. and {Schmid}, H.~M. and {Langlois}, M. and {Petit}, C. and {Baruffolo}, A. and {Feldt}, M. and {Milli}, J. and {Wahhaj}, Z. and {Abe}, L. and {Anselmi}, U. and {Antichi}, J. and {Barette}, R. and {Baudrand}, J. and {Baudoz}, P. and {Bazzon}, A. and {Bernardi}, P. and {Blanchard}, P. and {Brast}, R. and {Bruno}, P. and {Buey}, T. and {Carbillet}, M. and {Carle}, M. and {Cascone}, E. and {Chapron}, F. and {Charton}, J. and {Chauvin}, G. and {Claudi}, R. and {Costille}, A. and {De Caprio}, V. and {de Boer}, J. and {Delboulb{\'e}}, A. and {Desidera}, S. and {Dominik}, C. and {Downing}, M. and {Dupuis}, O. and {Fabron}, C. and {Fantinel}, D. and {Farisato}, G. and {Feautrier}, P. and {Fedrigo}, E. and {Fusco}, T. and {Gigan}, P. and {Ginski}, C. and {Girard}, J. and {Giro}, E. and {Gisler}, D. and {Gluck}, L. and {Gry}, C. and {Henning}, T. and {Hubin}, N. and {Hugot}, E. and {Incorvaia}, S. and {Jaquet}, M. and {Kasper}, M. and {Lagadec}, E. and {Lagrange}, A. -M. and {Le Coroller}, H. and {Le Mignant}, D. and {Le Ruyet}, B. and {Lessio}, G. and {Lizon}, J. -L. and {Llored}, M. and {Lundin}, L. and {Madec}, F. and {Magnard}, Y. and {Marteaud}, M. and {Martinez}, P. and {Maurel}, D. and {M{\'e}nard}, F. and {Mesa}, D. and {M{\"o}ller-Nilsson}, O. and {Moulin}, T. and {Moutou}, C. and {Orign{\'e}}, A. and {Parisot}, J. and {Pavlov}, A. and {Perret}, D. and {Pragt}, J. and {Puget}, P. and {Rabou}, P. and {Ramos}, J. and {Reess}, J. -M. and {Rigal}, F. and {Rochat}, S. and {Roelfsema}, R. and {Rousset}, G. and {Roux}, A. and {Saisse}, M. and {Salasnich}, B. and {Santambrogio}, E. and {Scuderi}, S. and {Segransan}, D. and {Sevin}, A. and {Siebenmorgen}, R. and {Soenke}, C. and {Stadler}, E. and {Suarez}, M. and {Tiph{\`e}ne}, D. and {Turatto}, M. and {Udry}, S. and {Vakili}, F. and {Waters}, L.~B.~F.~M. and {Weber}, L. and {Wildi}, F. and {Zins}, G. and {Zurlo}, A.},
        title = "{SPHERE: the exoplanet imager for the Very Large Telescope}",
      journal = {\aap},
         year = 2019,
        month = nov,
       volume = {631},
          eid = {A155},
        pages = {A155},
          doi = {10.1051/0004-6361/201935251},
archivePrefix = {arXiv},
       eprint = {1902.04080},
 primaryClass = {astro-ph.IM},
       adsurl = {https://ui.adsabs.harvard.edu/abs/2019A&A...631A.155B}
}

@ARTICLE{2007A&A...474..653V,
       author = {{van Leeuwen}, F.},
        title = "{Validation of the new Hipparcos reduction}",
      journal = {\aap},
         year = 2007,
        month = nov,
       volume = {474},
       number = {2},
        pages = {653-664},
          doi = {10.1051/0004-6361:20078357},
archivePrefix = {arXiv},
       eprint = {0708.1752},
 primaryClass = {astro-ph},
       adsurl = {https://ui.adsabs.harvard.edu/abs/2007A&A...474..653V}
}

@ARTICLE{2018ApJS..239...31B,
       author = {{Brandt}, Timothy D.},
        title = "{The Hipparcos-Gaia Catalog of Accelerations}",
      journal = {\apjs},
         year = 2018,
        month = dec,
       volume = {239},
       number = {2},
          eid = {31},
        pages = {31},
          doi = {10.3847/1538-4365/aaec06},
archivePrefix = {arXiv},
       eprint = {1811.07283},
 primaryClass = {astro-ph.SR},
       adsurl = {https://ui.adsabs.harvard.edu/abs/2018ApJS..239...31B}
}

@ARTICLE{2016ARA&A..54..271K,
       author = {{Kratter}, Kaitlin and {Lodato}, Giuseppe},
        title = "{Gravitational Instabilities in Circumstellar Disks}",
      journal = {\araa},
         year = 2016,
        month = sep,
       volume = {54},
        pages = {271-311},
          doi = {10.1146/annurev-astro-081915-023307},
archivePrefix = {arXiv},
       eprint = {1603.01280},
 primaryClass = {astro-ph.SR},
       adsurl = {https://ui.adsabs.harvard.edu/abs/2016ARA&A..54..271K}
}

@INCOLLECTION{2018haex.bookE.143M,
       author = {{Mordasini}, Christoph},
        title = "{Planetary Population Synthesis}",
    booktitle = {Handbook of Exoplanets},
         year = 2018,
       editor = {{Deeg}, Hans J. and {Belmonte}, Juan Antonio},
          eid = {143},
        pages = {143},
          doi = {10.1007/978-3-319-55333-7_143},
       adsurl = {https://ui.adsabs.harvard.edu/abs/2018haex.bookE.143M}
}

@Article{Dou2025,
    AUTHOR = {Dou, Jiangpei and Niu, Bingli and Zhao, Gang and Zhang, Xi and Wang, Gang and Yuan, Baoning and Wang, Di and Qian, Xingguang},
    TITLE = {Performance Calibration of the Wavefront Sensor’s EMCCD Detector for the Cool Planets Imaging Coronagraph Aboard CSST},
    JOURNAL = {Journal of Imaging},
    VOLUME = {11},
    YEAR = {2025},
    NUMBER = {6},
    pages = {203},
    URL = {https://www.mdpi.com/2313-433X/11/6/203},
    PubMedID = {40558802},
    ISSN = {2313-433X},
    DOI = {10.3390/jimaging11060203}
}

@ARTICLE{2010Natur.468.1080M,
       author = {{Marois}, Christian and {Zuckerman}, B. and {Konopacky}, Quinn M. and {Macintosh}, Bruce and {Barman}, Travis},
        title = "{Images of a fourth planet orbiting HR 8799}",
      journal = {\nat},
         year = 2010,
        month = dec,
       volume = {468},
       number = {7327},
        pages = {1080-1083},
          doi = {10.1038/nature09684},
archivePrefix = {arXiv},
       eprint = {1011.4918},
 primaryClass = {astro-ph.EP},
       adsurl = {https://ui.adsabs.harvard.edu/abs/2010Natur.468.1080M}
}

@ARTICLE{1995Natur.378..355M,
       author = {{Mayor}, Michel and {Queloz}, Didier},
        title = "{A Jupiter-mass companion to a solar-type star}",
      journal = {\nat},
         year = 1995,
        month = nov,
       volume = {378},
       number = {6555},
        pages = {355-359},
          doi = {10.1038/378355a0},
       adsurl = {https://ui.adsabs.harvard.edu/abs/1995Natur.378..355M}
}

@ARTICLE{2000ApJ...529L..45C,
       author = {{Charbonneau}, David and {Brown}, Timothy M. and {Latham}, David W. and {Mayor}, Michel},
        title = "{Detection of Planetary Transits Across a Sun-like Star}",
      journal = {\apjl},
         year = 2000,
        month = jan,
       volume = {529},
       number = {1},
        pages = {L45-L48},
          doi = {10.1086/312457},
archivePrefix = {arXiv},
       eprint = {astro-ph/9911436},
 primaryClass = {astro-ph},
       adsurl = {https://ui.adsabs.harvard.edu/abs/2000ApJ...529L..45C}
}

@ARTICLE{2004ApJ...606L.155B,
       author = {{Bond}, I.~A. and {Udalski}, A. and {Jaroszy{\'n}ski}, M. and {Rattenbury}, N.~J. and {Paczy{\'n}ski}, B. and {Soszy{\'n}ski}, I. and {Wyrzykowski}, L. and {Szyma{\'n}ski}, M.~K. and {Kubiak}, M. and {Szewczyk}, O. and {{\.Z}ebru{\'n}}, K. and {Pietrzy{\'n}ski}, G. and {Abe}, F. and {Bennett}, D.~P. and {Eguchi}, S. and {Furuta}, Y. and {Hearnshaw}, J.~B. and {Kamiya}, K. and {Kilmartin}, P.~M. and {Kurata}, Y. and {Masuda}, K. and {Matsubara}, Y. and {Muraki}, Y. and {Noda}, S. and {Okajima}, K. and {Sako}, T. and {Sekiguchi}, T. and {Sullivan}, D.~J. and {Sumi}, T. and {Tristram}, P.~J. and {Yanagisawa}, T. and {Yock}, P.~C.~M. and {OGLE Collaboration}},
        title = "{OGLE 2003-BLG-235/MOA 2003-BLG-53: A Planetary Microlensing Event}",
      journal = {\apjl},
         year = 2004,
        month = may,
       volume = {606},
       number = {2},
        pages = {L155-L158},
          doi = {10.1086/420928},
archivePrefix = {arXiv},
       eprint = {astro-ph/0404309},
 primaryClass = {astro-ph},
       adsurl = {https://ui.adsabs.harvard.edu/abs/2004ApJ...606L.155B}
}

@ARTICLE{2018JQSRT.217...86V,
       
       author = {G.L. Villanueva and M.D. Smith and S. Protopapa and S. Faggi and A.M. Mandell},
        title = "{Planetary Spectrum Generator: An accurate online radiative transfer suite for atmospheres, comets, small bodies and exoplanets}",
      journal = {JQSRT},
         year = 2018,
        month = sep,
       volume = {217},
        pages = {86-104},
          doi = {10.1016/j.jqsrt.2018.05.023},
archivePrefix = {arXiv},
       eprint = {1803.02008},
 primaryClass = {astro-ph.EP},
       adsurl = {https://ui.adsabs.harvard.edu/abs/2018JQSRT.217...86V}
}

@article{https://doi.org/10.1155/vib/5575921,
author = {Kong, Ling-Yi and Dou, Jiang-Pei and Guo, Wei and Xu, Ming-Ming and Jiang, Shu and Chen, Bo},
title = {Optimization Design and Analysis for the Mechanical Test Platform of Scientific Probe Module of the Cool Planet Imaging Coronagraph},
journal = {Shock VIB},
volume = {2025},
number = {1},
pages = {5575921},
doi = {https://doi.org/10.1155/vib/5575921},
url = {https://onlinelibrary.wiley.com/doi/abs/10.1155/vib/5575921},
eprint = {https://onlinelibrary.wiley.com/doi/pdf/10.1155/vib/5575921},
year = {2025}
}

@article{cjss2024012024yg01,
        author = {{Zhou}, Ji-Lin and {Xie}, Ji-Wei and {Ge}, Jian and {JI}, Jiang-Hui and {Dou}, Jiang-Pei and {Dong}, Su-Bo and {Liu}, Hui-Gen and {Wang}, Wei and {Guo}, Jian Heng and {Yu}, Cong and {Bai}, Xue-Ning and {Feng}, Fa-Bo and {Liu}, Bei-Bei},
        title = "{Progress on Exoplanet Detection and Research in Space}",
      journal = {CJSS},
         year = 2024,
        month = jan,
       volume = {44},
       number = {1},
        pages = {5-18},
          doi = {10.11728/cjss2024.01.2024-yg01},
       adsurl = {https://ui.adsabs.harvard.edu/abs/2024ChJSS..44....5Z}
}

@ARTICLE{2025Meas..25117206C,
       author = {{Cui}, Haodong and {Xu}, Mingming and {Dou}, Jiangpei},
        title = "{A novel six-dimensional micro-vibration measurement platform calibrated by transfer learning DNN}",
      journal = {Measurement},
         year = 2025,
        month = jun,
       volume = {251},
          eid = {117206},
        pages = {117206},
          doi = {10.1016/j.measurement.2025.117206},
       adsurl = {https://ui.adsabs.harvard.edu/abs/2025Meas..25117206C}
}

@ARTICLE{1990Icar...87..484B,
       author = {{Brown}, Robert A. and {Burrows}, Christopher J.},
        title = "{On the feasibility of detecting extrasolar planets by reflected starlight using the Hubble Space Telescope}",
      journal = {\icarus},
         year = 1990,
        month = oct,
       volume = {87},
       number = {2},
        pages = {484-497},
          doi = {10.1016/0019-1035(90)90150-8},
       adsurl = {https://ui.adsabs.harvard.edu/abs/1990Icar...87..484B}
}

@ARTICLE{2012OptEn..51a1002L,
       author = {{Lyon}, Richard G. and {Clampin}, Mark},
        title = "{Space telescope sensitivity and controls for exoplanet imaging}",
      journal = {OE},
         year = 2012,
        month = jan,
       volume = {51},
       number = {1},
          eid = {011002-011002-16},
        pages = {011002-011002-16},
          doi = {10.1117/1.OE.51.1.011002},
       adsurl = {https://ui.adsabs.harvard.edu/abs/2012OptEn..51a1002L}
}

@ARTICLE{2012RAA....12..591Z,
       author = {{Zhang}, Xi and {Ren}, De-Qing and {Zhu}, Yong-Tian and {Dou}, Jiang-Pei},
        title = "{An active coronagraph using a liquid crystal array for exoplanet imaging: principle and testing}",
      journal = {RAA},
         year = 2012,
        month = may,
       volume = {12},
       number = {5},
        pages = {591-600},
          doi = {10.1088/1674-4527/12/5/011},
       adsurl = {https://ui.adsabs.harvard.edu/abs/2012RAA....12..591Z}
}

@ARTICLE{2022RAA....22g2003J,
       author = {{Ji}, Jiang-Hui and {Li}, Hai-Tao and {Zhang}, Jun-Bo and {Fang}, Liang and {Li}, Dong and {Wang}, Su and {Cao}, Yang and {Deng}, Lei and {Li}, Bao-Quan and {Xian}, Hao and {Gao}, Xiao-Dong and {Zhang}, Ang and {Li}, Fei and {Liu}, Jia-Cheng and {Qi}, Zhao-Xiang and {Jin}, Sheng and {Liu}, Ya-Ning and {Chen}, Guo and {Li}, Ming-Tao and {Dong}, Yao and {Zhu}, Zi and {CHES Consortium}},
        title = "{CHES: A Space-borne Astrometric Mission for the Detection of Habitable Planets of the Nearby Solar-type Stars}",
      journal = {RAA},
         year = 2022,
        month = jul,
       volume = {22},
       number = {7},
          eid = {072003},
        pages = {072003},
          doi = {10.1088/1674-4527/ac77e4},
archivePrefix = {arXiv},
       eprint = {2205.05645},
 primaryClass = {astro-ph.EP},
       adsurl = {https://ui.adsabs.harvard.edu/abs/2022RAA....22g2003J}
}

@ARTICLE{2007ApJ...654..625W,
       author = {{Wittenmyer}, Robert A. and {Endl}, Michael and {Cochran}, William D.},
        title = "{Long-Period Objects in the Extrasolar Planetary Systems 47 Ursae Majoris and 14 Herculis}",
      journal = {\apj},
         year = 2007,
        month = jan,
       volume = {654},
       number = {1},
        pages = {625-632},
          doi = {10.1086/509110},
archivePrefix = {arXiv},
       eprint = {astro-ph/0609117},
 primaryClass = {astro-ph},
       adsurl = {https://ui.adsabs.harvard.edu/abs/2007ApJ...654..625W}
}

@ARTICLE{2021ApJ...922L..43B,
       author = {{Bardalez Gagliuffi}, Daniella C. and {Faherty}, Jacqueline K. and {Li}, Yiting and {Brandt}, Timothy D. and {Williams}, Lauryn and {Brandt}, G. Mirek and {Gelino}, Christopher R.},
        title = "{14 Her: A Likely Case of Planet-Planet Scattering}",
      journal = {\apjl},
         year = 2021,
        month = dec,
       volume = {922},
       number = {2},
          eid = {L43},
        pages = {L43},
          doi = {10.3847/2041-8213/ac382c},
archivePrefix = {arXiv},
       eprint = {2111.06004},
 primaryClass = {astro-ph.EP},
       adsurl = {https://ui.adsabs.harvard.edu/abs/2021ApJ...922L..43B}
}

@ARTICLE{2023AJ....166...27B,
       author = {{Benedict}, G.~F. and {McArthur}, B.~E. and {Nelan}, E.~P. and {Bean}, J.~L.},
        title = "{The 14 Her Planetary System: Companion Masses and Architecture from Radial Velocities and Astrometry}",
      journal = {\aj},
         year = 2023,
        month = jul,
       volume = {166},
       number = {1},
          eid = {27},
        pages = {27},
          doi = {10.3847/1538-3881/acd93a},
archivePrefix = {arXiv},
       eprint = {2305.11753},
 primaryClass = {astro-ph.EP},
       adsurl = {https://ui.adsabs.harvard.edu/abs/2023AJ....166...27B}
}

@ARTICLE{2010MNRAS.403..731G,
       author = {{Gregory}, Philip C. and {Fischer}, Debra A.},
        title = "{A Bayesian periodogram finds evidence for three planets in 47UrsaeMajoris}",
      journal = {\mnras},
         year = 2010,
        month = apr,
       volume = {403},
       number = {2},
        pages = {731-747},
          doi = {10.1111/j.1365-2966.2009.16233.x},
archivePrefix = {arXiv},
       eprint = {1003.5549},
 primaryClass = {astro-ph.EP},
       adsurl = {https://ui.adsabs.harvard.edu/abs/2010MNRAS.403..731G}
}

@ARTICLE{2014MNRAS.441..442N,
       author = {{Nelson}, Benjamin E. and {Ford}, Eric B. and {Wright}, Jason T. and {Fischer}, Debra A. and {von Braun}, Kaspar and {Howard}, Andrew W. and {Payne}, Matthew J. and {Dindar}, Saleh},
        title = "{The 55 Cancri planetary system: fully self-consistent N-body constraints and a dynamical analysis}",
      journal = {\mnras},
         year = 2014,
        month = jun,
       volume = {441},
       number = {1},
        pages = {442-451},
          doi = {10.1093/mnras/stu450},
archivePrefix = {arXiv},
       eprint = {1402.6343},
 primaryClass = {astro-ph.EP},
       adsurl = {https://ui.adsabs.harvard.edu/abs/2014MNRAS.441..442N}
}

@ARTICLE{2019AJ....157..149L,
       author = {{Luhn}, Jacob K. and {Bastien}, Fabienne A. and {Wright}, Jason T. and {Johnson}, John A. and {Howard}, Andrew W. and {Isaacson}, Howard},
        title = "{Retired A Stars and Their Companions. VIII. 15 New Planetary Signals around Subgiants and Transit Parameters for California Planet Search Planets with Subgiant Hosts}",
      journal = {\aj},
         year = 2019,
        month = apr,
       volume = {157},
       number = {4},
          eid = {149},
        pages = {149},
          doi = {10.3847/1538-3881/aaf5d0},
archivePrefix = {arXiv},
       eprint = {1811.03043},
 primaryClass = {astro-ph.EP},
       adsurl = {https://ui.adsabs.harvard.edu/abs/2019AJ....157..149L}
}

@ARTICLE{2010MNRAS.403.1703J,
       author = {{Jones}, Hugh R.~A. and {Butler}, R. Paul and {Tinney}, C.~G. and {O'Toole}, Simon and {Wittenmyer}, Rob and {Henry}, Gregory W. and {Meschiari}, Stefano and {Vogt}, Steve and {Rivera}, Eugenio and {Laughlin}, Greg and {Carter}, Brad D. and {Bailey}, Jeremy and {Jenkins}, James S.},
        title = "{A long-period planet orbiting a nearby Sun-like star}",
      journal = {\mnras},
         year = 2010,
        month = apr,
       volume = {403},
       number = {4},
        pages = {1703-1713},
          doi = {10.1111/j.1365-2966.2009.16232.x},
archivePrefix = {arXiv},
       eprint = {0912.2716},
 primaryClass = {astro-ph.SR},
       adsurl = {https://ui.adsabs.harvard.edu/abs/2010MNRAS.403.1703J}
}

@ARTICLE{2012ApJ...753..169W,
       author = {{Wittenmyer}, Robert A. and {Horner}, J. and {Tuomi}, Mikko and {Salter}, G.~S. and {Tinney}, C.~G. and {Butler}, R.~P. and {Jones}, H.~R.~A. and {O'Toole}, S.~J. and {Bailey}, J. and {Carter}, B.~D. and {Jenkins}, J.~S. and {Zhang}, Z. and {Vogt}, S.~S. and {Rivera}, Eugenio J.},
        title = "{The Anglo-Australian Planet Search. XXII. Two New Multi-planet Systems}",
      journal = {\apj},
         year = 2012,
        month = jul,
       volume = {753},
       number = {2},
          eid = {169},
        pages = {169},
          doi = {10.1088/0004-637X/753/2/169},
archivePrefix = {arXiv},
       eprint = {1205.2765},
 primaryClass = {astro-ph.EP},
       adsurl = {https://ui.adsabs.harvard.edu/abs/2012ApJ...753..169W}
}

@ARTICLE{2012A&A...545A..55B,
       author = {{Boisse}, I. and {Pepe}, F. and {Perrier}, C. and {Queloz}, D. and {Bonfils}, X. and {Bouchy}, F. and {Santos}, N.~C. and {Arnold}, L. and {Beuzit}, J. -L. and {D{\'\i}az}, R.~F. and {Delfosse}, X. and {Eggenberger}, A. and {Ehrenreich}, D. and {Forveille}, T. and {H{\'e}brard}, G. and {Lagrange}, A. -M. and {Lovis}, C. and {Mayor}, M. and {Moutou}, C. and {Naef}, D. and {Santerne}, A. and {S{\'e}gransan}, D. and {Sivan}, J. -P. and {Udry}, S.},
        title = "{The SOPHIE search for northern extrasolar planets. V. Follow-up of ELODIE candidates: Jupiter-analogs around Sun-like stars}",
      journal = {\aap},
         year = 2012,
        month = sep,
       volume = {545},
          eid = {A55},
        pages = {A55},
          doi = {10.1051/0004-6361/201118419},
archivePrefix = {arXiv},
       eprint = {1205.5835},
 primaryClass = {astro-ph.EP},
       adsurl = {https://ui.adsabs.harvard.edu/abs/2012A&A...545A..55B}
}

@ARTICLE{2016ApJ...818...34E,
       author = {{Endl}, Michael and {Brugamyer}, Erik J. and {Cochran}, William D. and {MacQueen}, Phillip J. and {Robertson}, Paul and {Meschiari}, Stefano and {Ramirez}, Ivan and {Shetrone}, Matthew and {Gullikson}, Kevin and {Johnson}, Marshall C. and {Wittenmyer}, Robert and {Horner}, Jonathan and {Ciardi}, David R. and {Horch}, Elliott and {Simon}, Attila E. and {Howell}, Steve B. and {Everett}, Mark and {Caldwell}, Caroline and {Castanheira}, Barbara G.},
        title = "{Two New Long-period Giant Planets from the McDonald Observatory Planet Search and Two Stars with Long-period Radial Velocity Signals Related to Stellar Activity Cycles}",
      journal = {\apj},
         year = 2016,
        month = feb,
       volume = {818},
       number = {1},
          eid = {34},
        pages = {34},
          doi = {10.3847/0004-637X/818/1/34},
archivePrefix = {arXiv},
       eprint = {1512.02965},
 primaryClass = {astro-ph.EP},
       adsurl = {https://ui.adsabs.harvard.edu/abs/2016ApJ...818...34E}
}

@ARTICLE{2015A&A...581A..38C,
       author = {{Courcol}, B. and {Bouchy}, F. and {Pepe}, F. and {Santerne}, A. and {Delfosse}, X. and {Arnold}, L. and {Astudillo-Defru}, N. and {Boisse}, I. and {Bonfils}, X. and {Borgniet}, S. and {Bourrier}, V. and {Cabrera}, N. and {Deleuil}, M. and {Demangeon}, O. and {D{\'\i}az}, R.~F. and {Ehrenreich}, D. and {Forveille}, T. and {H{\'e}brard}, G. and {Lagrange}, A.~M. and {Montagnier}, G. and {Moutou}, C. and {Rey}, J. and {Santos}, N.~C. and {S{\'e}gransan}, D. and {Udry}, S. and {Wilson}, P.~A.},
        title = "{The SOPHIE search for northern extrasolar planets. VII. A warm Neptune orbiting HD 164595}",
      journal = {\aap},
         year = 2015,
        month = sep,
       volume = {581},
          eid = {A38},
        pages = {A38},
          doi = {10.1051/0004-6361/201526329},
archivePrefix = {arXiv},
       eprint = {1506.07144},
 primaryClass = {astro-ph.EP},
       adsurl = {https://ui.adsabs.harvard.edu/abs/2015A&A...581A..38C}
}

@ARTICLE{2005ApJ...632..638V,
       author = {{Vogt}, Steven S. and {Butler}, R. Paul and {Marcy}, Geoffrey W. and {Fischer}, Debra A. and {Henry}, Gregory W. and {Laughlin}, Greg and {Wright}, Jason T. and {Johnson}, John A.},
        title = "{Five New Multicomponent Planetary Systems}",
      journal = {\apj},
         year = 2005,
        month = oct,
       volume = {632},
       number = {1},
        pages = {638-658},
          doi = {10.1086/432901},
       adsurl = {https://ui.adsabs.harvard.edu/abs/2005ApJ...632..638V}
}

@ARTICLE{2009ApJ...693.1084W,
       author = {{Wright}, J.~T. and {Upadhyay}, S. and {Marcy}, G.~W. and {Fischer}, D.~A. and {Ford}, Eric B. and {Johnson}, John Asher},
        title = "{Ten New and Updated Multiplanet Systems and a Survey of Exoplanetary Systems}",
      journal = {\apj},
         year = 2009,
        month = mar,
       volume = {693},
       number = {2},
        pages = {1084-1099},
          doi = {10.1088/0004-637X/693/2/1084},
archivePrefix = {arXiv},
       eprint = {0812.1582},
 primaryClass = {astro-ph},
       adsurl = {https://ui.adsabs.harvard.edu/abs/2009ApJ...693.1084W}
}

@ARTICLE{2008JKAS...41...59H,
       author = {{Ran}, In-Woo and {Lee}, Byeong-Cheol and {Kim}, Kang-Min and {Mkrtichian}, D.~E.},
        title = "{Confirmation of the Exoplanet around {\ensuremath{\beta}} GEM from the RV Observations Using Boes}",
      journal = {JKAS},
         year = 2008,
        month = jun,
       volume = {41},
       number = {3},
        pages = {59-64},
          doi = {10.5303/JKAS.2008.41.3.059},
       adsurl = {https://ui.adsabs.harvard.edu/abs/2008JKAS...41...59H}
}

@ARTICLE{2009ApJ...703.1545F,
       author = {{Fischer}, Debra and {Driscoll}, Peter and {Isaacson}, Howard and {Giguere}, Matt and {Marcy}, Geoffrey W. and {Valenti}, Jeff and {Wright}, Jason T. and {Henry}, Gregory W. and {Johnson}, John Asher and {Howard}, Andrew and {Peek}, Katherine and {McCarthy}, Chris},
        title = "{Five Planets and an Independent Confirmation of HD 196885Ab from Lick Observatory}",
      journal = {\apj},
         year = 2009,
        month = oct,
       volume = {703},
       number = {2},
        pages = {1545-1556},
          doi = {10.1088/0004-637X/703/2/1545},
archivePrefix = {arXiv},
       eprint = {0908.1596},
 primaryClass = {astro-ph.EP},
       adsurl = {https://ui.adsabs.harvard.edu/abs/2009ApJ...703.1545F}
}

@ARTICLE{2019A&A...625A..71R,
       author = {{Rickman}, E.~L. and {S{\'e}gransan}, D. and {Marmier}, M. and {Udry}, S. and {Bouchy}, F. and {Lovis}, C. and {Mayor}, M. and {Pepe}, F. and {Queloz}, D. and {Santos}, N.~C. and {Allart}, R. and {Bonvin}, V. and {Bratschi}, P. and {Cersullo}, F. and {Chazelas}, B. and {Choplin}, A. and {Conod}, U. and {Deline}, A. and {Delisle}, J. -B. and {Dos Santos}, L.~A. and {Figueira}, P. and {Giles}, H.~A.~C. and {Girard}, M. and {Lavie}, B. and {Martin}, D. and {Motalebi}, F. and {Nielsen}, L.~D. and {Osborn}, H. and {Ottoni}, G. and {Raimbault}, M. and {Rey}, J. and {Roger}, T. and {Seidel}, J.~V. and {Stalport}, M. and {Su{\'a}rez Mascare{\~n}o}, A. and {Triaud}, A. and {Turner}, O. and {Weber}, L. and {Wyttenbach}, A.},
        title = "{The CORALIE survey for southern extrasolar planets. XVIII. Three new massive planets and two low-mass brown dwarfs at greater than 5 AU separation}",
      journal = {\aap},
         year = 2019,
        month = may,
       volume = {625},
          eid = {A71},
        pages = {A71},
          doi = {10.1051/0004-6361/201935356},
archivePrefix = {arXiv},
       eprint = {1904.01573},
 primaryClass = {astro-ph.EP},
       adsurl = {https://ui.adsabs.harvard.edu/abs/2019A&A...625A..71R}
}

@ARTICLE{2021AJ....162..181L,
       author = {{Llop-Sayson}, Jorge and {Wang}, Jason J. and {Ruffio}, Jean-Baptiste and {Mawet}, Dimitri and {Blunt}, Sarah and {Absil}, Olivier and {Bond}, Charlotte and {Brinkman}, Casey and {Bowler}, Brendan P. and {Bottom}, Michael and {Chontos}, Ashley and {Dalba}, Paul A. and {Fulton}, B.~J. and {Giacalone}, Steven and {Hill}, Michelle and {Hirsch}, Lea A. and {Howard}, Andrew W. and {Isaacson}, Howard and {Karlsson}, Mikael and {Lubin}, Jack and {Madurowicz}, Alex and {Matthews}, Keith and {Morris}, Evan and {Perrin}, Marshall and {Ren}, Bin and {Rice}, Malena and {Rosenthal}, Lee J. and {Ruane}, Garreth and {Rubenzahl}, Ryan and {Sun}, He and {Wallack}, Nicole and {Xuan}, Jerry W. and {Ygouf}, Marie},
        title = "{Constraining the Orbit and Mass of epsilon Eridani b with Radial Velocities, Hipparcos IAD-Gaia DR2 Astrometry, and Multiepoch Vortex Coronagraphy Upper Limits}",
      journal = {\aj},
         year = 2021,
        month = nov,
       volume = {162},
       number = {5},
          eid = {181},
        pages = {181},
          doi = {10.3847/1538-3881/ac134a},
archivePrefix = {arXiv},
       eprint = {2108.02305},
 primaryClass = {astro-ph.EP},
       adsurl = {https://ui.adsabs.harvard.edu/abs/2021AJ....162..181L}
}

@ARTICLE{2019AJ....157...33M,
       author = {{Mawet}, Dimitri and {Hirsch}, Lea and {Lee}, Eve J. and {Ruffio}, Jean-Baptiste and {Bottom}, Michael and {Fulton}, Benjamin J. and {Absil}, Olivier and {Beichman}, Charles and {Bowler}, Brendan and {Bryan}, Marta and {Choquet}, Elodie and {Ciardi}, David and {Christiaens}, Valentin and {Defr{\`e}re}, Denis and {Gomez Gonzalez}, Carlos Alberto and {Howard}, Andrew W. and {Huby}, Elsa and {Isaacson}, Howard and {Jensen-Clem}, Rebecca and {Kosiarek}, Molly and {Marcy}, Geoff and {Meshkat}, Tiffany and {Petigura}, Erik and {Reggiani}, Maddalena and {Ruane}, Garreth and {Serabyn}, Eugene and {Sinukoff}, Evan and {Wang}, Ji and {Weiss}, Lauren and {Ygouf}, Marie},
        title = "{Deep Exploration of ɛ Eridani with Keck Ms-band Vortex Coronagraphy and Radial Velocities: Mass and Orbital Parameters of the Giant Exoplanet}",
      journal = {\aj},
         year = 2019,
        month = jan,
       volume = {157},
       number = {1},
          eid = {33},
        pages = {33},
          doi = {10.3847/1538-3881/aaef8a},
archivePrefix = {arXiv},
       eprint = {1810.03794},
 primaryClass = {astro-ph.EP},
       adsurl = {https://ui.adsabs.harvard.edu/abs/2019AJ....157...33M}
}

@ARTICLE{2007A&A...462..769P,
       author = {{Pepe}, F. and {Correia}, A.~C.~M. and {Mayor}, M. and {Tamuz}, O. and {Couetdic}, J. and {Benz}, W. and {Bertaux}, J. -L. and {Bouchy}, F. and {Laskar}, J. and {Lovis}, C. and {Naef}, D. and {Queloz}, D. and {Santos}, N.~C. and {Sivan}, J. -P. and {Sosnowska}, D. and {Udry}, S.},
        title = "{The HARPS search for southern extra-solar planets. VIII. <ASTROBJ>{\ensuremath{\mu}} Arae</ASTROBJ>, a system with four planets}",
      journal = {\aap},
         year = 2007,
        month = feb,
       volume = {462},
       number = {2},
        pages = {769-776},
          doi = {10.1051/0004-6361:20066194},
archivePrefix = {arXiv},
       eprint = {astro-ph/0608396},
 primaryClass = {astro-ph},
       adsurl = {https://ui.adsabs.harvard.edu/abs/2007A&A...462..769P}
}

@ARTICLE{2018A&A...619L..10G,
       author = {{Gandolfi}, D. and {Barrag{\'a}n}, O. and {Livingston}, J.~H. and {Fridlund}, M. and {Justesen}, A.~B. and {Redfield}, S. and {Fossati}, L. and {Mathur}, S. and {Grziwa}, S. and {Cabrera}, J. and {Garc{\'\i}a}, R.~A. and {Persson}, C.~M. and {Van Eylen}, V. and {Hatzes}, A.~P. and {Hidalgo}, D. and {Albrecht}, S. and {Bugnet}, L. and {Cochran}, W.~D. and {Csizmadia}, Sz. and {Deeg}, H. and {Eigm{\"u}ller}, Ph. and {Endl}, M. and {Erikson}, A. and {Esposito}, M. and {Guenther}, E. and {Korth}, J. and {Luque}, R. and {Monta{\~n}es Rodr{\'\i}guez}, P. and {Nespral}, D. and {Nowak}, G. and {P{\"a}tzold}, M. and {Prieto-Arranz}, J.},
        title = "{TESS's first planet. A super-Earth transiting the naked-eye star {\ensuremath{\pi}} Mensae}",
      journal = {\aap},
         year = 2018,
        month = nov,
       volume = {619},
          eid = {L10},
        pages = {L10},
          doi = {10.1051/0004-6361/201834289},
archivePrefix = {arXiv},
       eprint = {1809.07573},
 primaryClass = {astro-ph.EP},
       adsurl = {https://ui.adsabs.harvard.edu/abs/2018A&A...619L..10G}
}

@ARTICLE{2022AJ....163..223H,
       author = {{Hatzes}, Artie P. and {Gandolfi}, Davide and {Korth}, Judith and {Rodler}, Florian and {Sabotta}, Silvia and {Esposito}, Massimiliano and {Barrag{\'a}n}, Oscar and {Van Eylen}, Vincent and {Livingston}, John H. and {Serrano}, Luisa Maria and {Luque}, Rafael and {Smith}, Alexis M.~S. and {Redfield}, Seth and {Persson}, Carina M. and {P{\"a}tzold}, Martin and {Palle}, Enric and {Nowak}, Grzegorz and {Osborne}, Hannah L.~M. and {Narita}, Norio and {Mathur}, Savita and {Lam}, Kristine W.~F. and {Kab{\'a}th}, Petr and {Johnson}, Marshall C. and {Guenther}, Eike W. and {Grziwa}, Sascha and {Goffo}, Elisa and {Fridlund}, Malcolm and {Endl}, Michael and {Deeg}, Hans J. and {Csizmadia}, Szilard and {Cochran}, William D. and {Cuesta}, Luc{\'\i}a Gonz{\'a}lez and {Chaturvedi}, Priyanka and {Carleo}, Ilaria and {Cabrera}, Juan and {Beck}, Paul G. and {Albrecht}, Simon},
        title = "{A Radial Velocity Study of the Planetary System of {\ensuremath{\pi}} Mensae: Improved Planet Parameters for {\ensuremath{\pi}} Mensae c and a Third Planet on a 125 Day Orbit}",
      journal = {\aj},
         year = 2022,
        month = may,
       volume = {163},
       number = {5},
          eid = {223},
        pages = {223},
          doi = {10.3847/1538-3881/ac5dcb},
archivePrefix = {arXiv},
       eprint = {2203.01018},
 primaryClass = {astro-ph.EP},
       adsurl = {https://ui.adsabs.harvard.edu/abs/2022AJ....163..223H}
}

@ARTICLE{2002ApJ...567L.149B,
       author = {{Boss}, Alan P.},
        title = "{Stellar Metallicity and the Formation of Extrasolar Gas Giant Planets}",
      journal = {\apjl},
         year = 2002,
        month = mar,
       volume = {567},
       number = {2},
        pages = {L149-L153},
          doi = {10.1086/340108},
       adsurl = {https://ui.adsabs.harvard.edu/abs/2002ApJ...567L.149B}
}

@ARTICLE{2015ApJ...800...82P,
       author = {{Piso}, Ana-Maria A. and {Youdin}, Andrew N. and {Murray-Clay}, Ruth A.},
        title = "{Minimum Core Masses for Giant Planet Formation with Realistic Equations of State and Opacities}",
      journal = {\apj},
         year = 2015,
        month = feb,
       volume = {800},
       number = {2},
          eid = {82},
        pages = {82},
          doi = {10.1088/0004-637X/800/2/82},
archivePrefix = {arXiv},
       eprint = {1412.5185},
 primaryClass = {astro-ph.EP},
       adsurl = {https://ui.adsabs.harvard.edu/abs/2015ApJ...800...82P}
}

@ARTICLE{2021ApJ...915L..16B,
       author = {{Brandt}, G. Mirek and {Brandt}, Timothy D. and {Dupuy}, Trent J. and {Michalik}, Daniel and {Marleau}, Gabriel-Dominique},
        title = "{The First Dynamical Mass Measurement in the HR 8799 System}",
      journal = {\apjl},
         year = 2021,
        month = jul,
       volume = {915},
       number = {1},
          eid = {L16},
        pages = {L16},
          doi = {10.3847/2041-8213/ac0540},
archivePrefix = {arXiv},
       eprint = {2105.12820},
 primaryClass = {astro-ph.EP},
       adsurl = {https://ui.adsabs.harvard.edu/abs/2021ApJ...915L..16B}
}

@ARTICLE{2025NatAs...9.1184S,
       author = {{Sun}, L. and {Gu}, S. and {Wang}, X. and {Schmitt}, J.~H.~M.~M. and {Ioannidis}, P. and {Kouwenhoven}, M.~B.~N. and {Dou}, J. and {Zhao}, G.},
        title = "{A temperate 10-Earth-mass exoplanet around the Sun-like star Kepler-725}",
      journal = {Nature Astronomy},
         year = 2025,
        month = aug,
       volume = {9},
        pages = {1184-1194},
          doi = {10.1038/s41550-025-02565-z},
       adsurl = {https://ui.adsabs.harvard.edu/abs/2025NatAs...9.1184S}
}

@ARTICLE{2025AJ....169..342W,
       author = {{Wang}, Weilong and {Gu}, Shenghong and {Wang}, Xiaobin and {Sun}, Leilei and {Lee}, Byeong-Cheol and {Kwok}, Chi-Tai and {Hui}, Ho-Keung and {Dou}, Jiangpei and {Xiang}, Yue and {Cao}, Dongtao and {Xu}, Fukun},
        title = "{Observations and Studies on the Transiting Systems HAT-P-36, XO-2 and WASP-76}",
      journal = {\aj},
         year = 2025,
        month = jun,
       volume = {169},
       number = {6},
          eid = {342},
        pages = {342},
          doi = {10.3847/1538-3881/add1de},
       adsurl = {https://ui.adsabs.harvard.edu/abs/2025AJ....169..342W}
}

@ARTICLE{2022MNRAS.512.3113B,
       author = {{Bai}, Lu and {Gu}, Shenghong and {Wang}, Xiaobin and {Sun}, Leilei and {Kwok}, Chi-Tai and {Hui}, Ho-Keung},
        title = "{The study on transmission spectrum and TTV behaviour of the hot Jupiter WASP-12b}",
      journal = {\mnras},
         year = 2022,
        month = may,
       volume = {512},
       number = {3},
        pages = {3113-3123},
          doi = {10.1093/mnras/stac623},
       adsurl = {https://ui.adsabs.harvard.edu/abs/2022MNRAS.512.3113B}
}

@article{Parker_2016,
   title={DISCOVERY OF A MAKEMAKEAN MOON},
   volume={825},
   ISSN={2041-8213},
   url={http://dx.doi.org/10.3847/2041-8205/825/1/L9},
   DOI={10.3847/2041-8205/825/1/l9},
   number={1},
   journal={\apjl},
   publisher={American Astronomical Society},
   author={Parker, Alex H. and Buie, Marc W. and Grundy, Will M. and Noll, Keith S.},
   year={2016},
   month=jun, pages={L9} }

@ARTICLE{2025PrA....43..100H,
       author = {{Huang}, Jing and {Wang}, Xiaobin and {Dou}, Jiangpei},
        title = "{Photometry Analysis of Asteroid (2572) Annschnell - the Contact Binary Asteroids Model}",
      journal = {Progress in Astronomy},
         year = 2025,
        month = mar,
       volume = {43},
        pages = {100-113},
          doi = {10.3969/j.issn.1000-8349.2025.01.06},
       adsurl = {https://ui.adsabs.harvard.edu/abs/2025PrA....43..100H}
}

@ARTICLE{2023Natur.614..659R,
       author = {{Rustamkulov}, Z. and {Sing}, D.~K. and {Mukherjee}, S. and {May}, E.~M. and {Kirk}, J. and {Schlawin}, E. and {Line}, M.~R. and {Piaulet}, C. and {Carter}, A.~L. and {Batalha}, N.~E. and {Goyal}, J.~M. and {L{\'o}pez-Morales}, M. and {Lothringer}, J.~D. and {MacDonald}, R.~J. and {Moran}, S.~E. and {Stevenson}, K.~B. and {Wakeford}, H.~R. and {Espinoza}, N. and {Bean}, J.~L. and {Batalha}, N.~M. and {Benneke}, B. and {Berta-Thompson}, Z.~K. and {Crossfield}, I.~J.~M. and {Gao}, P. and {Kreidberg}, L. and {Powell}, D.~K. and {Cubillos}, P.~E. and {Gibson}, N.~P. and {Leconte}, J. and {Molaverdikhani}, K. and {Nikolov}, N.~K. and {Parmentier}, V. and {Roy}, P. and {Taylor}, J. and {Turner}, J.~D. and {Wheatley}, P.~J. and {Aggarwal}, K. and {Ahrer}, E. and {Alam}, M.~K. and {Alderson}, L. and {Allen}, N.~H. and {Banerjee}, A. and {Barat}, S. and {Barrado}, D. and {Barstow}, J.~K. and {Bell}, T.~J. and {Blecic}, J. and {Brande}, J. and {Casewell}, S. and {Changeat}, Q. and {Chubb}, K.~L. and {Crouzet}, N. and {Daylan}, T. and {Decin}, L. and {D{\'e}sert}, J. and {Mikal-Evans}, T. and {Feinstein}, A.~D. and {Flagg}, L. and {Fortney}, J.~J. and {Harrington}, J. and {Heng}, K. and {Hong}, Y. and {Hu}, R. and {Iro}, N. and {Kataria}, T. and {Kempton}, E.~M. -R. and {Krick}, J. and {Lendl}, M. and {Lillo-Box}, J. and {Louca}, A. and {Lustig-Yaeger}, J. and {Mancini}, L. and {Mansfield}, M. and {Mayne}, N.~J. and {Miguel}, Y. and {Morello}, G. and {Ohno}, K. and {Palle}, E. and {Petit dit de la Roche}, D.~J.~M. and {Rackham}, B.~V. and {Radica}, M. and {Ramos-Rosado}, L. and {Redfield}, S. and {Rogers}, L.~K. and {Shkolnik}, E.~L. and {Southworth}, J. and {Teske}, J. and {Tremblin}, P. and {Tucker}, G.~S. and {Venot}, O. and {Waalkes}, W.~C. and {Welbanks}, L. and {Zhang}, X. and {Zieba}, S.},
        title = "{Early Release Science of the exoplanet WASP-39b with JWST NIRSpec PRISM}",
      journal = {\nat},
         year = 2023,
        month = feb,
       volume = {614},
       number = {7949},
        pages = {659-663},
          doi = {10.1038/s41586-022-05677-y},
archivePrefix = {arXiv},
       eprint = {2211.10487},
 primaryClass = {astro-ph.EP},
       adsurl = {https://ui.adsabs.harvard.edu/abs/2023Natur.614..659R}
}

@ARTICLE{2015ApJ...809...93D,
       author = {{Dong}, Ruobing and {Zhu}, Zhaohuan and {Whitney}, Barbara},
        title = "{Observational Signatures of Planets in Protoplanetary Disks I. Gaps Opened by Single and Multiple Young Planets in Disks}",
      journal = {\apj},
         year = 2015,
        month = aug,
       volume = {809},
       number = {1},
          eid = {93},
        pages = {93},
          doi = {10.1088/0004-637X/809/1/93},
archivePrefix = {arXiv},
       eprint = {1411.6063},
 primaryClass = {astro-ph.EP},
       adsurl = {https://ui.adsabs.harvard.edu/abs/2015ApJ...809...93D}
}

@ARTICLE{2015ApJ...809L...5D,
       author = {{Dong}, Ruobing and {Zhu}, Zhaohuan and {Rafikov}, Roman R. and {Stone}, James M.},
        title = "{Observational Signatures of Planets in Protoplanetary Disks: Spiral Arms Observed in Scattered Light Imaging Can be Induced by Planets}",
      journal = {\apjl},
         year = 2015,
        month = aug,
       volume = {809},
       number = {1},
          eid = {L5},
        pages = {L5},
          doi = {10.1088/2041-8205/809/1/L5},
archivePrefix = {arXiv},
       eprint = {1507.03596},
 primaryClass = {astro-ph.EP},
       adsurl = {https://ui.adsabs.harvard.edu/abs/2015ApJ...809L...5D}
}

@ARTICLE{2015ApJ...815L..21F,
       author = {{Fung}, Jeffrey and {Dong}, Ruobing},
        title = "{Inferring Planet Mass from Spiral Structures in Protoplanetary Disks}",
      journal = {\apjl},
         year = 2015,
        month = dec,
       volume = {815},
       number = {2},
          eid = {L21},
        pages = {L21},
          doi = {10.1088/2041-8205/815/2/L21},
archivePrefix = {arXiv},
       eprint = {1511.01178},
 primaryClass = {astro-ph.EP},
       adsurl = {https://ui.adsabs.harvard.edu/abs/2015ApJ...815L..21F}
}

@ARTICLE{2017ApJ...835...38D,
       author = {{Dong}, Ruobing and {Fung}, Jeffrey},
        title = "{How Bright are Planet-induced Spiral Arms in Scattered Light?}",
      journal = {\apj},
         year = 2017,
        month = jan,
       volume = {835},
       number = {1},
          eid = {38},
        pages = {38},
          doi = {10.3847/1538-4357/835/1/38},
archivePrefix = {arXiv},
       eprint = {1612.00446},
 primaryClass = {astro-ph.EP},
       adsurl = {https://ui.adsabs.harvard.edu/abs/2017ApJ...835...38D}
}

@ARTICLE{2017ApJ...835..146D,
       author = {{Dong}, Ruobing and {Fung}, Jeffrey},
        title = "{What is the Mass of a Gap-opening Planet?}",
      journal = {\apj},
         year = 2017,
        month = feb,
       volume = {835},
       number = {2},
          eid = {146},
        pages = {146},
          doi = {10.3847/1538-4357/835/2/146},
archivePrefix = {arXiv},
       eprint = {1612.04821},
 primaryClass = {astro-ph.EP},
       adsurl = {https://ui.adsabs.harvard.edu/abs/2017ApJ...835..146D}
}

@INPROCEEDINGS{2023ASPC..534..423B,
       author = {{Bae}, J. and {Isella}, A. and {Zhu}, Z. and {Martin}, R. and {Okuzumi}, S. and {Suriano}, S.},
        title = "{Structured Distributions of Gas and Solids in Protoplanetary Disks}",
    booktitle = {},
         year = 2023,
       editor = {},
       series = {ASP Conference Series},
       volume = {534},
        month = jul,
        pages = {423},
          doi = {10.48550/arXiv.2210.13314},
archivePrefix = {arXiv},
       eprint = {2210.13314},
 primaryClass = {astro-ph.EP},
       adsurl = {https://ui.adsabs.harvard.edu/abs/2023ASPC..534..423B}
}

@INPROCEEDINGS{2021SPIE11823E..0FM,
       author = {{Morgan}, Rhonda and {Vlahakis}, Sophia K. and {Pogorelyuk}, Leonid and {Gubner}, Jenny and {Fitzgerald}, Riley and {Wang}, Sophia and {Cahoy}, Kerri},
        title = "{Planet matching and orbit determination in multi-planet systems for exoplanet direct imaging}",
    booktitle = {Proc.SPIE},
         year = 2021,
       editor = {},
       series = {},
       volume = {11823},
        month = sep,
          eid = {118230F},
        pages = {118230F},
          doi = {10.1117/12.2594998},
       adsurl = {https://ui.adsabs.harvard.edu/abs/2021SPIE11823E..0FM}
}

@ARTICLE{2024arXiv240102039P,
       author = {{Plavchan}, Peter and {Berberian}, Jr, John E. and {Kane}, Stephen R and {Morgan}, Rhonda and {Peretz}, Eliad and {Economon}, Sophia},
        title = "{Analytic relations assessing the impact of precursor knowledge and key mission parameters on direct imaging survey yield}",
      journal = {arXiv e-prints},
         year = 2024,
        month = jan,
          eid = {arXiv:2401.02039},
        pages = {arXiv:2401.02039},
          doi = {10.48550/arXiv.2401.02039},
archivePrefix = {arXiv},
       eprint = {2401.02039},
 primaryClass = {astro-ph.EP},
       adsurl = {https://ui.adsabs.harvard.edu/abs/2024arXiv240102039P}
}

@ARTICLE{2025arXiv250916761B,
       author = {{Bao}, Chunhui and {Ji}, Jianghui and {Zhao}, Gang and {Zhu}, Yiming and {Dou}, Jiangpei and {Wang}, Su and {Dong}, Yao},
        title = "{Direct Imaging for the Debris Disk around $ε$ Eridani with the Cool-Planet Imaging Coronagraph}",
      journal = {arXiv e-prints},
         year = 2025,
        month = sep,
          eid = {arXiv:2509.16761},
        pages = {arXiv:2509.16761},
          doi = {10.48550/arXiv.2509.16761},
archivePrefix = {arXiv},
       eprint = {2509.16761},
 primaryClass = {astro-ph.EP},
       adsurl = {https://ui.adsabs.harvard.edu/abs/2025arXiv250916761B}
}

@ARTICLE{2008Sci...322.1345K,
       author = {{Kalas}, Paul and {Graham}, James R. and {Chiang}, Eugene and {Fitzgerald}, Michael P. and {Clampin}, Mark and {Kite}, Edwin S. and {Stapelfeldt}, Karl and {Marois}, Christian and {Krist}, John},
        title = "{Optical Images of an Exosolar Planet 25 Light-Years from Earth}",
      journal = {Science},
         year = 2008,
        month = nov,
       volume = {322},
       number = {5906},
        pages = {1345},
          doi = {10.1126/science.1166609},
archivePrefix = {arXiv},
       eprint = {0811.1994},
 primaryClass = {astro-ph},
       adsurl = {https://ui.adsabs.harvard.edu/abs/2008Sci...322.1345K}
}

@ARTICLE{2011ApJ...741...55S,
       author = {{Soummer}, R{\'e}mi and {Hagan}, J. Brendan and {Pueyo}, Laurent and {Thormann}, Adrien and {Rajan}, Abhijith and {Marois}, Christian},
        title = "{Orbital Motion of HR 8799 b, c, d Using Hubble Space Telescope Data from 1998: Constraints on Inclination, Eccentricity, and Stability}",
      journal = {\apj},
         year = 2011,
        month = nov,
       volume = {741},
       number = {1},
          eid = {55},
        pages = {55},
          doi = {10.1088/0004-637X/741/1/55},
archivePrefix = {arXiv},
       eprint = {1110.1382},
 primaryClass = {astro-ph.EP},
       adsurl = {https://ui.adsabs.harvard.edu/abs/2011ApJ...741...55S}
}

@ARTICLE{2022A&A...667A.165B,
       author = {{Boccaletti}, A. and {Cossou}, C. and {Baudoz}, P. and {Lagage}, P.~O. and {Dicken}, D. and {Glasse}, A. and {Hines}, D.~C. and {Aguilar}, J. and {Detre}, O. and {Nickson}, B. and {Noriega-Crespo}, A. and {G{\'a}sp{\'a}r}, A. and {Labiano}, A. and {Stark}, C. and {Rouan}, D. and {Reess}, J.~M. and {Wright}, G.~S. and {Rieke}, G. and {Garcia Marin}, M. and {Lajoie}, C. and {Girard}, J. and {Perrin}, M. and {Soummer}, R. and {Pueyo}, L.},
        title = "{JWST/MIRI coronagraphic performances as measured on-sky}",
      journal = {\aap},
         year = 2022,
        month = nov,
       volume = {667},
          eid = {A165},
        pages = {A165},
          doi = {10.1051/0004-6361/202244578},
archivePrefix = {arXiv},
       eprint = {2207.11080},
 primaryClass = {astro-ph.IM},
       adsurl = {https://ui.adsabs.harvard.edu/abs/2022A&A...667A.165B}
}

@INPROCEEDINGS{2007SPIE.6693E..0HK,
       author = {{Krist}, John E. and {Beichman}, Charles A. and {Trauger}, John T. and {Rieke}, Marcia J. and {Somerstein}, Steve and {Green}, Joseph J. and {Horner}, Scott D. and {Stansberry}, John A. and {Shi}, Fang and {Meyer}, Michael R. and {Stapelfeldt}, Karl R. and {Roellig}, Thomas L.},
        title = "{Hunting planets and observing disks with the JWST NIRCam coronagraph}",
    booktitle = {Proc.SPIE},
         year = 2007,
       editor = {},
       series = {},
       volume = {6693},
        month = sep,
          eid = {66930H},
        pages = {66930H},
          doi = {10.1117/12.734873},
       adsurl = {https://ui.adsabs.harvard.edu/abs/2007SPIE.6693E..0HK}
}

@INPROCEEDINGS{2020SPIE11443E..1UK,
       author = {{Kasdin}, N. Jeremy and {Bailey}, Vanessa P. and {Mennesson}, Bertrand and {Zellem}, Robert T. and {Ygouf}, Marie and {Rhodes}, Jason and {Luchik}, Thomas and {Zhao}, Feng and {Riggs}, A.~J. Eldorado and {Seo}, Byoung-Joon and {Krist}, John and {Kern}, Brian and {Tang}, Hong and {Nemati}, Bijan and {Groff}, Tyler D. and {Zimmerman}, Neil and {Macintosh}, Bruce and {Turnbull}, Margaret and {Debes}, John and {Douglas}, Ewan S. and {Lupu}, Roxana E.},
        title = "{The Nancy Grace Roman Space Telescope Coronagraph Instrument (CGI) technology demonstration}",
    booktitle = {Proc.SPIE},
         year = 2020,
       editor = {},
       series = {},
       volume = {11443},
        month = dec,
          eid = {114431U},
        pages = {114431U},
          doi = {10.1117/12.2562997},
archivePrefix = {arXiv},
       eprint = {2103.01980},
 primaryClass = {astro-ph.IM},
       adsurl = {https://ui.adsabs.harvard.edu/abs/2020SPIE11443E..1UK}
}

@ARTICLE{2020AJ....159..177E,
       author = {{Ertel}, S. and {Defr{\`e}re}, D. and {Hinz}, P. and {Mennesson}, B. and {Kennedy}, G.~M. and {Danchi}, W.~C. and {Gelino}, C. and {Hill}, J.~M. and {Hoffmann}, W.~F. and {Mazoyer}, J. and {Rieke}, G. and {Shannon}, A. and {Stapelfeldt}, K. and {Spalding}, E. and {Stone}, J.~M. and {Vaz}, A. and {Weinberger}, A.~J. and {Willems}, P. and {Absil}, O. and {Arbo}, P. and {Bailey}, V.~P. and {Beichman}, C. and {Bryden}, G. and {Downey}, E.~C. and {Durney}, O. and {Esposito}, S. and {Gaspar}, A. and {Grenz}, P. and {Haniff}, C.~A. and {Leisenring}, J.~M. and {Marion}, L. and {McMahon}, T.~J. and {Millan-Gabet}, R. and {Montoya}, M. and {Morzinski}, K.~M. and {Perera}, S. and {Pinna}, E. and {Pott}, J.-U. and {Power}, J. and {Puglisi}, A. and {Roberge}, A. and {Serabyn}, E. and {Skemer}, A.~J. and {Su}, K.~Y.~L. and {Vaitheeswaran}, V. and {Wyatt}, M.~C.},
        title = "{The HOSTS Survey for Exozodiacal Dust: Observational Results from the Complete Survey}",
      journal = {\aj},
         year = 2020,
        month = apr,
       volume = {159},
       number = {4},
          eid = {177},
        pages = {177},
          doi = {10.3847/1538-3881/ab7817},
archivePrefix = {arXiv},
       eprint = {2003.03499},
 primaryClass = {astro-ph.SR},
       adsurl = {https://ui.adsabs.harvard.edu/abs/2020AJ....159..177E}
}

@INPROCEEDINGS{2010SPIE.7734E..0LA,
       author = {{Absil}, O. and {Defr{\`e}re}, Denis and {Roberge}, A. and {Augereau}, J.-C. and {Coud{\'e} du Foresto}, V. and {Hanot}, C. and {Stark}, C. and {Surdej}, J.},
        title = "{Direct imaging of Earth-like planets: why we care about exozodis}",
    booktitle = {Proc.SPIE},
         year = 2010,
       editor = {},
       series = {},
       volume = {7734},
        month = jul,
          eid = {77340L},
        pages = {77340L},
          doi = {10.1117/12.858257},
       adsurl = {https://ui.adsabs.harvard.edu/abs/2010SPIE.7734E..0LA}
}

@ARTICLE{2023ApJ...951L..20C,
       author = {{Carter}, Aarynn L. and {Hinkley}, Sasha and {Kammerer}, Jens and {Skemer}, Andrew and {Biller}, Beth A. and {Leisenring}, Jarron M. and {Millar-Blanchaer}, Maxwell A. and {Petrus}, Simon and {Stone}, Jordan M. and {Ward-Duong}, Kimberly and {Wang}, Jason J. and {Girard}, Julien H. and {Hines}, Dean C. and {Perrin}, Marshall D. and {Pueyo}, Laurent and {Balmer}, William O. and {Bonavita}, Mariangela and {Bonnefoy}, Mickael and {Chauvin}, Gael and {Choquet}, Elodie and {Christiaens}, Valentin and {Danielski}, Camilla and {Kennedy}, Grant M. and {Matthews}, Elisabeth C. and {Miles}, Brittany E. and {Patapis}, Polychronis and {Ray}, Shrishmoy and {Rickman}, Emily and {Sallum}, Steph and {Stapelfeldt}, Karl R. and {Whiteford}, Niall and {Zhou}, Yifan and {Absil}, Olivier and {Boccaletti}, Anthony and {Booth}, Mark and {Bowler}, Brendan P. and {Chen}, Christine H. and {Currie}, Thayne and {Fortney}, Jonathan J. and {Grady}, Carol A. and {Greebaum}, Alexandra Z. and {Henning}, Thomas and {Hoch}, Kielan K.~W. and {Janson}, Markus and {Kalas}, Paul and {Kenworthy}, Matthew A. and {Kervella}, Pierre and {Kraus}, Adam L. and {Lagage}, Pierre-Olivier and {Liu}, Michael C. and {Macintosh}, Bruce and {Marino}, Sebastian and {Marley}, Mark S. and {Marois}, Christian and {Matthews}, Brenda C. and {Mawet}, Dimitri and {McElwain}, Michael W. and {Metchev}, Stanimir and {Meyer}, Michael R. and {Molliere}, Paul and {Moran}, Sarah E. and {Morley}, Caroline V. and {Mukherjee}, Sagnick and {Pantin}, Eric and {Quirrenbach}, Andreas and {Rebollido}, Isabel and {Ren}, Bin B. and {Schneider}, Glenn and {Vasist}, Malavika and {Worthen}, Kadin and {Wyatt}, Mark C. and {Briesemeister}, Zackery W. and {Bryan}, Marta L. and {Calissendorff}, Per and {Cantalloube}, Faustine and {Cugno}, Gabriele and {De Furio}, Matthew and {Dupuy}, Trent J. and {Factor}, Samuel M. and {Faherty}, Jacqueline K. and {Fitzgerald}, Michael P. and {Franson}, Kyle and {Gonzales}, Eileen C. and {Hood}, Callie E. and {Howe}, Alex R. and {Kuzuhara}, Masayuki and {Lagrange}, Anne-Marie and {Lawson}, Kellen and {Lazzoni}, Cecilia and {Lew}, Ben W.~P. and {Liu}, Pengyu and {Llop-Sayson}, Jorge and {Lloyd}, James P. and {Martinez}, Raquel A. and {Mazoyer}, Johan and {Palma-Bifani}, Paulina and {Quanz}, Sascha P. and {Redai}, Jea Adams and {Samland}, Matthias and {Schlieder}, Joshua E. and {Tamura}, Motohide and {Tan}, Xianyu and {Uyama}, Taichi and {Vigan}, Arthur and {Vos}, Johanna M. and {Wagner}, Kevin and {Wolff}, Schuyler G. and {Ygouf}, Marie and {Zhang}, Xi and {Zhang}, Keming and {Zhang}, Zhoujian},
        title = "{The JWST Early Release Science Program for Direct Observations of Exoplanetary Systems I: High-contrast Imaging of the Exoplanet HIP 65426 b from 2 to 16 {\ensuremath{\mu}}m}",
      journal = {\apjl},
         year = 2023,
        month = jul,
       volume = {951},
       number = {1},
          eid = {L20},
        pages = {L20},
          doi = {10.3847/2041-8213/acd93e},
archivePrefix = {arXiv},
       eprint = {2208.14990},
 primaryClass = {astro-ph.EP},
       adsurl = {https://ui.adsabs.harvard.edu/abs/2023ApJ...951L..20C}
}

@ARTICLE{2025arXiv250905880B,
       author = {{Bamberger}, Daniel},
        title = "{A preliminary orbit for the satellite of dwarf planet (136472) Makemake}",
      journal = {arXiv e-prints},
         year = 2025,
        month = sep,
          eid = {arXiv:2509.05880},
        pages = {arXiv:2509.05880},
          doi = {10.48550/arXiv.2509.05880},
archivePrefix = {arXiv},
       eprint = {2509.05880},
 primaryClass = {astro-ph.EP},
       adsurl = {https://ui.adsabs.harvard.edu/abs/2025arXiv250905880B}
}

@ARTICLE{2025ApJ...991L..34P,
       author = {{Protopapa}, Silvia and {Wong}, Ian and {Lellouch}, Emmanuel and {Johnson}, Perianne E. and {Grundy}, William M. and {Glein}, Christopher R. and {M{\"u}ller}, Thomas and {Kiss}, Csaba and {Emery}, Joshua P. and {Brunetto}, Rosario and {Holler}, Bryan J. and {Parker}, Alex H. and {Stansberry}, John A. and {Hammel}, Heidi B. and {Milam}, Stefanie N. and {Guilbert-Lepoutre}, Aur{\'e}lie and {Santos-Sanz}, Pablo and {Pinilla-Alonso}, Noem{\'\i}},
        title = "{JWST Detection of Hydrocarbon Ices and Methane Gas on Makemake}",
      journal = {\apjl},
         year = 2025,
        month = oct,
       volume = {991},
       number = {2},
          eid = {L34},
        pages = {L34},
          doi = {10.3847/2041-8213/adfe63},
archivePrefix = {arXiv},
       eprint = {2509.06772},
 primaryClass = {astro-ph.EP},
       adsurl = {https://ui.adsabs.harvard.edu/abs/2025ApJ...991L..34P}
}

@ARTICLE{2025arXiv251109862Z,
       author = {{Zhu}, Yi-Ming and {Zhao}, Gang and {Dou}, Jiang-Pei and {Lv}, Zhong-Hua and {Chen}, Yi-Li and {Ma}, Bo and {Yan}, Zhao-Jun and {Tang}, Jing and {Li}, Ran},
        title = "{Mock Observations for the CSST Mission: CPI-C -- Targets for High Contrast Imaging}",
      journal = {arXiv e-prints},
         year = 2025,
        month = nov,
          eid = {arXiv:2511.09862},
        pages = {arXiv:2511.09862},
          doi = {10.48550/arXiv.2511.09862},
archivePrefix = {arXiv},
       eprint = {2511.09862},
 primaryClass = {astro-ph.IM},
       adsurl = {https://ui.adsabs.harvard.edu/abs/2025arXiv251109862Z},
       url={http://iopscience.iop.org/article/10.1088/1674-4527/ae2100}
}

@ARTICLE{2025arXiv251108886Z,
       author = {{Zhao}, Gang and {Zhu}, Yiming and {Dou}, Jiangpei and {Chen}, Yili and {Lv}, Zhonghua and {Niu}, Bingli and {Yan}, Zhaojun and {Ma}, Bo and {Li}, Ran},
        title = "{Mock Observations for the CSST Mission: CPI-C -- Instrument Simulation}",
      journal = {arXiv e-prints},
         year = 2025,
        month = nov,
          eid = {arXiv:2511.08886},
        pages = {arXiv:2511.08886},
          doi = {10.48550/arXiv.2511.08886},
archivePrefix = {arXiv},
       eprint = {2511.08886},
 primaryClass = {astro-ph.IM},
       adsurl = {https://ui.adsabs.harvard.edu/abs/2025arXiv251108886Z},
       url={https://iopscience.iop.org/article/10.1088/1674-4527/ae2102}
}

@ARTICLE{2024ChJSS..44..193J,
       author = {{Ji}, Jiang-Hui and {Li}, Hai-Tao and {Zhang}, Jun-Bo and {Li}, Dong and {Fang}, Liang and {Wang}, Su and {Deng}, Lei and {Chen}, Guo and {Li}, Fei and {Dong}, Yao and {Li}, Baoquan and {Gao}, Xiaodong and {Xian}, Hao},
        title = "{Closeby Habitable Exoplanet Survey (CHES): an Astrometry Mission for Probing Nearby Habitable Planets}",
      journal = {CJSS},
         year = 2024,
        month = mar,
       volume = {44},
       number = {2},
        pages = {193-214},
          doi = {10.11728/cjss2024.02.yg03},
       adsurl = {https://ui.adsabs.harvard.edu/abs/2024ChJSS..44..193J}
}

\label{lastpage}

\end{document}